\documentclass[a4paper]{article}
\pdfoutput=1
\usepackage[round]{natbib}
\usepackage{amsmath,amsthm,amssymb}
\usepackage{float}
\usepackage{physics}
\usepackage{graphicx}
\usepackage{setspace}
\usepackage{subcaption}
\usepackage{fullpage}
\usepackage{authblk}
\usepackage{lineno}
\usepackage{hyperref}
\hypersetup{
    colorlinks=true,
    citecolor=blue,
}

\begin{document}
\title{Ability of a pore network model to predict fluid flow and drag in saturated granular materials}
\author[1]{Adnan Sufian\footnote{Corresponding author, ORCid: \href{https://orcid.org/0000-0003-2816-3250}{0000-0003-2816-3250}}}
\author[1]{Chris Knight}
\author[1]{Catherine O'Sullivan} 
\author[2]{Berend van Wachem}
\author[3]{Daniele Dini}
\affil[1]{Department of Civil and Environmental Engineering, Imperial College London, London, United Kingdom}
\affil[2]{Chair of Mechanical Process Engineering, Otto-von-Guericke-Universität Magdeburg, Magdeburg, Germany}
\affil[3]{Department of Mechanical Engineering, Imperial College London, London, United Kingdom}
\date{\today}
\setcounter{Maxaffil}{0}
\renewcommand\Affilfont{\itshape\small}

\maketitle

\begin{abstract}
The local flow field and seepage induced drag obtained from Pore Network Models (PNM) is compared to Immersed Boundary Method (IBM) simulations, for a range of linear graded and bimodal samples. PNM were generated using a weighted Delaunay Tessellation (DT), along with the Modified Delaunay Tessellation (MDT) which considers the merging of tetrahedral Delaunay cells. The local pressure field was very accurately captured in all linear graded and bimodal samples. Local flux (flow rate) exhibited more scatter, but the PNM based on the MDT clearly provided a better correlation with the IBM. There was close similarity in the network shortest paths obtained from PNM and IBM, indicating that the PNM captures dominant flow channels. Further, by overlaying the PNM on a streamline profile, it was demonstrated that local pressure drops coincided with the pore constrictions. A rigorous validation was undertaken for the drag force calculated from the PNM by comparing with analytical solutions for ordered array of spheres. This method was subsequently applied to all linear graded and bimodal samples, and the calculated force was compared with the IBM data. Linear graded samples were able to calculate the force with reasonable accuracy, while the bimodal sample exhibited slightly more scatter.
\end{abstract}


\newcommand{\uf}{\mathbf{u}}
\newcommand{\Fp}{\mathbf{F}}
\newcommand{\Fm}{\mathrm{F}}
\newcommand{\DT}{\mathrm{DT}}
\newcommand{\MDT}{\mathrm{MDT}}

\section{Introduction}
A detailed understanding of the fluid flow at the level of the individual particles and the drag induced by the seepage flow in saturated sands is important in applications including sand production in hydrocarbon extraction \citep{climent2014}, filter design \citep{huang2014}, piping erosion \citep{tao2017} and suffusion \citep{kawano2018,hosn2018} in dams and flood embankments and fundamental studies of liquefaction \citep{elshamy2010}. The topology of the void space between the particles can now be determined relatively accurately using micro-computed tomography. Moreover, model soils can be developed using Discrete Element Method (DEM) simulations, from which particle position data and contact locations are readily available. This topology data can be used to develop fully resolved flow simulations using Computational Fluid Dynamics (CFD), in which the fluid motion is described at a scale that is smaller than the individual particles, with particle surfaces considered as boundaries to the fluid flow. In these simulations, the Navier-Stokes equations can be solved using the finite element method \citep{glowinski2001}, the immersed boundary method \citep{peskin2002,mittal2005}, or the lattice Boltzmann method \citep{chen1998,succi2001}. However, the high resolution fluid flow and drag data associated with these simulations result in a high computational cost. Consequently, there have been relatively few documented studies within geomechanics using these fully resolved coupled models. Coupled, coarse-grained, unresolved simulations following the \citet{anderson1967} formulation are more tractable \citep{kawaguchi1998,kafui2002,zeghal2004}. Typically, these unresolved simulations use CFD to model the fluid phase coupled with DEM to model the solid phase, such that the fluid phase is discretised at a scale that is of the order of several particle diameters. However, such unresolved CFD-DEM cannot accurately capture local flow field effects and seepage induced drag, as quantities are averaged over a fluid cell comprising several particles. Furthermore, the drag coefficients commonly used in unresolved CFD-DEM in soil mechanics research do not account for the effect of polydispersity which are known to influence the fluid-particle interaction force \citep{vanderhoef2005}. 

Pore Network Models (PNM) present an alternate numerical technique to model fluid flow through saturated sands. In these models, the nodes representing individual pores are connected by edges representing the narrow constrictions or pore throats. Conceptually, pore network models resolve the fluid flow at length-scale that is intermediate between the fully resolved and unresolved simulations approaches mentioned above. They have been employed in a wide range of fields including petroleum engineering \citep{silin2006,thompson2008}, nuclear waste disposal \citep{xiong2015}, ionic diffusion \citep{mohajeri2010} and carbon dioxide sequestration \citep{li2006}. Within soil mechanics and geotechnical engineering, PNM have been employed to explore single-phase permeability \citep{bryant1993,vanderlinden2018}, unsaturated soil characteristics and fluid retention curves \citep{held2001,ferraro2017} and filtration properties \citep{shire2017}. The seepage induced drag force can also be obtained from PNM using the method developed in \citet{chareyre2012}.

While PNM are a computationally efficient approach to model fluid flow and fluid-particle interactions, they greatly simplify the geometry of the pore space and the complex physics associated with flow through a granular, porous medium \citep{miao2017}.  Network models are typically verified by comparing the macroscopic permeability with experimental data \citep{bryant1993} or fully resolved numerical simulations \citep{vanderlinden2018}. The ability of PNM to capture local flow field effects has only been considered in a few studies. \citet{helland2008} and \citet{miao2017} compared local conductivity in PNM against finite element simulations using rock imaging data, while the lattice-Boltzmann method was employed by \citet{sholokhova2009} for a sandstone sample. However, these studies were limited to considering local conductivity in isolated pore constrictions, and hence, did not capture the influence of neighbouring pores and constrictions on the local flow field behaviour. Low permeability rock samples were considered rather than granular ``unconsolidated'' sand samples. Moreover, PNM simplifications affect the ability to predict the seepage induced drag forces. The method proposed in \citet{chareyre2012} was verified against finite element simulations for a simple cubic unit cell (comprising 8 particles) with an additional particle of variable size in the centre, while \citet{catalano2014} extended the verification by comparing with the analytical solution for one-dimensional consolidation. Nevertheless, a more detailed validation of this method is required to establish whether PNM can accurately capture seepage induced drag. Therefore, this study aims to address the following key questions:
\begin{enumerate}
    \item How accurately can a pore network model predict the local pressures at individual pores and the fluxes through individual constrictions across a broad range of packing fractions?  
    \item How sensitive are the results to the choice of local conductance models and the way the void space is partitioned?
    \item How accurately can existing pore network models predict the seepage induced drag on individual particles?
\end{enumerate}

The study compares the data generated from PNM with data from fully-resolved numerical simulations employing the Immersed Boundary Method (IBM) for a wide range of linearly graded and bimodal samples. The IBM simulations on entire samples capture the effect of neighbouring pores and constrictions, which is absent in prior studies. Further, the seepage induced drag determined from PNM is compared with analytical solutions for various ordered packings (including simple cubic, body centred cubic and face centred cubic arrays) provided in \citet{zickhomsy1982}. This analytical validation, in conjunction with the comparative study of drag forces obtained from IBM, provided much stronger support for the use of PNM to model fluid-particle systems.

\section{Immersed Boundary Method}

The equations of motion for an incompressible fluid in a multi-phase (fluid-particle) system are given by the Navier-Stokes equations:
\begin{equation} \label{eqn:mass-conservation}
    \nabla \cdot \uf = 0
\end{equation}
\begin{equation} \label{eqn:momentum-conservation}
    \pdv{\uf}{t} + \left( \uf \cdot \nabla \right) \uf = -\frac{1}{\rho} \nabla p + \nu \nabla^2\uf + \mathbf{f}_{fp}
\end{equation}
where $\uf$ is the fluid velocity, $p$ is the pressure, $\mathbf{f}_{fp}$ is the force to account for the presence of the local particle surface segments, $\rho$ is the fluid density and $\nu$ is the kinematic viscosity. These equations express the mass and momentum conservation of the fluid.

The solution to Eqs. \ref{eqn:mass-conservation}-\ref{eqn:momentum-conservation} can be obtained using Computational Fluid Dynamics (CFD), a set of methods which numerically solves the Navier-Stokes equations by discretising the continuous fluid domain into a contiguous space-filling set of finite-volume fluid cells. The equations of motion for fluid cells form a system of non-linear equations, which are solved for the fluid velocity, $\uf$, and the pressure, $p$, at each fluid cell within the domain. Appropriate boundary conditions must be specified for fluid velocity and pressure at the domain boundaries, along with initial values for each fluid cell.

\begin{figure}[ht]
    \centering
    \includegraphics[width=0.7\textwidth]{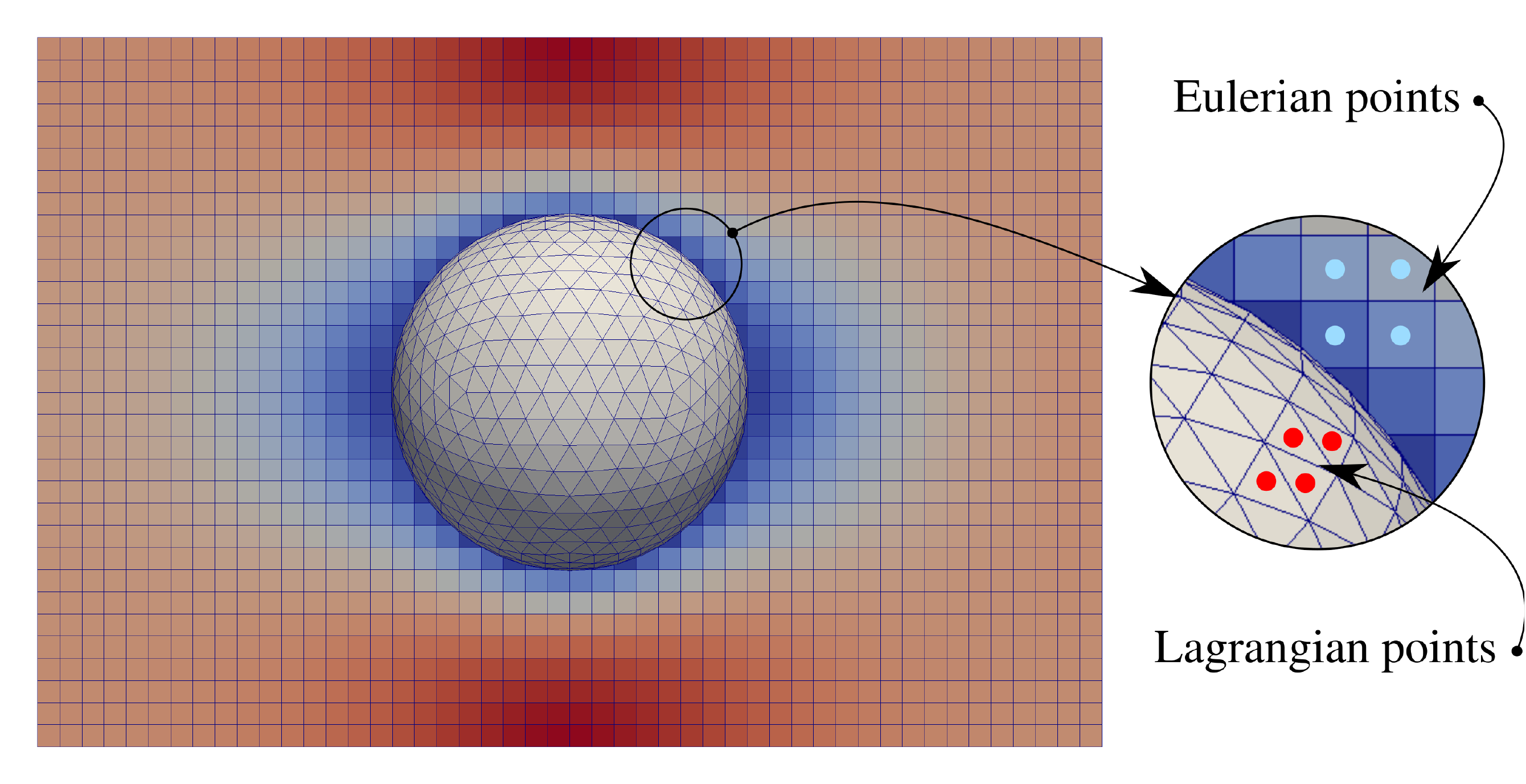}
    \caption{\label{fig:ibm-lag-eul-mesh}Fluid velocity field around a sphere at low $Re$ produced by the IBM, where red indicates areas of high velocity. The mesh of Lagrangian points on the surface of the sphere is immersed in the Cartesian Eulerian mesh of fluid cells.}
\end{figure}

The Immersed Boundary Method (IBM) is a class of CFD methods in which the fluid equations are modified by the introduction of additional boundary conditions inside the fluid domain. Points representing the surfaces of solid objects (i.e. particle surfaces) are inserted within the fluid domain and no-slip, no-penetration boundary conditions are imposed by the addition of source terms representing the fluid-particle interaction force, $\mathbf{f}_{fp}$, in Eq. \ref{eqn:momentum-conservation}. This approach enables simulation of flow through the pore space in granular, porous media. In the IBM, the fluid mesh grid spacing is typically at least an order of magnitude smaller than the individual particles to enable the local flow field around the particles to be captured. The points defining the fluid cells are referred to as Eulerian points, as they remain fixed in space as fluid flows between them, while the mesh points representing particle surfaces are referred to as Lagrangian points, as they track the motion of the particles across the Eulerian fluid mesh. An illustration of the Lagrangian and Eulerian meshes is shown in Fig.~\ref{fig:ibm-lag-eul-mesh}.

This study used \texttt{MultiFlow} \citep{denner2014}, a fully-coupled fluid solver which solves the Navier-Stokes equations using the iterative biconjugate gradient stabilised method \citep{vandervorst1992} with a collocated variable arrangement. In this study, all IBM simulations were conducted with structured Cartesian grids. A multi-direct forcing type IBM is implemented in \texttt{MultiFlow} similar to that described by \citet{Luo2007}. Assuming the particle properties are known at step $N$, at the next fluid time step, $N+1$, the velocity of Lagrangian point $l$ at coordinates $\mathbf{X}^l_i$ on the surface of particle $i$ whose centroid is at a position $\mathbf{x}^p_i$ is given by
\begin{equation}
  \mathbf{U}^{N+1}(\mathbf{X}^{l}_{i}) = \mathbf{v}_{i} + \mathbf{\omega}_{i}\times(\mathbf{X}^{l}_{i} - \mathbf{x}^{p}_{i})
  \label{eqn:object-surface-vel}
\end{equation}
where $\mathbf{v}_i$ and $\mathbf{\omega}_i$ are the translational and rotational velocity vectors of particle $i$. In order to satisfy the no-slip condition at the next fluid time step, the force from Lagrangian point $l$ to be applied to the neighbouring fluid at time step $N$ is
\begin{equation}
  \mathbf{F}^{N}_{fp}(\mathbf{X}^{l}_{i}) = \frac{1}{\Delta{t}}\left[\mathbf{U}^{N+1}(\mathbf{X}^{l}_{i}) - \mathbf{\hat{U}}^{N}(\mathbf{X}^{l}_{i})\right]
  \label{eqn:Lag-force-intermediate-vel}
\end{equation}
where $\mathbf{\hat{U}}^{N}(\mathbf{X}^{l}_{i})$ is an intermediate fluid velocity at $\mathbf{X}^{l}_{i}$ and time step $N$ which satisfies Eq. \ref{eqn:momentum-conservation} in the absence of the forcing due to the solid phase. The use of upper case symbols in Eq. \ref{eqn:object-surface-vel} and \ref{eqn:Lag-force-intermediate-vel} indicates variables defined at the Lagrangian points. Fluid velocities must be interpolated from the Eulerian mesh to the Lagrangian points such that
\begin{equation}
  \mathbf{\hat{U}}^{N}(\mathbf{X}^{l}_{i}) = \sum_{\mathbf{x}_{m}\in\Omega}\mathbf{\hat{u}}^{N}(\mathbf{x}_{m})\delta_{h}(\mathbf{X}^{l}_{i} - \mathbf{x}_{m})\Delta{x}\Delta{y}\Delta{z},
  \label{eqn:interpolate-velocity}
\end{equation}
where $\mathbf{x}_m$ is the position of Eulerian cell $m$, $\delta_{h}$ is the regularised Dirac delta function (DDF) of \citet{Roma1999}, and $\Delta x$, $\Delta y$, and $\Delta z$ are the Eulerian mesh spacings in the $x$, $y$, and $z$ directions. Correspondingly, forces calculated at the Lagrangian points must be distributed across all Eulerian cells within the support region of each point, which is three Eulerian cells wide for the DDF, such that
\begin{equation}
  \mathbf{f}^{N}_{fp}(\mathbf{x}_{m}) = \sum^{N_{p}}_{i=1}\sum^{N_{L}}_{l=1}\mathbf{F}^{N}_{fp}(\mathbf{X}^{l}_{i})\delta_{h}(\mathbf{X}^{l}_{i} - \mathbf{x}_{m})\Delta{V}^{l}_{i},
  \label{eqn:spread-forces}
\end{equation}
where the sums are taken over the total number of immersed objects, $N_{P}$, and the total number of Lagrangian points, $N_{L}$, on each object and $\Delta{V}^{l}_{i}$ is the volume of the Lagrangian point. $\Delta{V}^{l}_{i}$ is related to the Lagrangian point area density and should match closely with the Eulerian cell volume for correct application of the DDF.

The requirement of no-slip between the particle surface and neighbouring fluid dictates that the fluid velocity at each Lagrangian point be equal to the rigid body velocity of the point as given by Eq. \ref{eqn:object-surface-vel}. Due to overlapping of the support regions, each fluid cell is influenced by multiple Lagrangian points and hence the values of $\uf$ obtained at time step $N+1$ will not satisfy the no-slip condition well. Multiple rounds of forcing are applied until the velocity field at time step $N+1$ satisfies the no-slip condition to within a specified tolerance as measured by the difference between the target and interpolated fluid velocities at each Lagrangian point. The interpolation and spreading of variables also causes an enhancement of particle size from the perspective of the fluid, which leads to an overprediction of $\mathbf{f}_{fp}$. The radius retraction scheme \citep{Breugem2012} is used to mitigate this effect.

\subsection{Sample Generation}

Samples were generated by producing a non-interacting cloud of spherical particles with a specified particle size distribution (PSD) and subsequently isotropically compressed to a target confining pressure in a cubic periodic cell. The particle coordinates following compression were used to construct the input for the IBM and PNM simulations.

\begin{figure}[ht]
    \centering
    \includegraphics{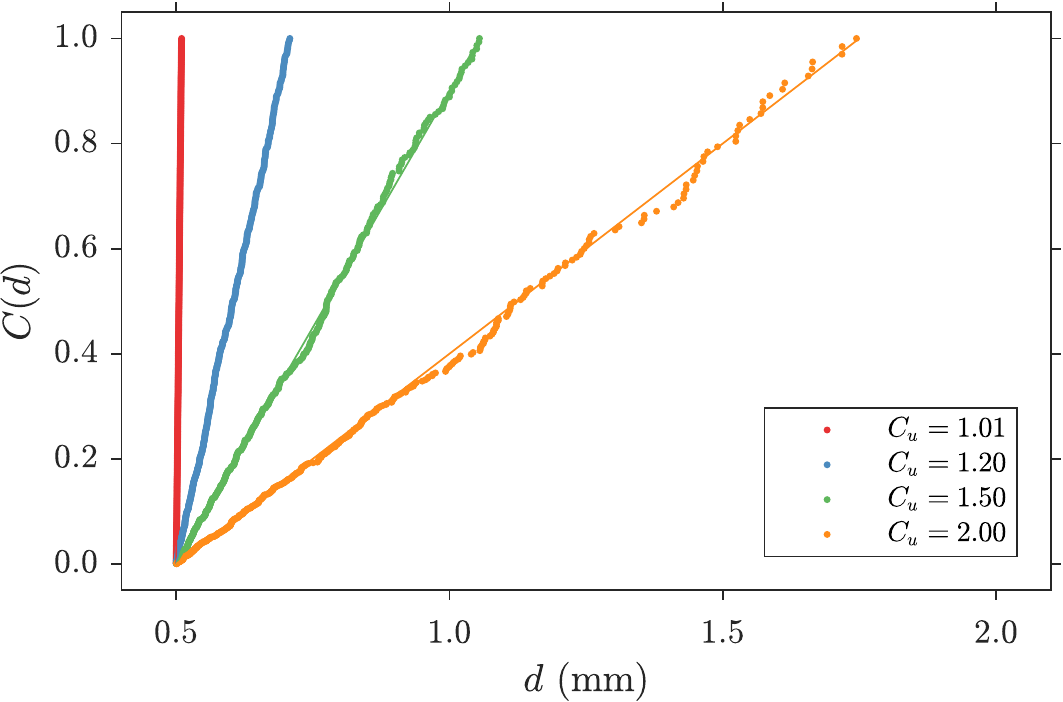}
    \caption{\label{fig:particle-size-dist}Particle size distribution for the linear graded samples considered in this study. Solid lines represent the target gradation.}
\end{figure}

This study considered linear graded samples with coefficient of uniformity, $C_u = 1.01, \: 1.20, \: 1.50, \: 2.00$ (Fig.~\ref{fig:particle-size-dist}) and bimodal samples with a particle size ratio of maximum to minimum diameter of $\chi = 1.99, \: 3.97$ and various fines fractions (volume fraction of smaller particles) of $\psi \approx 0.10, \: 0.25, \:0.35, \: 0.50$. In all cases, the minimum particle diameter was taken as $d_{min} = 5.0 \times 10^{-4}$m. A Monte-Carlo procedure was implemented in \texttt{MultiFlow} to generate the non-interacting particle assembly. The PSD was split into several size classes, and the number of particles and their diameters for each size class was calculated to match the target PSD. Particle diameters were distributed uniformly within each size class. Particles were placed within the domain in order of decreasing diameters, as the probability of placing larger particles is decreased if the space is already occupied by the smaller particles. Trial positions are only accepted if there were no overlap with other particles.

After the non-interacting particle assembly was generated, each sample was isotropically compressed (with no inter-particle friction) to a confining pressure of 100kPa using servo-controlled periodic boundaries and the \texttt{MultiFlow} DEM functionality. Snapshots were taken throughout the isotropic compression phase to obtain six configurations for each linear graded samples and two configurations for each bimodal sample. For the linear graded samples a wide range of packing fractions, $\phi = \frac{V_s}{V}$ ($V_s$ is the total particle volume and $V$ is the total sample volume), of interest to both chemical engineering and geotechnical engineering applications was considered. The least dense cases $\left( \phi \approx 0.40 \right)$ correspond to heavily laden particle suspensions found in conveying and materials processing applications, whilst the most dense cases $\left( \phi \approx 0.65 \right)$ correspond to percolating stress transmitting networks encountered in soil mechanics. For each sample, similar density values were considered to enable direct comparison. Table \ref{tab:linear-samples} lists the co-efficient of uniformity $\left( C_u \right)$, the ratio of maximum to minimum particle diameter $\left( \chi = d_{max}/d_{min} \right)$, the number of particles $\left( N_p \right)$, packing fraction $\left( \phi \right)$, and void ratio $\left( e \right)$, of the linear graded samples considered in this study. The bimodal samples focused on stress percolating assemblies at two different densities in order to vary the PSD shape and explore the influence of fines content. Table \ref{tab:bimodal-samples} lists the particle size ratio $\left( \chi \right)$, fines fraction $\left( \psi \right)$, number of particles $\left( N_p \right)$, number of fine particles $\left( N_p^\textrm{F} \right)$, number of coarse particles $\left( N_p^\textrm{C} \right)$, packing fraction $\left( \phi \right)$ and void ratio $\left( e \right)$.

Following isotropic compression, the particle positions and diameters for each configuration were used as the initial particle positions for the IBM simulation. In the IBM simulations, the particles were fixed in position and a flow with low Reynolds number was allowed to develop to a steady state solution. Periodic boundaries were applied in the $y$ and $z$ directions only. In the $x$ direction, the inlet boundary conditions

\begin{align*}
    \left( u_x, u_y, u_z \right) &= \left(U, 0, 0 \right) \\
    \dv{p}{x} &= 0
\end{align*}

\noindent were applied at the lower end of the domain.  The inlet velocity was set to $U = 2 \times 10^{-4} \: \textrm{m/s}$ to simulate laminar flow (low Reynolds number, $Re \approx 0.4$). The outlet boundary conditions

\begin{align*}
    \left( \pdv{u_x}{x}, \pdv{u_y}{y}, \pdv{u_z}{z} \right) &= \left(0, 0, 0 \right) \\
    p &= 0
\end{align*}

\noindent were applied at the upper end of the domain.

All the Lagrangian points on the particle surface must lie inside the domain. Placing Lagrangian points too close to the inlet will result in a non-physical solution as the no-slip condition imposed by these points will be in conflict with the constant velocity enforced by the inlet boundary condition.

\begin{table}[ht!]
    \caption{\label{tab:linear-samples}Linear graded samples used in IBM and PNM simulations}
    \centering
    \begin{tabular}{|c|c|c|c|c|}
        \hline
        $C_u$ & $\chi$ & $N_p$ & $\phi$ & $e$ \\
        \hline
        \hline
        1.01 & 1.02 & 629 & 0.396 & 1.525 \\
             &      &     & 0.460 & 1.174 \\
             &      &     & 0.519 & 0.927 \\
             &      &     & 0.567 & 0.764 \\
             &      &     & 0.621 & 0.610 \\
             &      &     & 0.651 & 0.536 \\
        \hline
        1.20 & 1.42 & 629 & 0.396 & 1.525 \\
             &      &     & 0.460 & 1.174 \\
             &      &     & 0.518 & 0.931 \\
             &      &     & 0.567 & 0.764 \\
             &      &     & 0.620 & 0.613 \\
             &      &     & 0.643 & 0.555 \\
        \hline
        1.50 & 2.11 & 491 & 0.408 & 1.451 \\
             &      &     & 0.460 & 1.174 \\
             &      &     & 0.519 & 0.927 \\
             &      &     & 0.568 & 0.761 \\
             &      &     & 0.621 & 0.610 \\
             &      &     & 0.654 & 0.529 \\
        \hline
        2.00 & 3.48 & 497 & 0.398 & 1.513 \\
             &      &     & 0.449 & 1.227 \\
             &      &     & 0.522 & 0.916 \\
             &      &     & 0.571 & 0.751 \\
             &      &     & 0.624 & 0.603 \\
             &      &     & 0.681 & 0.468 \\
        \hline
    \end{tabular}
\end{table}

\begin{table}[ht!]
  \caption{\label{tab:bimodal-samples}Bimodal samples used in IBM and PNM simulations}
  \centering
  \begin{tabular}{|c|c|c|c|c|c|c|}
    \hline
    $\chi$ & $\psi$ & $N_p$ & $N_p^\textrm{F}$ & $N_p^\textrm{C}$ & $\phi$ & $e$ \\
    \hline
    \hline
    1.99 & 0.10 & 383  & 180 & 203 & 0.621 & 0.610 \\
         &      &      &     &     & 0.656 & 0.524 \\
    \hline
         & 0.25 & 615  & 445 & 170 & 0.622 & 0.608 \\
         &      &      &     &     & 0.670 & 0.493 \\
    \hline
         & 0.35 & 771  & 624 & 147 & 0.622 & 0.608 \\
         &      &      &     &     & 0.673 & 0.486 \\
    \hline
         & 0.50 & 1003 & 890 & 113 & 0.622 & 0.608 \\
         &      &      &     &     & 0.665 & 0.504 \\
    \hline
    \hline
    3.97 & 0.11 & 317  & 279 &  38 & 0.645 & 0.550 \\
         &      &      &     &     & 0.685 & 0.460 \\
    \hline
         & 0.25 & 696  & 664 &  32 & 0.648 & 0.543 \\
         &      &      &     &     & 0.751 & 0.332 \\
    \hline
         & 0.35 & 958  & 930 &  28 & 0.651 & 0.536 \\
         &      &      &     &     & 0.740 & 0.351 \\
    \hline
         & 0.51 & 1363 & 1342 & 21 & 0.621 & 0.610 \\
         &      &      &     &     & 0.715 & 0.399 \\
     \hline
  \end{tabular}
\end{table}

\begin{figure}[ht]
    \centering
    \includegraphics[width=0.6\textwidth]{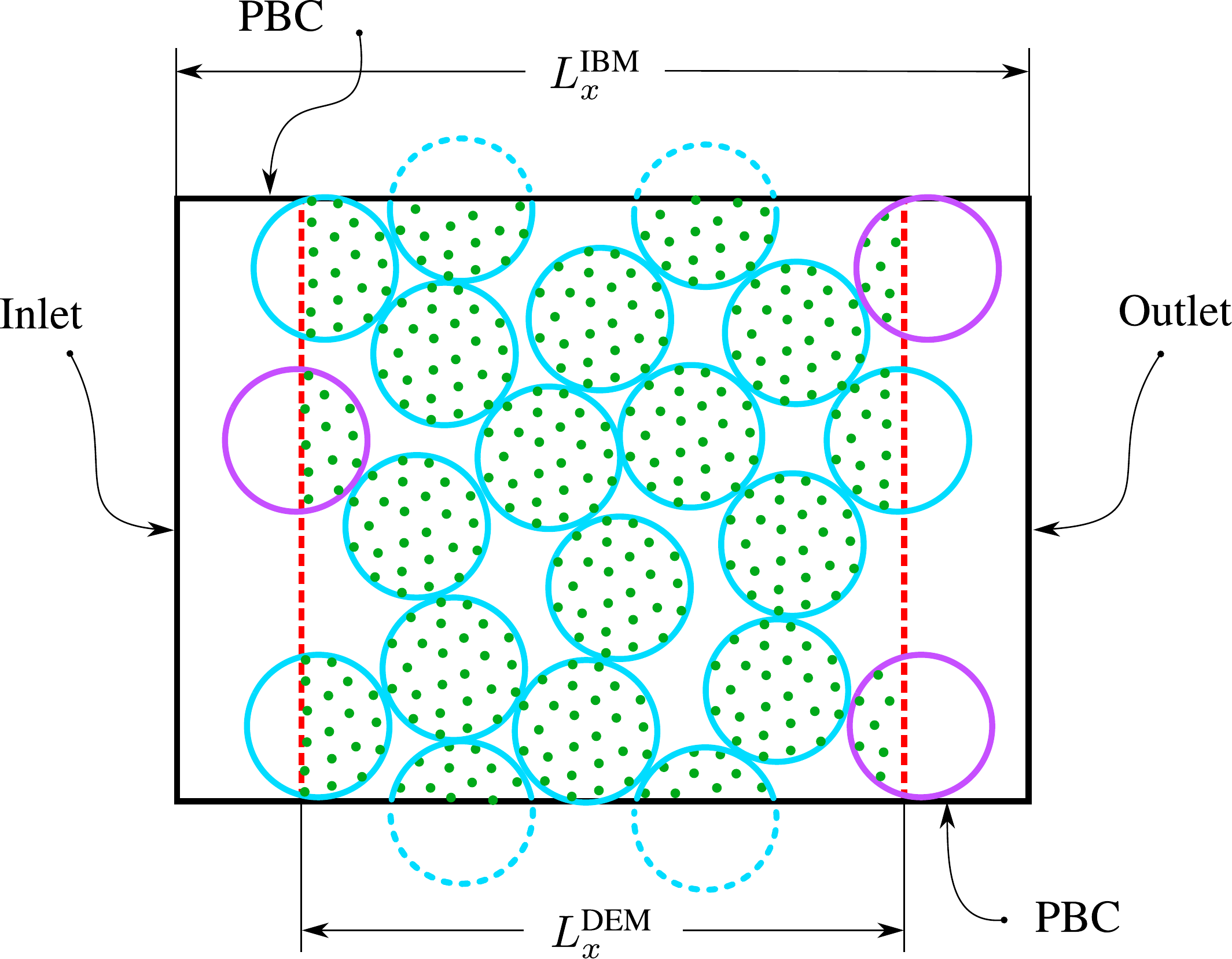}
    \caption{\label{fig:IBM-simulation-setup}Illustration of an IBM case with periodic boundary conditions in the $y$ and $z$ directions and inlet/outlet boundaries in the $x$ direction. The original extent of the DEM simulation are indicated by red dotted lines. Any particles which cross the original DEM extents are duplicated as ghost particles (purple circles) at the opposite side of the domain. Only Lagrangian points inside the original DEM extents (green dots) are considered when extracting the fluid-particle forces.}
\end{figure}

Validation studies using ordered packings revealed that $\mathbf{f}_{fp}$ was slightly over-predicted for particles at the inlet region and slightly under-predicted for those at the outlet region. In order to avoid these issues, the procedure illustrated in Fig.~\ref{fig:IBM-simulation-setup} was adopted, as explained below. Particles which crossed the DEM simulation boundaries along the $x$ direction, are duplicated and translated to the opposite end of the domain. The domain is extended in the $x$ direction, such that all Lagrangian points are separated by at least $2\Delta x$ from the inlet and outlet boundaries (where $\Delta x$ is the size of the fluid cell). When extracting the fluid-particle interaction forces, only those Lagrangian points which lie inside the original extents of the DEM simulation domain were considered. For a particle which is present at the inlet and also as a ghost particle duplicated at the outlet, part of the contribution to $\mathbf{f}_{fp}$ is due to the flow field in the inlet region, and the contribution from the remainder of the Lagrangian points is due to the flow field in the outlet region. This procedure effectively mimics periodic boundary conditions along the $x$ direction on $u_x$, $u_y$, $u_z$ and $\dv{p}{x}$ for the flow inside the original extents of the DEM simulation domain. The fact that $\mathbf{f}_{fp}$ depends on the values of $\nabla p$ across the particles as opposed to absolute pressure, $p$, is a sufficient justification for the use of this approach.

Tables~\ref{tab:linear-samples}-\ref{tab:bimodal-samples} show that the assemblies considered in this study comprise very small number of particles. The IBM provides high resolution detail of the interstitial flow at great computational cost. To accurately capture details of the interstitial flow field, the smallest constrictions present must be spanned by a sufficient number of fluid cells. Here a resolution of $\frac{d_{min}}{\Delta x} = 24$ was adopted, where $d_{min}$ is the minimum particle diameter. As the range of particle sizes increases with increasing $C_u$ for linear graded samples or with increasing $\chi$ for bimodal samples, the overall domain size increases with more of the particle phase volume concentrated in a small number of large particles. This increase in domain size coupled with the mesh resolution criterion meant that the computational resources increased dramatically with $C_u$ and $\chi$. This limited the sample sizes and particle gradations that could be considered in this study. A typical IBM case with approximately 600 particles embedded in an Eulerian mesh of 60 million fluid elements took approximately 16 hours to run on 400 processing cores. This imposed limits on the size and number of systems which could be considered.

\section{Pore Network Models}

\subsection{Delaunay-based Network Generation}

Pore Network Models (PNM) provide a simplified representation of flow in granular, porous media. The first stage in constructing a PNM is to partition the void space and identify the pores (PNM nodes) and pore throats or constrictions (PNM edges). Where assemblies of spheres are considered, a space tessellation method can be used, which identifies a set of unique, non-overlapping and space-filling cells. The classical Delaunay Tessellation (DT) has been used in a number of prior studies of fluid flow in porous media \citep{mason1971,bryant1992,thompson1997}. The DT produces a set of tetrahedral cells, where the vertices of the tetrahedron are the particle centroids. Each tetrahedron represents a node in the network model, while the faces of the tetrahedron are pore constrictions representing the edges in the network model. While the classical DT provides an adequate delineation of the pore space in monodisperse assemblies, it poses issues for assemblies of different sized spheres. Most notably, the faces of the tetrahedron may intersect adjacent particles \citep{chareyre2012}. A weighted Delaunay Tessellation can overcome this issue by including consideration of the particle radius $\left( r_p \right)$.

The weighted DT is applied to the samples generated for IBM simulations listed in Tables~\ref{tab:linear-samples}-\ref{tab:bimodal-samples}. This was conducted using the C++ package \texttt{TetGen} \citep{si2015}, where the weighting considered was $r_p^2$. In order to avoid degenerate Delaunay cells that may develop on planar boundaries, a sub-sampling procedure was applied, taking into account the periodic domain employed in sample generation. In this sub-sampling process, the primary sample domain was replicated in all 26 adjacent directions and the DT was applied to this extended domain. Then, only those Delaunay cells completely within the primary domain were extracted and used in the network generation.

While the DT provides a simple and computationally efficient method to tessellate the pore space, \citet{alraoush2003} suggested that it can overly subdivide the pore space and therefore not capture the existence of large pores. The DT also has a fixed network structure with all Delaunay cells comprising exactly four faces, and hence, four constrictions or edges in the network. To overcome these issues, \citet{alraoush2003} suggested that tetrahedral Delaunay cells should be merged according to a merging criteria to form polyhedral pore cells, and termed this the Modified Delaunay Tessellation (MDT). The MDT results in a more physically representative tessellation of the pore space by capturing the large variation in pore sizes (that is, the existence of small and large pore bodies and constrictions) and allowing for variable edge connectivity in the network. 

For the MDT considered here the merging criteria based on inscribed sphere \citep{alraoush2003,reboul2008,sufian2019} was used. An inscribed sphere of a Delaunay cell is the sphere which is internally tangent to the four particles forming the tetrahedron. A local optimisation procedure may be required if the calculated inscribed sphere is not contained entirely within the pore space. This is particularly prevalent in polydisperse assemblies. The local optimisation procedure outlined in \citet{sufian2015} is employed in this study, which takes a combinatorics approach to determine the maximal inscribed sphere contained entirely within the pore space. Delaunay cells are merged if there respective inscribed spheres overlap.

The current study examines the sensitivity of PNM results to the way the void space is partitioned and so both DT and MDT data are used to construct the PNM. The geometric properties of the unit cells in the DT and MDT can be readily calculated analytically with the assumption of non-overlapping particles; an appropriate assumption for the simulations considered in this study. The geometric properties of interest are the pore volume, $V_p$, the curved particle surface area within the unit cell, $S_p$, and the pore area of the constriction, $A_p$, which are used in the subsequent sections to define the network conductance model and calculate the fluid-particle interaction force. In the DT, $S_p$ is calculated from the spherical triangle formula, while $V_p$ is the determined by subtracting the volume of the particle sectors from the volume of the tetrahedral Delaunay cell, and $A_p$ is given by the area of the triangular face minus the area of the intersecting particle sectors (further details can be found in \citet{sufian2015}). The polyhedral pores in the MDT are a union of non-overlapping and space-filling Delaunay cells. Hence, their geometric properties are given by the summation of corresponding Delaunay cells.

\subsection{Local Pressure and Flow Field} \label{sub:local-pressure-and-flow-field}

Local fluid flow properties are obtained by simulating flow in the pore network using a Stokes flow algorithm. This is conducted using the open-source Python package, \texttt{OpenPNM} \citep{gostick2016}. As in the IBM simulations, particles are fixed in position when conducting the flow simulation.

The Stokes flow algorithm ensures conservation of mass in the pore network, which at steady state conditions and in the absence of any source/sink terms (such as volume change in the pores) is given by:
\begin{equation} \label{eqn:network-continuity}
    \sum_{i \rightarrow j}^n q_{ij} = \sum_{i \rightarrow j}^n g_{ij} \left( p_i - p_j \right) = 0
\end{equation}
where $q_{ij}$ is the flux (flow rate) from node $i$ to node $j$, $g_{ij}$ is the conductance of the $ij$ edge, $\{p_i, \: p_j\}$ are the nodal pressures, and the summation is over all $n$ edges connected to node $i$. Models used to define the conductance of a network edge are detailed in Sec.~\ref{sub:conductance-models}. Application of Eq.~\ref{eqn:network-continuity} to all nodes in the network, taking the pressure at the inlet and outlet nodes as boundary conditions, forms a system of linear equations which can be solved using matrix-inversion algorithms to obtain the pressure at each node and consequently the flux along each edge.

Dirichlet boundary conditions are imposed in the solution to Eq.~\ref{eqn:network-continuity}, where pressure is prescribed at the inlet and outlet nodes. Inlet and outlet nodes are identified as those unit cells that have at least one face that is not shared with any other unit cells (which can only occur for those unit cells at the domain boundary) and are located in close proximity to the domain boundaries in the direction of flow. Note that an inlet flow velocity is considered for the boundary conditions in the IBM simulations. For consistency with the IBM simulations, the pressure at the DEM boundaries (indicated by the red dotted line in Fig.~\ref{fig:IBM-simulation-setup}) are considered in the PNM to define the Dirichlet boundary conditions at the inlet and outlet nodes. A linear interpolation is performed to obtain the prescribed pressure at the centroid of the inlet and outlet nodes. The imposed pressure gradient in the PNM is equivalent and consistent with the imposed flow velocity (or flux) in the IBM, thereby enabling direct comparison between the two numerical techniques.

\subsection{Conductance Models}\label{sub:conductance-models}

The conductance of a network edge is defined by the geometry of the constrictions and/or pore bodies and the viscosity of the fluid. Two approaches are compared in this paper: (i) a hierarchical model based on pore-throat-pore series elements closely following \citet{vanderlinden2018} (Sec.~\ref{subsub:pore-throat-pore-model}); and (ii) equivalent hydraulic radius model based on single element closely following \citet{chareyre2012} (Sec.~\ref{subsub:hydraulic-radius-model}). Both conductance models are applicable to networks generated based on the DT and MDT.    

\subsubsection{Pore-Throat-Pore Series Model}\label{subsub:pore-throat-pore-model}

This model considers a simplified representation of the pore space as a collection of cylindrical elements connected in series and parallel. Flow through a cylindrical element is defined by the Hagen-Poiseuille equation:
\begin{equation} \label{eqn:Hagen-Poiseuille}
    q = g \Delta p = \left( \frac{\pi r^4}{8 \mu l} \right) \Delta p
\end{equation}
where $q$ is the flux, $r$ is the radius of the cylindrical element, $l$ is the length of the cylinder, $\mu$ is the dynamic viscosity of the fluid, $g$ is the conductance of the cylindrical element and $\Delta p$ is the pressure difference across the length of the element. The Hagen-Poiseuille equation is applicable for an incompressible fluid experiencing laminar flow, which is in line with fluid flow conditions in the IBM simulations.

\begin{figure}[ht]
    \centering
    \begin{subfigure}[b]{0.32\textwidth}
        \includegraphics[width=\textwidth]{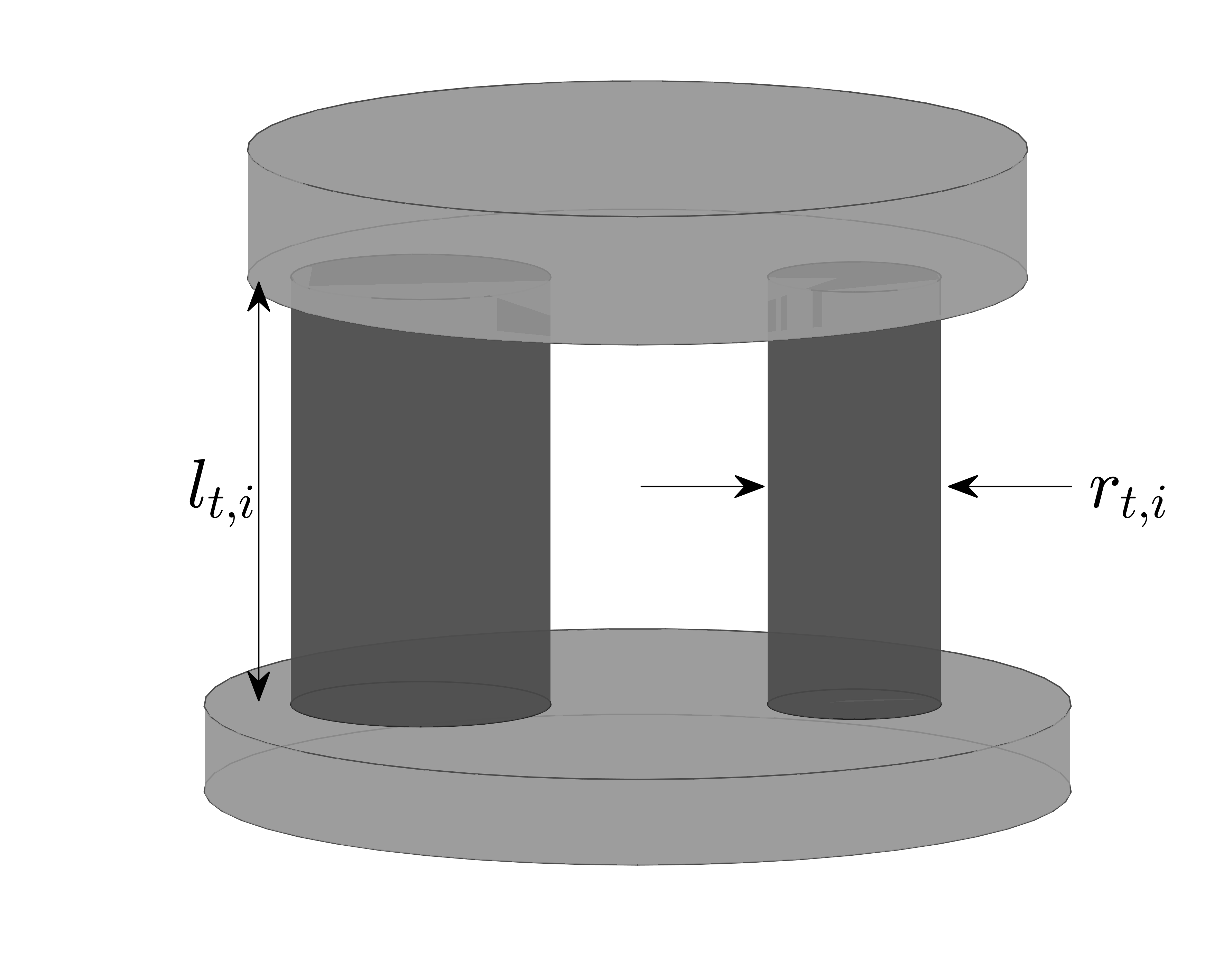}
        \caption{\label{fig:conductance-model-level1}}
    \end{subfigure}
    \hfill
    \begin{subfigure}[b]{0.32\textwidth}
        \includegraphics[width=\textwidth]{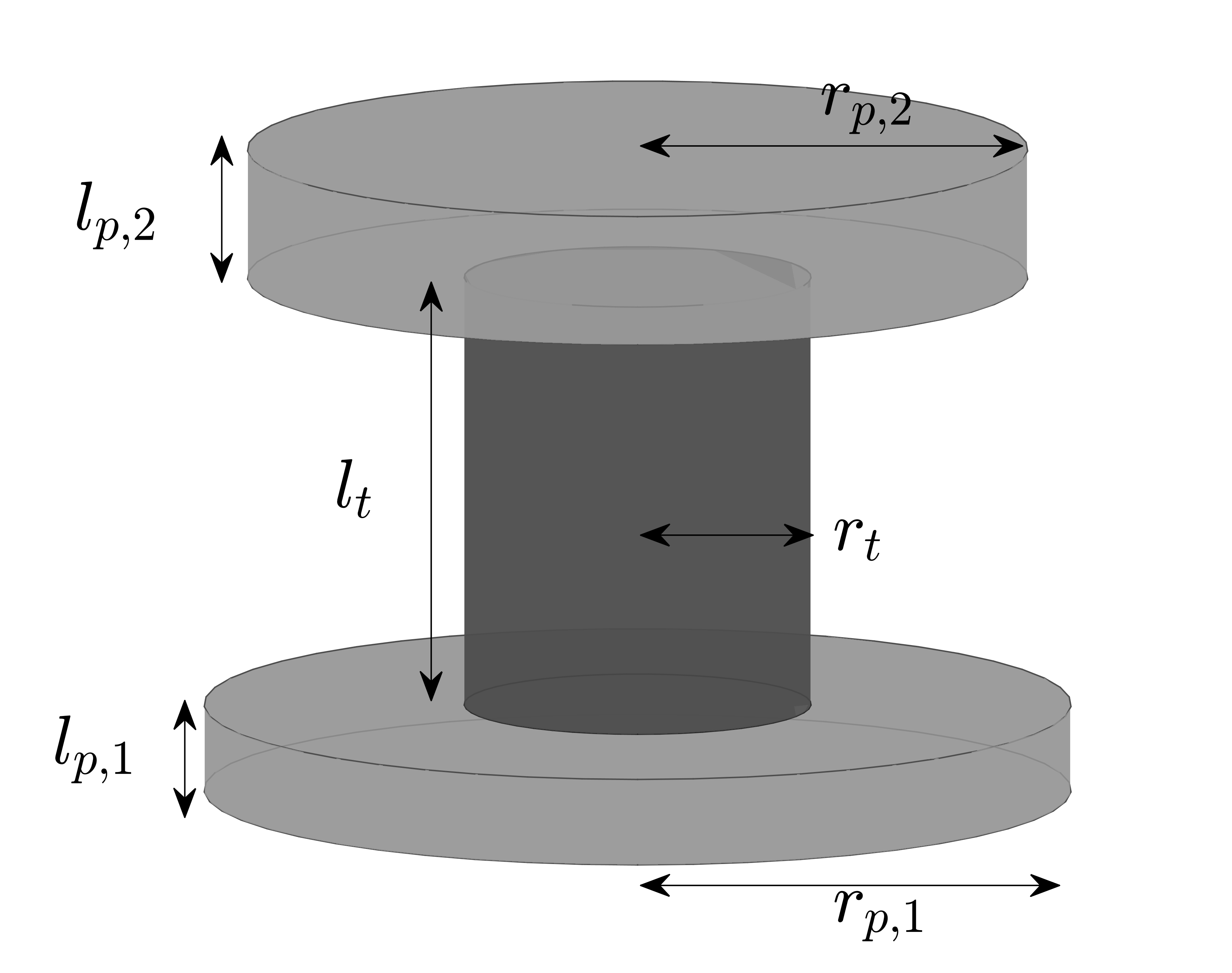}
        \caption{\label{fig:conductance-model-level2}}
    \end{subfigure}
    \hfill
    \begin{subfigure}[b]{0.32\textwidth}
        \includegraphics[width=\textwidth]{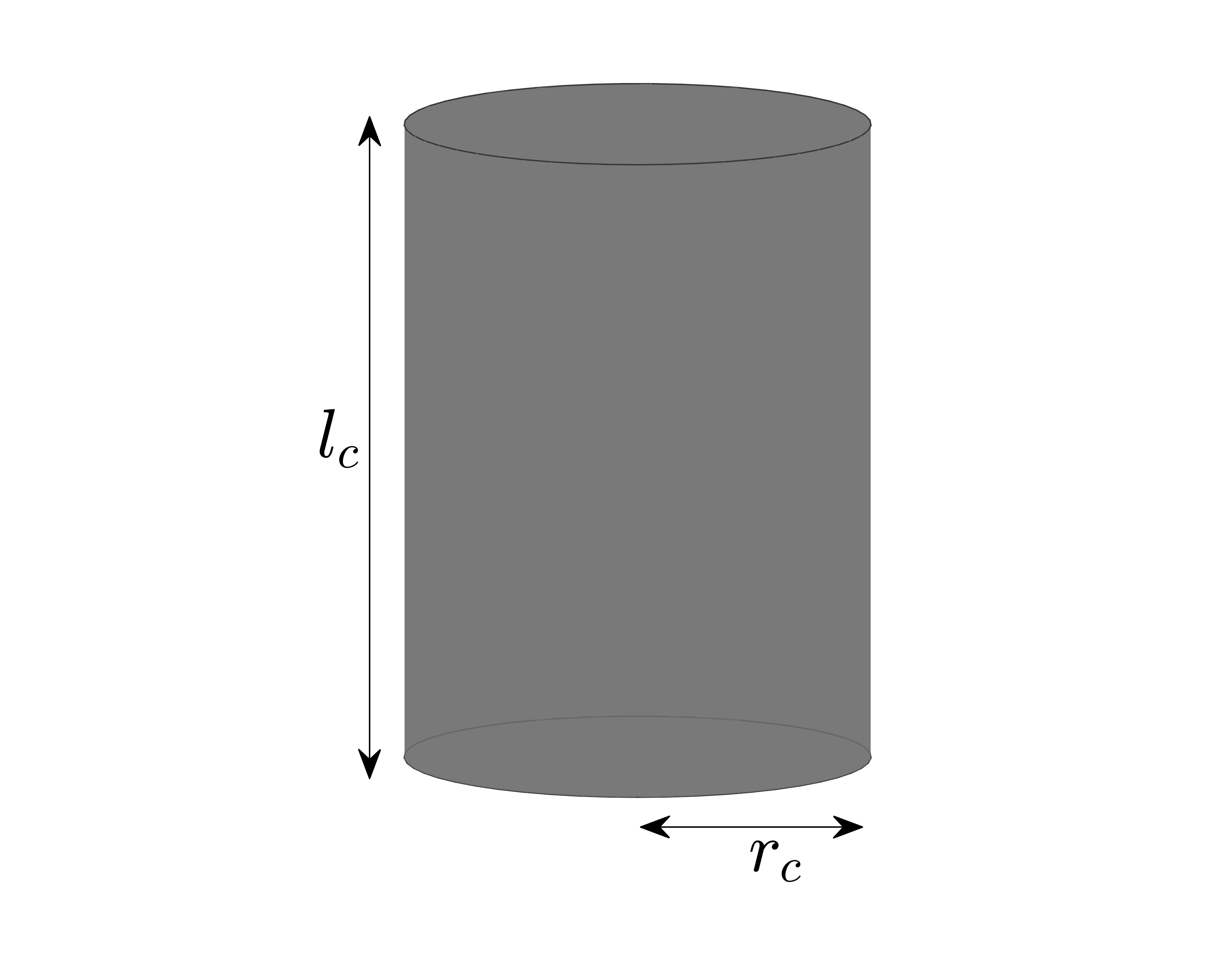}
        \caption{\label{fig:conductance-model-level3}}
    \end{subfigure}
    \caption{\label{fig:conductance-model}Schematic illustration of the conductance model: (a) two pore (node) elements connected by multiple constriction (edge) elements; (b) pore-throat-pore series element, where multiple throats are upscaled to an equivalent constriction; (c) network edge element formed by upscaling the pore-throat-pore element.}
\end{figure}

This conductance model represents the network edge element as a \textit{pore-throat-pore} series, comprising two cylindrical pore elements connected in series by one or more cylindrical constriction element (Fig.~\ref{fig:conductance-model-level1}-\ref{fig:conductance-model-level2}). The radius of the pore element, $r_p$, is defined as the radius of an equivalent sphere of pore volume, $V_p$. The length of the pore element, $l_p$, is the length of an equivalent cylinder with volume, $V_p$, and radius, $r_p$, calculated as above.

In the DT, each pair of pores is connected by a single constriction, as there is only one shared triangular face between adjacent cells. However, multiple triangular faces may connect adjacent cells in the MDT. These can be regarded as multiple constrictions, where radii of the $i^\textrm{th}$ constriction is $r_{t,i}$ and the length is $l_{t,i}$. The radius of the constriction elements, $r_{t,i}$, is given by the radius of a circle whose area equals the pore constriction area, $A_{p,i}$. The length of the throat element, $l_{t,i}$, is given by the distance between the centroids of the two corresponding Delaunay cells to the shared triangular face. These multiple constrictions are connected in parallel, and can be expressed as a single equivalent element (Fig.~\ref{fig:conductance-model-level2}) by noting that flux in the single equivalent element is equal to the sum of flux in each constriction, $q_t = \sum_i q_{t,i}$ and that the pressure drop is the same across all constrictions, $\Delta p_t = \Delta p_{t,i}$. If it is assumed that all constrictions connecting two pores have the same length defined by a weighted arithmetic mean,

\begin{equation*}
    l_t = \frac{\sum_i A_{t,i}l_{t_i}}{\sum_i A_{t,i}}
\end{equation*}

\noindent then the radius of the single equivalent constriction element is given by:
\begin{equation} \label{eqn:radius-throat}
    r_t^4 = \sum_i r_{t,i}^4
\end{equation}

The pressure drop across a network edge element is the sum of pressure drops across each component in the corresponding pore-throat-pore series element, and the flux through the edge equals the pore-throat-pore element flux. If it is assumed that a network edge element has a length, $l_c$, defined as the centre-to-centre distance of the respective pore bodies (this differs from the centre-to-centre distance of corresponding Delaunay cells in the definition of $l_{t,i}$ for the MDT), then it can be shown that:
\begin{equation} \label{eqn:radius-edge}
    \frac{l_c}{r_c^4} = \frac{l_{p,1}}{r_{p,1}^4} + \frac{l_{t}}{r_{t}^4} + \frac{l_{p,2}}{r_{p,2}^4}
\end{equation}
where $r_c$ is the radius of the network edge element.

Therefore the conductance of a network edge is given by:
\begin{equation} \label{eqn:conductance-pore-throat-pore}
    g = \frac{\pi r_c^4}{8 \mu l_c}
\end{equation}

\subsubsection{Hydraulic Radius Model}\label{subsub:hydraulic-radius-model}

While the pore-throat-pore series model in Sec.~\ref{subsub:pore-throat-pore-model} used an hierarchical up-scaling technique to obtain an equivalent conductance for the network edge, the hydraulic radius model simply specifies the conductance of the network edge based on the geometry of the pores and constriction. In this model, the conductance is given by:
\begin{equation}\label{eqn:conductance-hydraulic-radius}
    g = \frac{\alpha A_t R_h^2}{\mu l_c}
\end{equation}
where $A_t$ is the total pore constriction area ($A_t = \sum_i A_{t,i}$ for the MDT), $R_h$ is the hydraulic radius, $l_c$ is the centre-to-centre distance of the respective pore bodies (as used in Sec.~\ref{subsub:pore-throat-pore-model}) and $\alpha$ is a non-dimensional factor to account for the cross-sectional shape of constrictions. Following \citet{chareyre2012}, $\alpha = 0.5$, which was deemed to be appropriate for assemblies of spheres. The hydraulic radius, $R_h$, is defined as the ratio of volume to surface area of the two pore bodies corresponding to the shared constriction. Note that both conductance models (Eq.~\ref{eqn:conductance-pore-throat-pore} and Eq.~\ref{eqn:conductance-hydraulic-radius}) utilise the geometry of pores and constrictions in defining the local conductivity and this is preferred over conductance models which exclusively consider constriction geometry (e.g. \citet{bryant1992}) as the local flow field is influenced by the properties of both pores and constrictions. 

The conductance calculated from Eq.~\ref{eqn:conductance-hydraulic-radius} is equivalent to the Hagen-Poiseuille form in Eq.~\ref{eqn:conductance-pore-throat-pore} when $r_c = 2R_h$ and $\alpha = 0.5$. Therefore, the hydraulic radius approach with varying shape factors could be applied to each of the elements in the pore-throat-pore series and subsequently up-scaled to obtain an equivalent conductance. This is beyond the scope of this paper but demonstrates that the two conductance models are closely linked.    

The conductance specified in Eq.~\ref{eqn:conductance-pore-throat-pore} or Eq.~\ref{eqn:conductance-hydraulic-radius} is supplied to \texttt{OpenPNM} and used in the mass conservation equation in Eq.~\ref{eqn:network-continuity} to determine the local pressure and flux in the pore network.

\subsection{Fluid-Particle Interaction Force} \label{sub:Fluid-Particle-Interaction-Force}

\citet{chareyre2012} proposed calculating the fluid-particle interaction force, $\Fp_{fp}$, by considering the geometric properties of the constrictions, along with the local pressure gradient obtained from the Stokes flow algorithm. $\Fp_{fp}$ can be split into contribution of the pressure, $\Fp_{fp}^p$, and viscous shear, $\Fp_{fp}^v$, components (the subscript $fp$ is omitted hereafter for clarity):
\begin{equation}
    \Fp^k = \Fp^{p,k} + \Fp^{v,k}
\end{equation}
where the superscript $k$ indicates the force experienced by an individual particle. The pressure contribution to particle $k$ from flow in element ${ij}$ of the network is given by:
\begin{equation}
    \Fp_{ij}^{p,k} = A_{ij}^k\left( p_i - p_j \right)\mathbf{n}_{ij}
\end{equation}
where $A_{ij}^k$ is the area of the sector of particle $k$ intersecting the triangular face of  constriction ${ij}$, $\{ {p_i}, \: {p_j} \}$ are the corresponding nodal pressures and $\mathbf{n}_{ij}$ is the unit vector pointing from $p_i$ to $p_j$. In a pore network representation, the pressure everywhere within a pore cell is assumed to be constant. In this study, the nodal pressures are defined at the cell centroid locations, which are located within the pore space.

\begin{figure}[ht!]
    \centering
    \includegraphics[width=55mm]{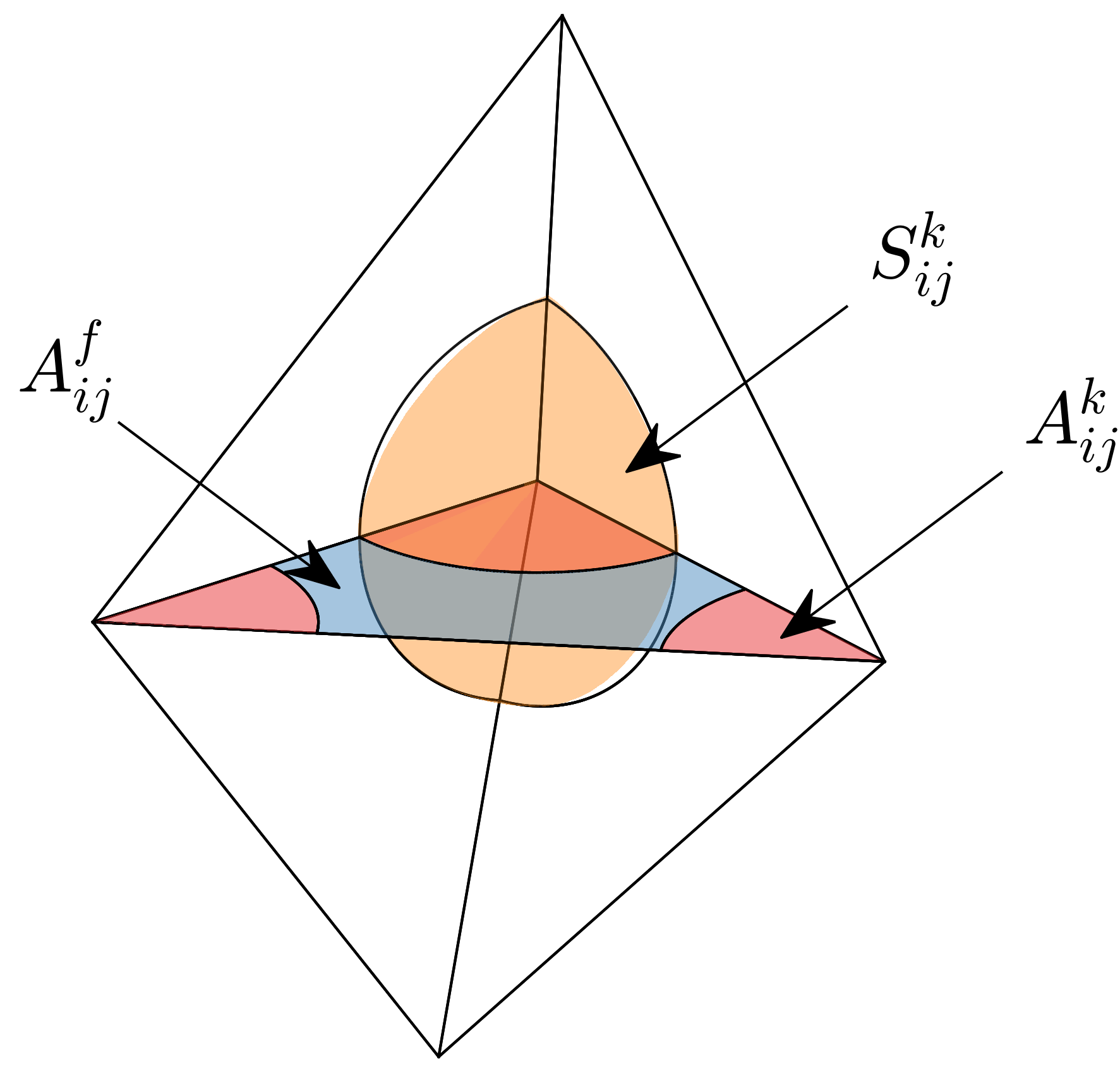}
    \caption{\label{fig:force-diagram}Schematic illustration of the terms required for the calculation of the fluid-particle interaction force. $A_{ij}^k$ is the particle area intersecting the constriction, shown in red. $A_{ij}^f$ is the pore or fluid area of the constriction, shown in blue. $S_{ij}^k$ is the particle surface area within the union of the unit cells adjacent to the constriction, shown in orange. Note that the particle surface area for only one particle is shown for clarity.}
\end{figure}

The viscous shear contribution due to edge ${ij}$ in the network is given by:
\begin{equation}
    \Fp_{ij}^v = A_{ij}^f\left( p_i - p_j \right)\mathbf{n}_{ij}
\end{equation}
where $A_{ij}^f$ is the pore area in the ${ij}$ constriction. The total viscous shear force is distributed to the particles intersecting the constriction by a ratio of their surface areas within the local domain:
\begin{equation} \label{eqn:viscous-particle}
    \Fp_{ij}^{v,k} = \Fp_{ij}^v \times \frac{S_{ij}^k}{\sum_{k=1}^3 S_{ij}^k}
\end{equation}
where $S_{ij}^k$ is the surface area of the particle $k$ within the local domain formed by nodes $i$ and $j$, as illustrated in Fig.~\ref{fig:force-diagram}. The summation in the denominator to Eq.~\ref{eqn:viscous-particle} is the total surface area within the local domain of the three particles forming the pore constriction.

\section{Comparison of Pore Network Models and the Immersed Boundary Method}

\subsection{Overall Flow Behaviour}

\begin{figure}[!ht]
    \centering
    \begin{subfigure}[b]{\textwidth}
        \centering
        \includegraphics[scale=0.95]{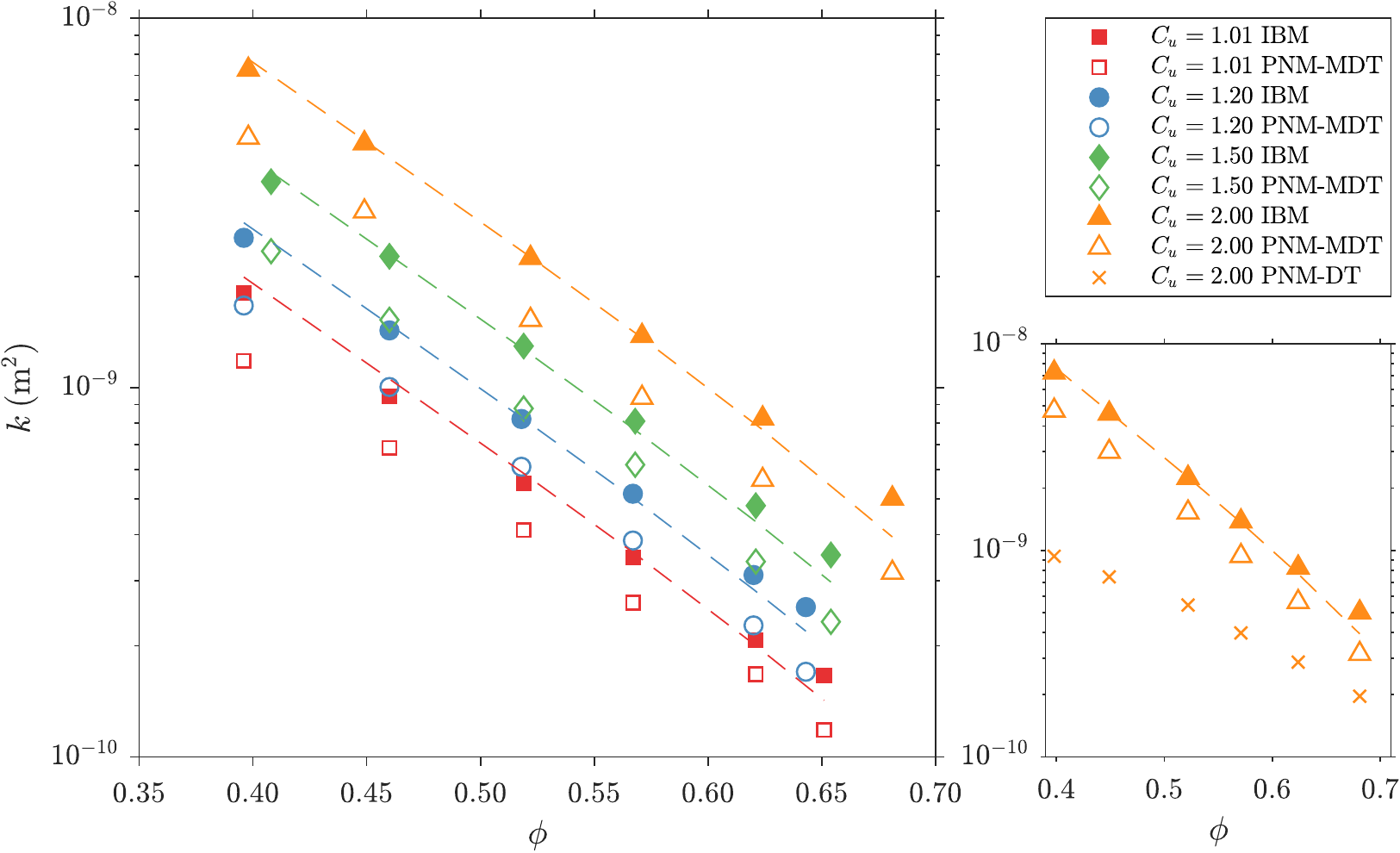}
        \caption{\label{fig:permeability-pore-throat-pore}Pore-throat-pore series model}
    \end{subfigure}
    \vfill
    \begin{subfigure}[b]{\textwidth}
        \centering
        \includegraphics[scale=0.95]{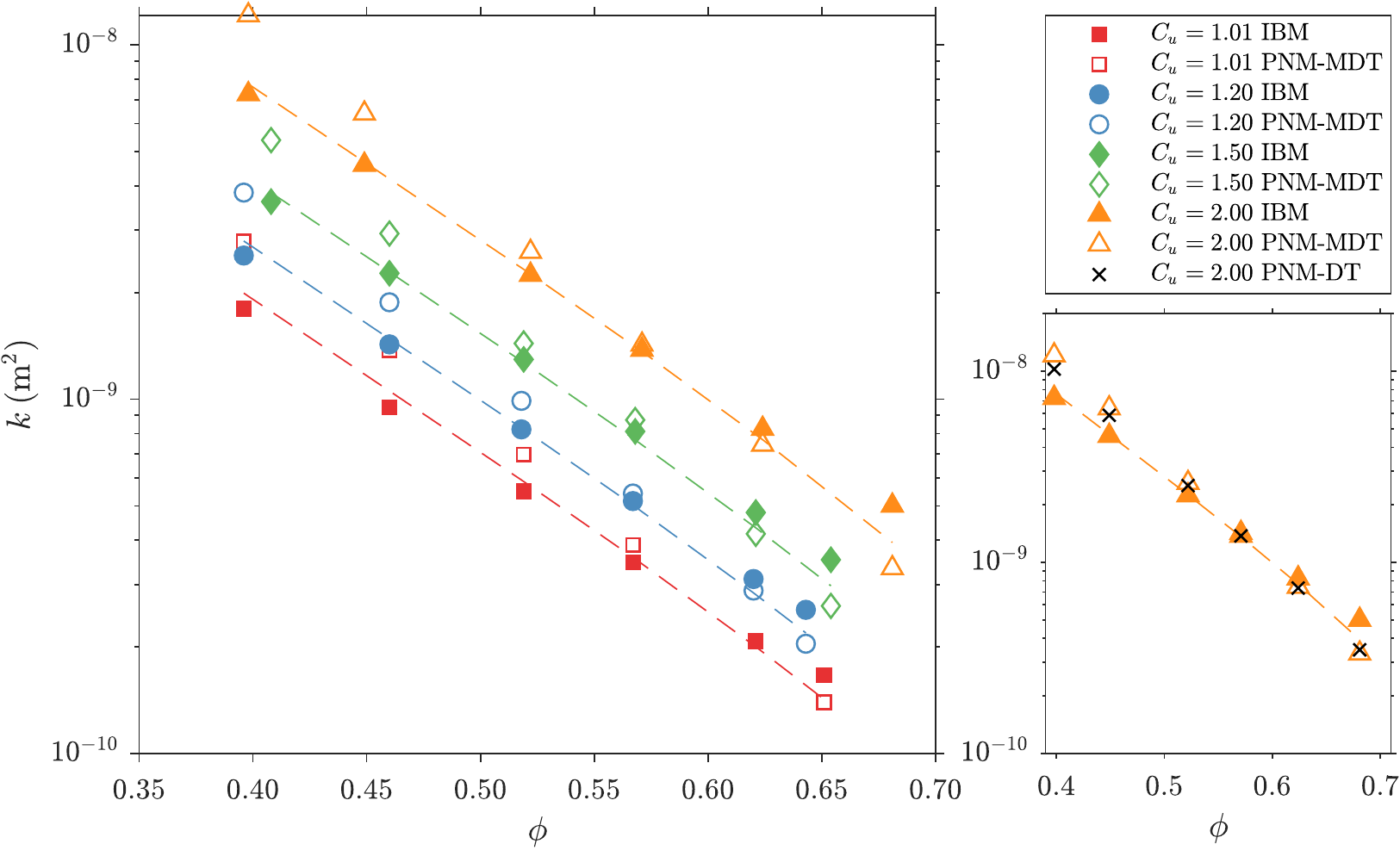}
        \caption{\label{fig:permeability-hydraulic-radius}Hydraulic radius model}
    \end{subfigure}
    \caption{\label{fig:permeability}Permeability, $k$, obtained from PNM simulations using the MDT in linear graded samples for both conductance models presented in Sec.~\ref{sub:conductance-models}. Results for DT are shown on the right for $C_u = 2.00$ (similar trends for other samples). The dashed lines represent the permeability estimate based on Kozeny-Carman.}
\end{figure}

A comparison of the overall flow behaviour is presented in Fig.~\ref{fig:permeability}, which shows the sample permeability obtained from the IBM and PNM. The PNM was generated using both the DT and MDT to explore the sensitivity to pore space partitioning. Both conductance models presented in Sec.~\ref{sub:conductance-models} are also compared. The permeability, $k$, is obtained from the PNM by:
\begin{equation} \label{eqn:pnm-permeability}
    k = \frac{Q^* \mu l}{A \Delta p}
\end{equation}
where $Q^*$ is the summed fluxes at the inlet or outlet nodes, $\mu$ is the dynamic viscosity of the fluid, $l$ is the domain length, $A$ is the domain cross-sectional area and $\Delta p$ is the pressure difference across the domain length. Fig.~\ref{fig:permeability} also plots the Kozeny-Carman estimate of permeability, defined as:
\begin{equation} \label{eqn:kozeny-carman}
    k = \frac{1 - \phi}{5 \phi^2 S_v^2}
\end{equation}
where $S_v$ is the specific internal surface area to volume ratio obtained from the complete PSD by:

\begin{equation*}
    S_v = \frac{\sum_i \pi d_i^2}{\sum_i \frac{\pi}{6}d_i^3}
\end{equation*}

\noindent where the summation is performed over all particles in the system. 

Fig.~\ref{fig:permeability} demonstrates that the permeability obtained from the IBM simulations closely match the Kozeny-Carman expression. The permeability obtained from the PNM for the pore-throat-pore series conductance model (Sec.~\ref{subsub:pore-throat-pore-model}) is shown in Fig.~\ref{fig:permeability-pore-throat-pore}, while Fig.~\ref{fig:permeability-hydraulic-radius} shows the permeability obtained using the hydraulic radius model (Sec.~\ref{subsub:hydraulic-radius-model}). The hydraulic radius model better captures the overall flow response across the range of linear gradations and packing fractions considered in this study. The permeability obtained from both DT and MDT closely matches the IBM simulations, with slight overestimation at low packing fractions. In the pore-throat-pore series model, the permeability obtained from the PNM based on the MDT is in reasonable agreement with the value calculated in the IBM simulations, with permeability slightly underestimated. This suggests that partitioning the pore space more accurately using the MDT can result is a better estimate of permeability, irrespective of the conductance model. However, significant differences are noted for the permeability obtained from the DT when considering the pore-throat-pore series conductance model (see right-side figure in Fig.~\ref{fig:permeability-pore-throat-pore}), as the least dense assembly showed an order of magnitude error, with the error reducing with increasing density. 

\subsection{Correlation in Local Pressure and Flux}

Local pressure and flux was obtained from the IBM simulations using the DT and MDT geometries that were input to the PNM, thereby allowing for direct comparison between the two numerical techniques. Pressures at the centroid of the DT cells were interpolated from the pressure fields obtained from the IBM simulations. A volume weighted sum of the pressures of the constituent DT cells was used to obtain each of the MDT cell pressures. These pressures calculated from the IBM simulations are compared against the nodal pressures obtained from the PNM. To calculate fluxes from the IBM data, each triangular face of the DT was discretised into a set of sub-triangles using a Delaunay triangulation such that the average area of the sub-triangles was approximately equal to the average area of the Eulerian cell faces. Fluid velocities were interpolated at the centres of each of these sub-triangles, allowing for the flux at each sub-triangle to be calculated. Fluxes through all sub-triangles were summed to obtain the total flux through each constriction in the DT. These DT fluxes were then summed to obtain the overall fluxes through the mutual faces between adjacent cells in the MDT. The IBM flux is compared against the flux through the corresponding edge in the PNM.

\begin{figure}[H]
    \centering
    \begin{subfigure}[b]{0.49\textwidth}
        \includegraphics[scale=0.95]{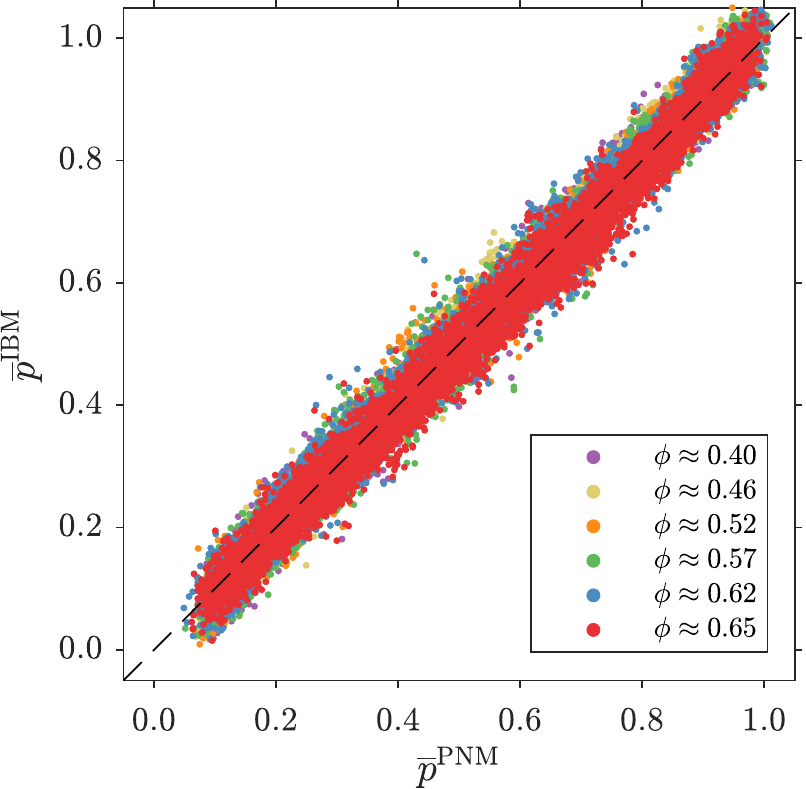}
        \caption{DT}
    \end{subfigure}
    \hfill
    \begin{subfigure}[b]{0.49\textwidth}
        \includegraphics[scale=0.95]{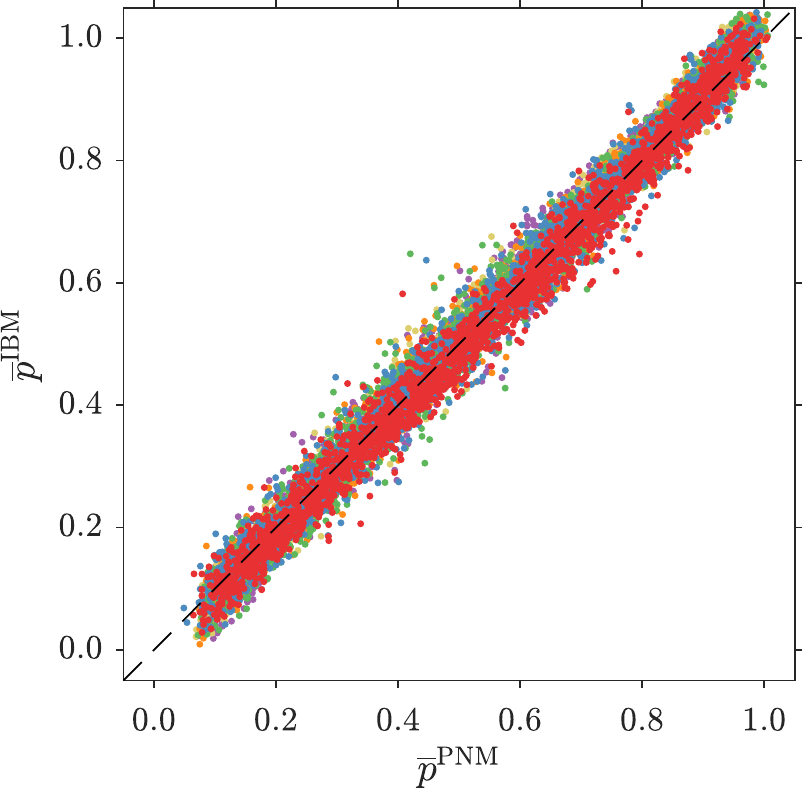}
        \caption{MDT}
    \end{subfigure}
    \caption{\label{fig:pressure-corr-linear-ptp}Correlation in normalised local pressure in the PNM ($\overline{p}^{\mathrm{PNM}}$) and the IBM ($\overline{p}^{\mathrm{IBM}}$) for linear graded samples using pore-throat-pore series conductance model in Sec.~\ref{subsub:pore-throat-pore-model}. Normalisation is conducted by $\overline{p} = \frac{p}{p_{\mathrm{inlet}} - p_{\mathrm{outlet}}}$, where $p$ is the nodal pressure, $p_{\mathrm{inlet}}$ is the inlet boundary pressure and $p_{\mathrm{outlet}}$ is the outlet boundary pressure.}
\end{figure}

\begin{figure}[H]
    \centering
    \begin{subfigure}[b]{0.49\textwidth}
        \includegraphics[scale=0.95]{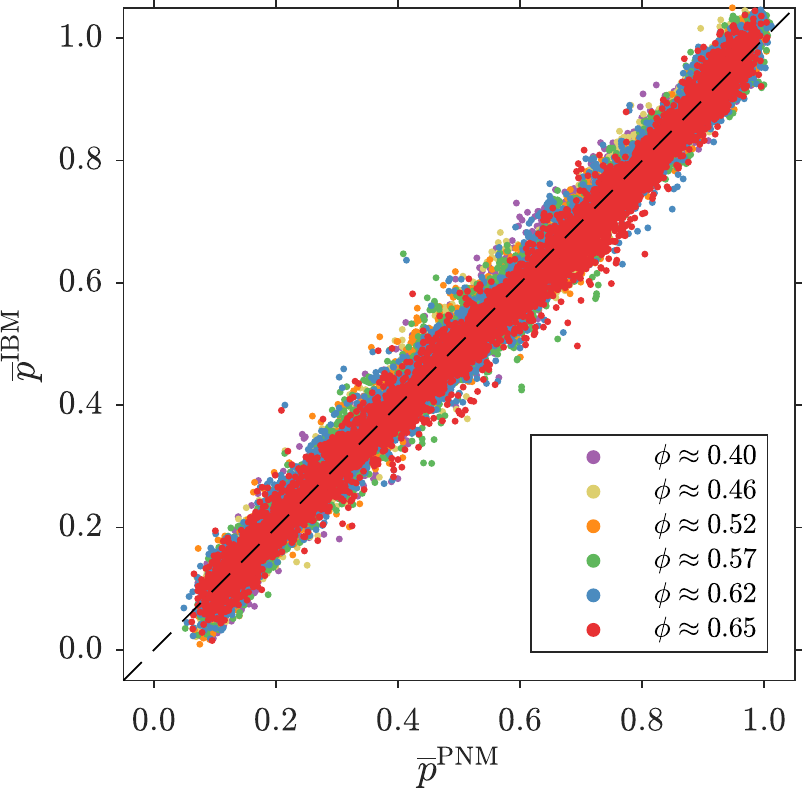}
        \caption{DT}
    \end{subfigure}
    \hfill
    \begin{subfigure}[b]{0.49\textwidth}
        \includegraphics[scale=0.95]{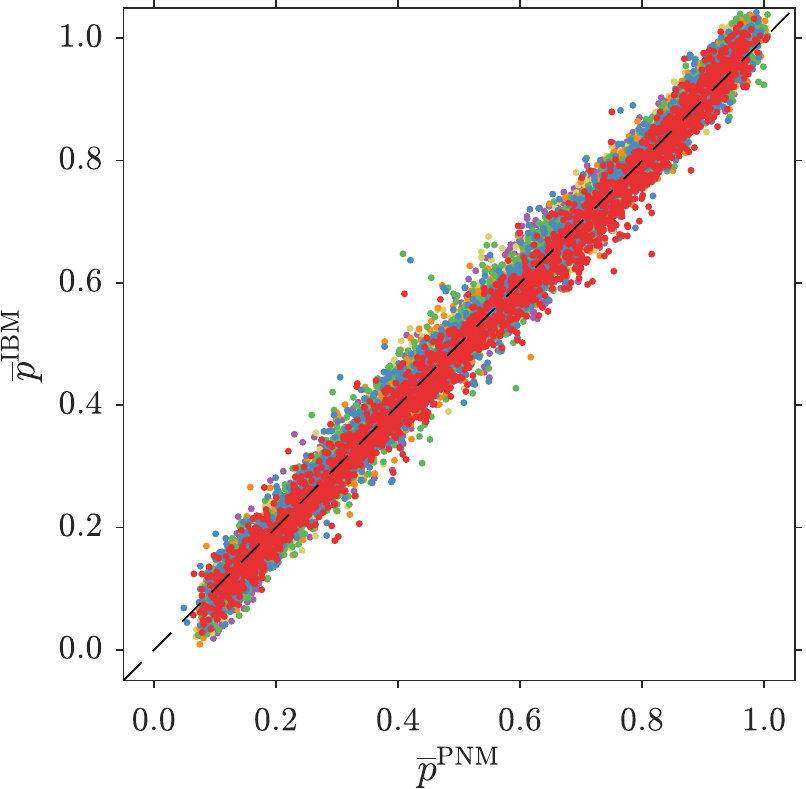}
        \caption{MDT}
    \end{subfigure}
    \caption{\label{fig:pressure-corr-linear-hr}Correlation in normalised local pressure in the PNM ($\overline{p}^{\mathrm{PNM}}$) and the IBM ($\overline{p}^{\mathrm{IBM}}$) for linear graded samples using hydraulic radius conductance model in Sec.~\ref{subsub:hydraulic-radius-model}. Normalisation is conducted as per Fig.~\ref{fig:pressure-corr-linear-ptp}.}
\end{figure}

\begin{table}[H]
    \caption{\label{tab:corr-coeff-linear}Pearson Correlation Coefficient in Linear Graded Samples for the Pore-Throat-Pore (PTP) and Hydraulic Radius (HR) Conductance Model}
    \centering
    \resizebox{\textwidth}{!}{
    \begin{tabular}{|c|c||c c|c c||c c|c c||c c|c c|c c|}
        \hline
         & &
            \multicolumn{4}{|c|}{Pressure} & 
            \multicolumn{4}{|c|}{Flux} & 
            \multicolumn{6}{|c|}{Force} \\
        \hline
         & & 
            \multicolumn{2}{|c|}{PTP} & \multicolumn{2}{|c|}{HR} &
            \multicolumn{2}{|c|}{PTP} & \multicolumn{2}{|c|}{HR} &
            \multicolumn{2}{|c|}{PTP} & \multicolumn{2}{|c|}{HR} & \multicolumn{2}{|c|}{IBM} \\
        \hline
        $C_u$ & $\phi$ & 
            $\rho_p^{\textrm{DT}}$ & $\rho_p^{\textrm{MDT}}$ & $\rho_p^{\textrm{DT}}$ & $\rho_p^{\textrm{MDT}}$ & 
            $\rho_q^{\textrm{DT}}$ & $\rho_q^{\textrm{MDT}}$ & $\rho_q^{\textrm{DT}}$ & $\rho_q^{\textrm{MDT}}$ & 
            $\rho_\Fm^{\textrm{DT}}$ & $\rho_\Fm^{\textrm{MDT}}$ & $\rho_\Fm^{\textrm{DT}}$ & $\rho_\Fm^{\textrm{MDT}}$ & $\rho_\Fm^{\textrm{DT}}$ & $\rho_\Fm^{\textrm{MDT}}$\\
        \hline
        \hline
        1.01 & 0.396 & 0.995 & 0.995 & 0.995 & 0.995 & 0.659 & 0.669 & 0.765 & 0.678 & 0.434 & 0.693 & 0.725 & 0.656 & 0.909 & 0.844 \\
             & 0.460 & 0.996 & 0.996 & 0.997 & 0.996 & 0.631 & 0.684 & 0.742 & 0.698 & 0.397 & 0.689 & 0.786 & 0.669 & 0.937 & 0.854 \\
             & 0.519 & 0.996 & 0.996 & 0.996 & 0.996 & 0.705 & 0.749 & 0.805 & 0.755 & 0.503 & 0.692 & 0.809 & 0.687 & 0.939 & 0.842 \\
             & 0.567 & 0.998 & 0.998 & 0.998 & 0.998 & 0.757 & 0.774 & 0.794 & 0.785 & 0.620 & 0.796 & 0.811 & 0.775 & 0.928 & 0.894 \\
             & 0.621 & 0.997 & 0.998 & 0.997 & 0.997 & 0.798 & 0.782 & 0.813 & 0.784 & 0.651 & 0.804 & 0.802 & 0.764 & 0.932 & 0.905 \\
             & 0.651 & 0.997 & 0.997 & 0.997 & 0.997 & 0.788 & 0.774 & 0.786 & 0.781 & 0.659 & 0.776 & 0.732 & 0.711 & 0.932 & 0.934 \\
        \hline
        1.20 & 0.396 & 0.996 & 0.995 & 0.996 & 0.995 & 0.666 & 0.636 & 0.762 & 0.668 & 0.839 & 0.880 & 0.890 & 0.858 & 0.955 & 0.923 \\
             & 0.460 & 0.996 & 0.996 & 0.996 & 0.996 & 0.649 & 0.667 & 0.739 & 0.661 & 0.881 & 0.857 & 0.923 & 0.853 & 0.970 & 0.927 \\
             & 0.518 & 0.996 & 0.996 & 0.996 & 0.996 & 0.642 & 0.722 & 0.743 & 0.718 & 0.864 & 0.892 & 0.906 & 0.894 & 0.963 & 0.955 \\
             & 0.567 & 0.995 & 0.996 & 0.995 & 0.995 & 0.681 & 0.747 & 0.716 & 0.740 & 0.860 & 0.889 & 0.909 & 0.878 & 0.972 & 0.949 \\
             & 0.620 & 0.997 & 0.998 & 0.997 & 0.997 & 0.729 & 0.787 & 0.726 & 0.791 & 0.887 & 0.936 & 0.931 & 0.934 & 0.969 & 0.974 \\
             & 0.643 & 0.998 & 0.997 & 0.997 & 0.997 & 0.721 & 0.723 & 0.737 & 0.754 & 0.914 & 0.939 & 0.941 & 0.938 & 0.966 & 0.975 \\
        \hline
        1.50 & 0.408 & 0.993 & 0.994 & 0.994 & 0.994 & 0.619 & 0.581 & 0.743 & 0.587 & 0.928 & 0.941 & 0.964 & 0.945 & 0.985 & 0.967 \\
             & 0.460 & 0.993 & 0.994 & 0.994 & 0.993 & 0.645 & 0.695 & 0.746 & 0.697 & 0.937 & 0.956 & 0.967 & 0.956 & 0.985 & 0.976 \\
             & 0.519 & 0.992 & 0.994 & 0.993 & 0.993 & 0.642 & 0.703 & 0.755 & 0.703 & 0.945 & 0.959 & 0.964 & 0.962 & 0.985 & 0.976 \\
             & 0.568 & 0.994 & 0.995 & 0.995 & 0.996 & 0.654 & 0.698 & 0.717 & 0.713 & 0.960 & 0.956 & 0.974 & 0.957 & 0.989 & 0.987 \\
             & 0.621 & 0.995 & 0.997 & 0.996 & 0.997 & 0.711 & 0.782 & 0.756 & 0.773 & 0.964 & 0.979 & 0.983 & 0.979 & 0.988 & 0.991 \\
             & 0.654 & 0.995 & 0.996 & 0.996 & 0.996 & 0.748 & 0.766 & 0.723 & 0.771 & 0.966 & 0.971 & 0.973 & 0.973 & 0.991 & 0.992 \\
        \hline
        2.00 & 0.398 & 0.995 & 0.995 & 0.995 & 0.996 & 0.645 & 0.599 & 0.760 & 0.603 & 0.975 & 0.967 & 0.985 & 0.969 & 0.990 & 0.976 \\
             & 0.449 & 0.994 & 0.995 & 0.995 & 0.995 & 0.623 & 0.690 & 0.757 & 0.670 & 0.978 & 0.979 & 0.984 & 0.981 & 0.990 & 0.988 \\
             & 0.522 & 0.995 & 0.996 & 0.995 & 0.996 & 0.658 & 0.675 & 0.678 & 0.648 & 0.980 & 0.989 & 0.989 & 0.990 & 0.984 & 0.988 \\
             & 0.571 & 0.996 & 0.996 & 0.996 & 0.996 & 0.694 & 0.734 & 0.708 & 0.743 & 0.983 & 0.986 & 0.993 & 0.985 & 0.985 & 0.986 \\
             & 0.624 & 0.996 & 0.996 & 0.996 & 0.996 & 0.739 & 0.751 & 0.720 & 0.718 & 0.985 & 0.989 & 0.993 & 0.989 & 0.982 & 0.990 \\
             & 0.681 & 0.996 & 0.996 & 0.995 & 0.996 & 0.788 & 0.773 & 0.732 & 0.718 & 0.993 & 0.992 & 0.995 & 0.993 & 0.986 & 0.991 \\
        \hline
    \end{tabular}
    }
\end{table}

Figs.~\ref{fig:pressure-corr-linear-ptp}-\ref{fig:pressure-corr-linear-hr} compares the local pressure field in the PNM and IBM simulations for the linear graded samples. Both DT and MDT are considered, along with the conductance models in Sec.~\ref{sub:conductance-models}). For all packing fractions considered in this study, the local pressure field exhibited a near-perfect correlation between the IBM and PNM. This was evident for both the DT and MDT partitioning, along with both conductance models. This is further reflected in the Pearson correlation coefficient, $\rho_x$, which is given by:
\begin{equation} \label{eqn:coeff-pearson}
    \rho_x = \frac{\sum_i \left( x^\mathrm{PNM}_i - \overline{x^\mathrm{PNM}} \right)\left( x^\mathrm{IBM}_i - \overline{x^\mathrm{IBM}} \right)}{ \left\{ \sum_i \left( x^\mathrm{PNM}_i - \overline{x^\mathrm{PNM}} \right)^2 \sum_i \left( x^\mathrm{IBM}_i - \overline{x^\mathrm{IBM}} \right)^2 \right\}^{1/2}}
\end{equation}
where $x$ is the property of interest (that is, pressure, flux or force), $x^\mathrm{PNM}_i$ is the value in the PNM, $x^\mathrm{IBM}_i$ is the value in the IBM, $\overline{x^\mathrm{PNM}}$ is the mean value of that property in the PNM, and $\overline{x^\mathrm{IBM}}$ is the mean value of that property in the IBM. The Pearson correlation coefficient has values $\rho_x \in \left[ -1, \: 1 \right]$, where $\rho_x = 1$ indicates perfect correspondence between PNM and IBM values.  In Table~\ref{tab:corr-coeff-linear}, the Pearson correlation coefficient for pressure ($\rho_p^{\textrm{DT}}$ and $\rho_p^{\textrm{MDT}}$) is above $0.99$ for all linear graded samples, which confirms the visual observation of strong correlation. The correct identification of inlet and outlet nodes, along with the interpolated pressure at inlet and outlet nodes (as discussed in Sec.~\ref{sub:local-pressure-and-flow-field}) was found to have a significant effect on the strong correlation in local pressure.

\begin{figure}[H]
    \centering
    \begin{subfigure}[b]{0.49\textwidth}
        \includegraphics[scale=0.95]{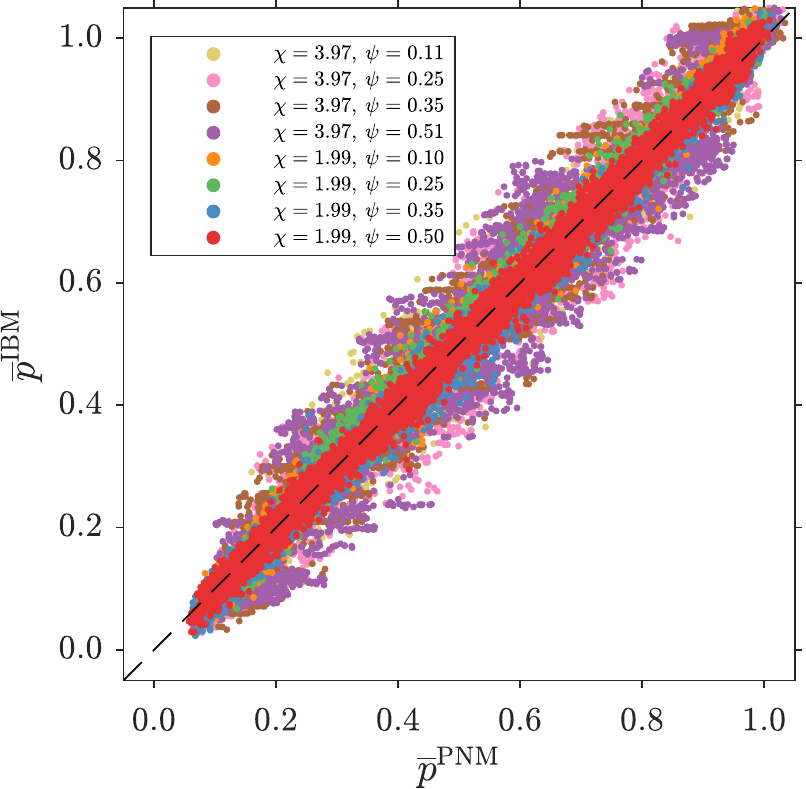}
        \caption{DT}
    \end{subfigure}
    \hfill
    \begin{subfigure}[b]{0.49\textwidth}
        \includegraphics[scale=0.95]{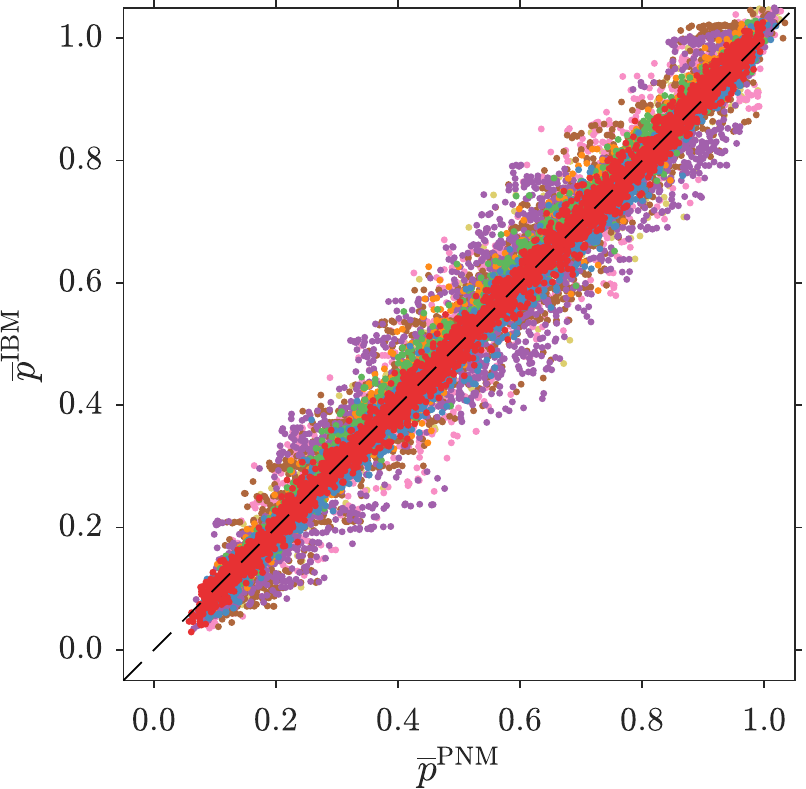}
        \caption{MDT}
    \end{subfigure}
    \caption{\label{fig:pressure-corr-bimodal-ptp}Correlation in normalised local pressure in the PNM ($\overline{p}^{\mathrm{PNM}}$) and the IBM ($\overline{p}^{\mathrm{IBM}}$) for bimodal samples using pore-throat-pore series conductance model in Sec.~\ref{subsub:pore-throat-pore-model}. Normalisation is conducted as per Fig.~\ref{fig:pressure-corr-linear-ptp}.}
\end{figure}

\begin{figure}[H]
    \centering
    \begin{subfigure}[b]{0.49\textwidth}
        \includegraphics[scale=0.95]{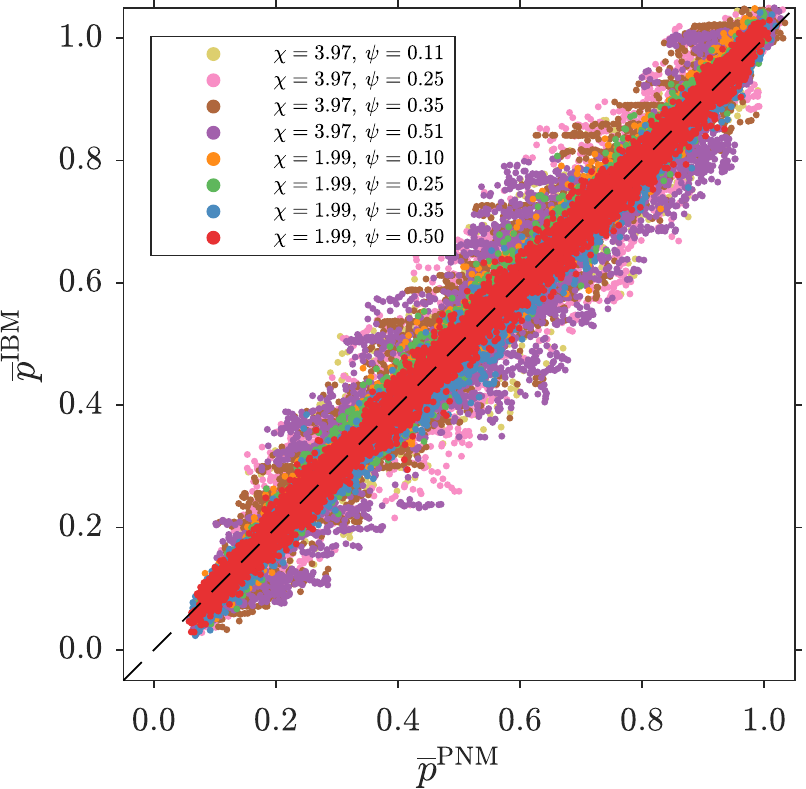}
        \caption{DT}
    \end{subfigure}
    \hfill
    \begin{subfigure}[b]{0.49\textwidth}
        \includegraphics[scale=0.95]{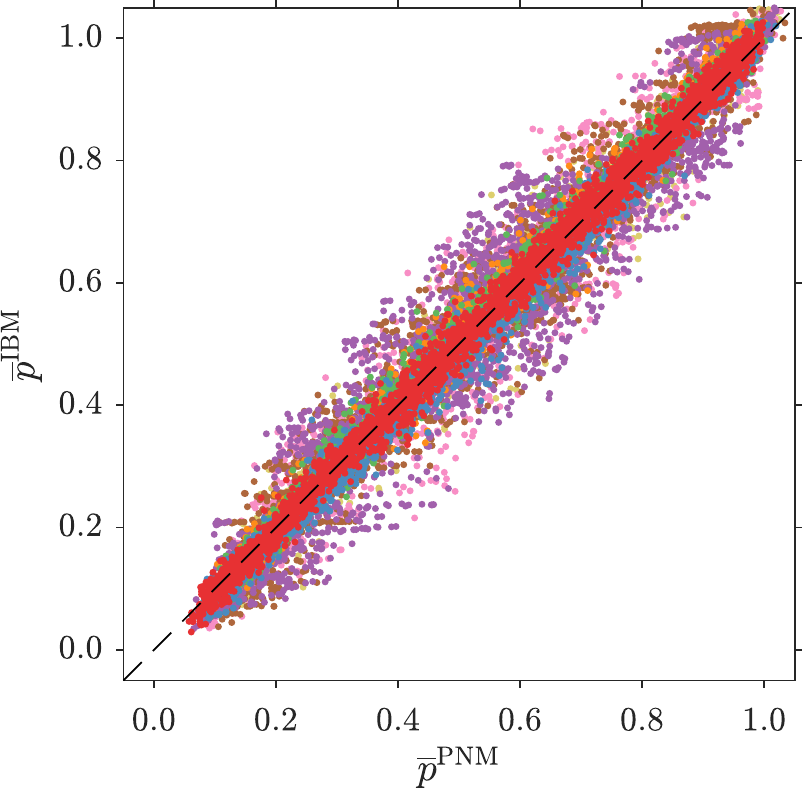}
        \caption{MDT}
    \end{subfigure}
    \caption{\label{fig:pressure-corr-bimodal-hr}Correlation in normalised local pressure in the PNM ($\overline{p}^{\mathrm{PNM}}$) and the IBM ($\overline{p}^{\mathrm{IBM}}$) for bimodal samples using hydraulic radius conductance model in Sec.~\ref{subsub:hydraulic-radius-model}. Normalisation is conducted as per Fig.~\ref{fig:pressure-corr-linear-ptp}.}
\end{figure}

A reasonably strong correlation is maintained in the bimodal samples (Figs.~\ref{fig:pressure-corr-bimodal-ptp}-\ref{fig:pressure-corr-bimodal-hr}), although slightly more scatter is observed. This is reflected in the Pearson correlation coefficients shown in Table~\ref{tab:corr-coeff-bimodal}, which remains greater than $0.97$ for all bimodal samples. It is acknowledged that the bimodal samples with $\chi = 3.97$ comprise very few coarse particles ($N_p^\mathrm{C} < 40$ in Table~\ref{tab:bimodal-samples}), which may compromise the statistical significance of the results. Nevertheless, this computational limitation does not detract from the observation that the PNM can accurately capture the local pressure field irrespective of the method of tessellation or the specified conductance model.

\clearpage

\begin{table}[ht]
    \caption{\label{tab:corr-coeff-bimodal}Pearson Correlation Coefficient in Bimodal Samples for the Pore-Throat-Pore (PTP) and Hydraulic Radius (HR) Conductance Model}
    \centering
    \resizebox{\textwidth}{!}{
    \begin{tabular}{|c|c|c||c c|c c||c c|c c||c c|c c|c c|}
        \hline
         & & &
            \multicolumn{4}{|c|}{Pressure} & 
            \multicolumn{4}{|c|}{Flux} & 
            \multicolumn{6}{|c|}{Force} \\
        \hline
         & & &
            \multicolumn{2}{|c|}{PTP} & \multicolumn{2}{|c|}{HR} &
            \multicolumn{2}{|c|}{PTP} & \multicolumn{2}{|c|}{HR} &
            \multicolumn{2}{|c|}{PTP} & \multicolumn{2}{|c|}{HR} & \multicolumn{2}{|c|}{IBM} \\
        \hline
        $\chi$ & $\psi$ & $\phi$ & 
            $\rho_p^{\textrm{DT}}$ & $\rho_p^{\textrm{MDT}}$ & $\rho_p^{\textrm{DT}}$ & $\rho_p^{\textrm{MDT}}$ & 
            $\rho_q^{\textrm{DT}}$ & $\rho_q^{\textrm{MDT}}$ & $\rho_q^{\textrm{DT}}$ & $\rho_q^{\textrm{MDT}}$ & 
            $\rho_\Fm^{\textrm{DT}}$ & $\rho_\Fm^{\textrm{MDT}}$ & $\rho_\Fm^{\textrm{DT}}$ & $\rho_\Fm^{\textrm{MDT}}$ & $\rho_\Fm^{\textrm{DT}}$ & $\rho_\Fm^{\textrm{MDT}}$ \\
        \hline
        \hline
        1.99 & 0.10 & 0.621 & 0.995 & 0.994 & 0.995 & 0.995 & 0.714 & 0.753 & 0.716 & 0.768 & 0.984 & 0.977 & 0.988 & 0.981 & 0.992 & 0.989 \\
             &      & 0.656 & 0.996 & 0.993 & 0.995 & 0.995 & 0.766 & 0.819 & 0.769 & 0.806 & 0.983 & 0.978 & 0.986 & 0.986 & 0.989 & 0.991 \\
        \hline
             & 0.25 & 0.622 & 0.996 & 0.995 & 0.995 & 0.996 & 0.741 & 0.768 & 0.740 & 0.758 & 0.982 & 0.984 & 0.989 & 0.987 & 0.983 & 0.988 \\
             &      & 0.670 & 0.996 & 0.996 & 0.996 & 0.996 & 0.774 & 0.778 & 0.710 & 0.779 & 0.987 & 0.987 & 0.990 & 0.990 & 0.991 & 0.994 \\
        \hline
             & 0.35 & 0.622 & 0.997 & 0.996 & 0.997 & 0.997 & 0.703 & 0.739 & 0.745 & 0.723 & 0.986 & 0.988 & 0.991 & 0.991 & 0.986 & 0.992 \\
             &      & 0.673 & 0.997 & 0.997 & 0.997 & 0.997 & 0.783 & 0.806 & 0.779 & 0.795 & 0.992 & 0.990 & 0.993 & 0.992 & 0.990 & 0.993 \\
        \hline
             & 0.50 & 0.622 & 0.997 & 0.998 & 0.997 & 0.997 & 0.715 & 0.750 & 0.714 & 0.751 & 0.988 & 0.990 & 0.992 & 0.991 & 0.986 & 0.990 \\
             &      & 0.665 & 0.998 & 0.998 & 0.997 & 0.998 & 0.785 & 0.782 & 0.789 & 0.779 & 0.992 & 0.991 & 0.993 & 0.992 & 0.987 & 0.990 \\
        \hline
        3.97 & 0.11 & 0.645 & 0.981 & 0.987 & 0.977 & 0.985 & 0.698 & 0.774 & 0.502 & 0.654 & 0.970 & 0.996 & 0.994 & 0.996 & 0.987 & 0.996 \\
             &      & 0.685 & 0.980 & 0.983 & 0.976 & 0.982 & 0.736 & 0.709 & 0.628 & 0.647 & 0.996 & 0.993 & 0.995 & 0.996 & 0.975 & 0.987 \\
        \hline
             & 0.25 & 0.648 & 0.984 & 0.980 & 0.977 & 0.975 & 0.732 & 0.730 & 0.665 & 0.662 & 0.997 & 0.999 & 0.999 & 0.999 & 0.972 & 0.990 \\
             &      & 0.751 & 0.979 & 0.980 & 0.976 & 0.977 & 0.846 & 0.806 & 0.737 & 0.765 & 0.998 & 0.998 & 0.997 & 0.997 & 0.970 & 0.988 \\
        \hline
             & 0.35 & 0.651 & 0.984 & 0.984 & 0.980 & 0.983 & 0.562 & 0.363 & 0.381 & 0.330 & 0.987 & 0.981 & 0.986 & 0.990 & 0.968 & 0.981 \\
             &      & 0.740 & 0.982 & 0.982 & 0.981 & 0.982 & 0.708 & 0.594 & 0.566 & 0.574 & 0.995 & 0.996 & 0.994 & 0.995 & 0.971 & 0.983 \\
        \hline
             & 0.51 & 0.621 & 0.990 & 0.985 & 0.986 & 0.983 & 0.711 & 0.743 & 0.733 & 0.706 & 0.996 & 0.995 & 0.996 & 0.996 & 0.801 & 0.928 \\
             &      & 0.715 & 0.985 & 0.980 & 0.982 & 0.978 & 0.819 & 0.761 & 0.698 & 0.729 & 0.997 & 0.998 & 0.997 & 0.997 & 0.732 & 0.887 \\
        \hline
    \end{tabular}
    }
\end{table}

\begin{figure}[H]
    \centering
    \includegraphics[scale=0.94]{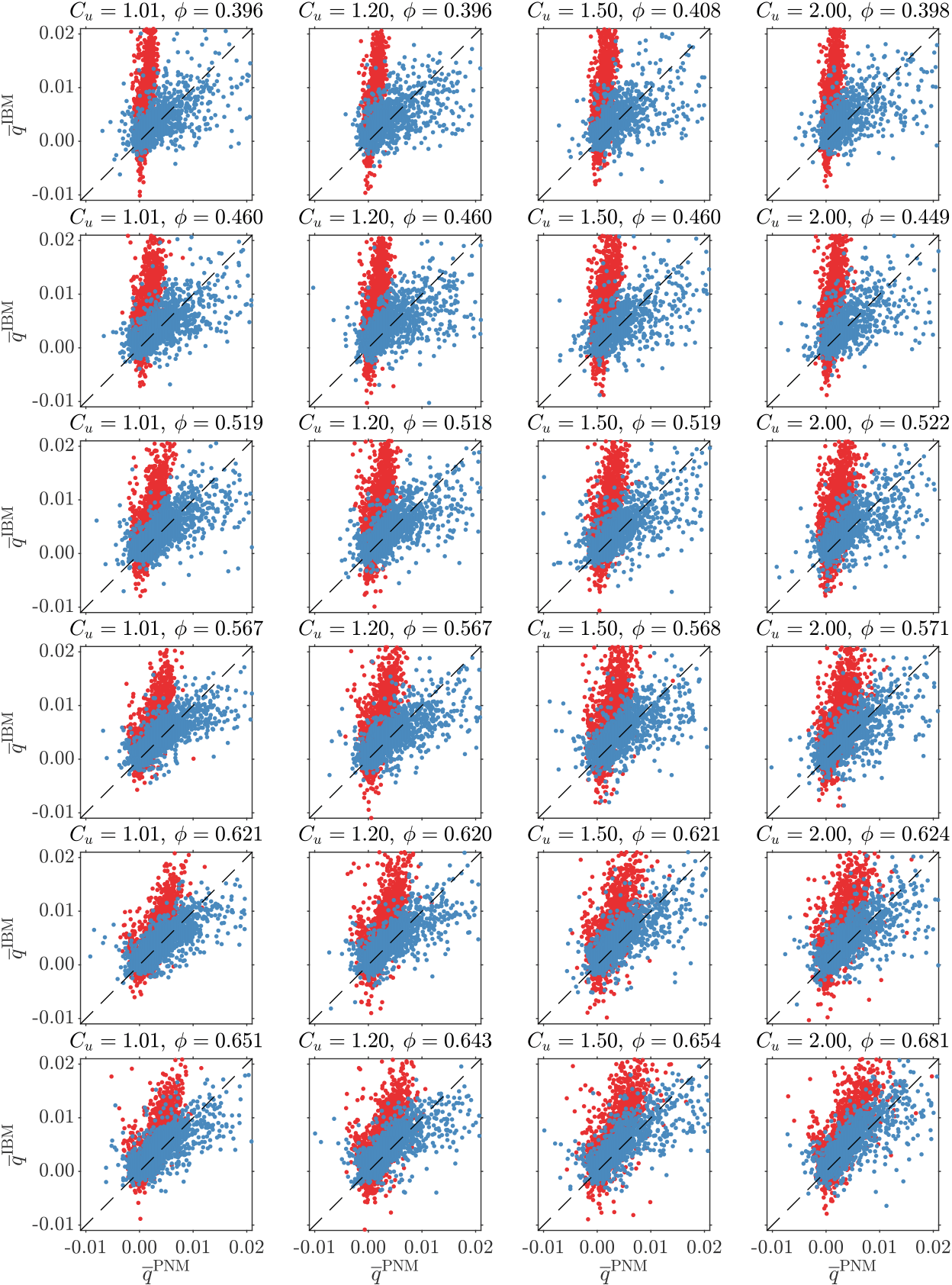}
    \caption{\label{fig:flux-corr-linear-ptp}Correlation in normalised local flux (flow rate) in the PNM ($\overline{q}^{\mathrm{PNM}}$) and the IBM ($\overline{q}^{\mathrm{IBM}}$) for linear graded samples using the pore-throat-pore series conductance model in Sec.~\ref{subsub:pore-throat-pore-model}. Normalisation is conducted by $\overline{q} = \frac{q}{q_{\mathrm{ave}}}$, where $q$ is the flux through network edge element and $q_{\mathrm{ave}} = UA_x$ is the average sample flux with $U = 2 \times 10^{-4}$m/s is the boundary inlet velocity and $A_x$ is the cross-sectional area of the sample. Red markers indicate the data points for DT, while blue markers indicate data points for MDT.}
\end{figure}

\begin{figure}[H]
    \centering
    \includegraphics[scale=0.94]{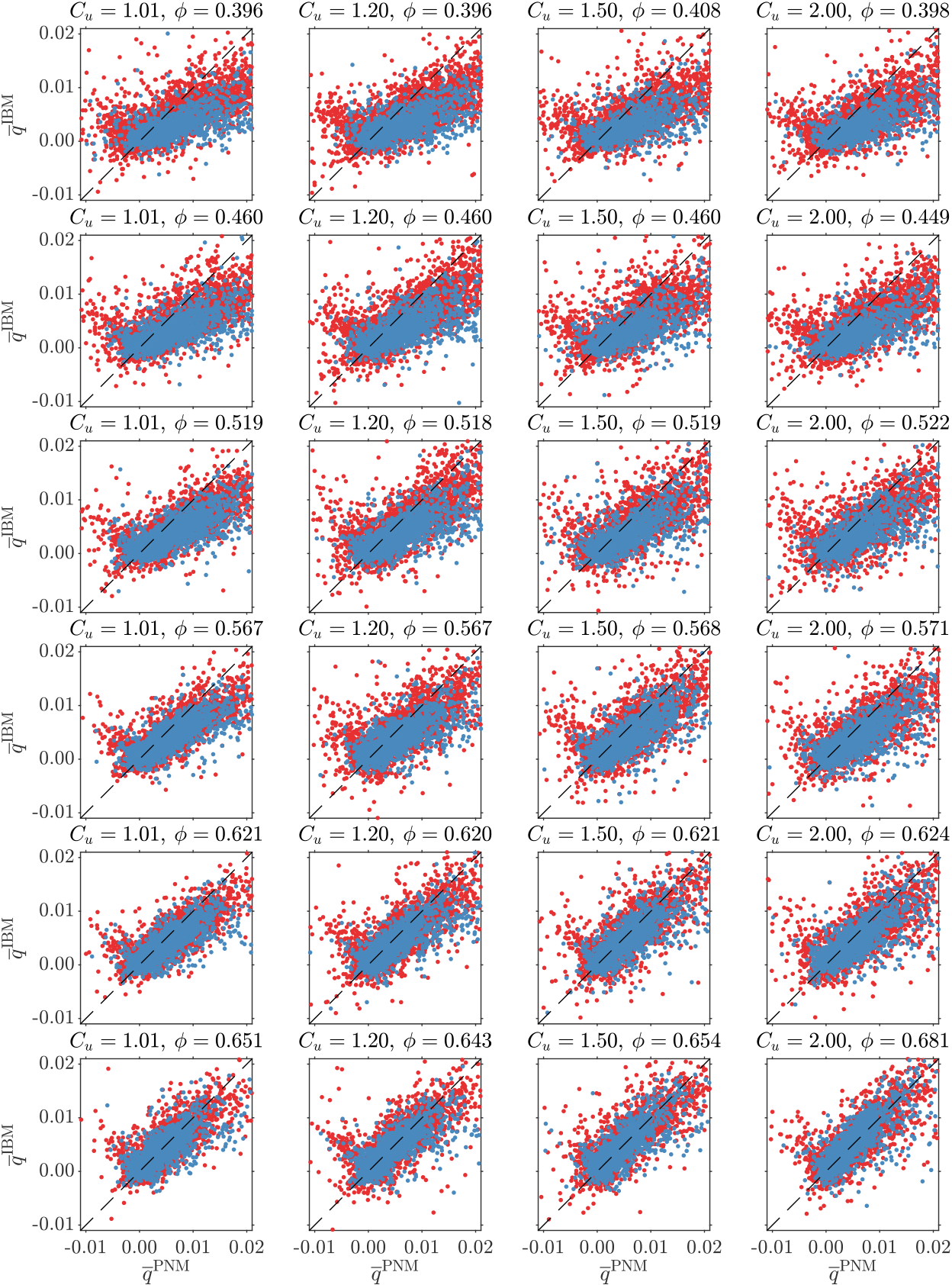}
    \caption{\label{fig:flux-corr-linear-hr}Correlation in normalised local flux (flow rate) in the PNM ($\overline{q}^{\mathrm{PNM}}$) and the IBM ($\overline{q}^{\mathrm{IBM}}$) for linear graded samples using the hydraulic radius conductance model in Sec.~\ref{subsub:hydraulic-radius-model}. Normalisation is conducted as per Fig.~\ref{fig:flux-corr-linear-ptp}. Red markers indicate the data points for DT, while blue markers indicate data points for MDT.}
\end{figure}

\begin{figure}[H]
    \centering
    \includegraphics[scale=0.94]{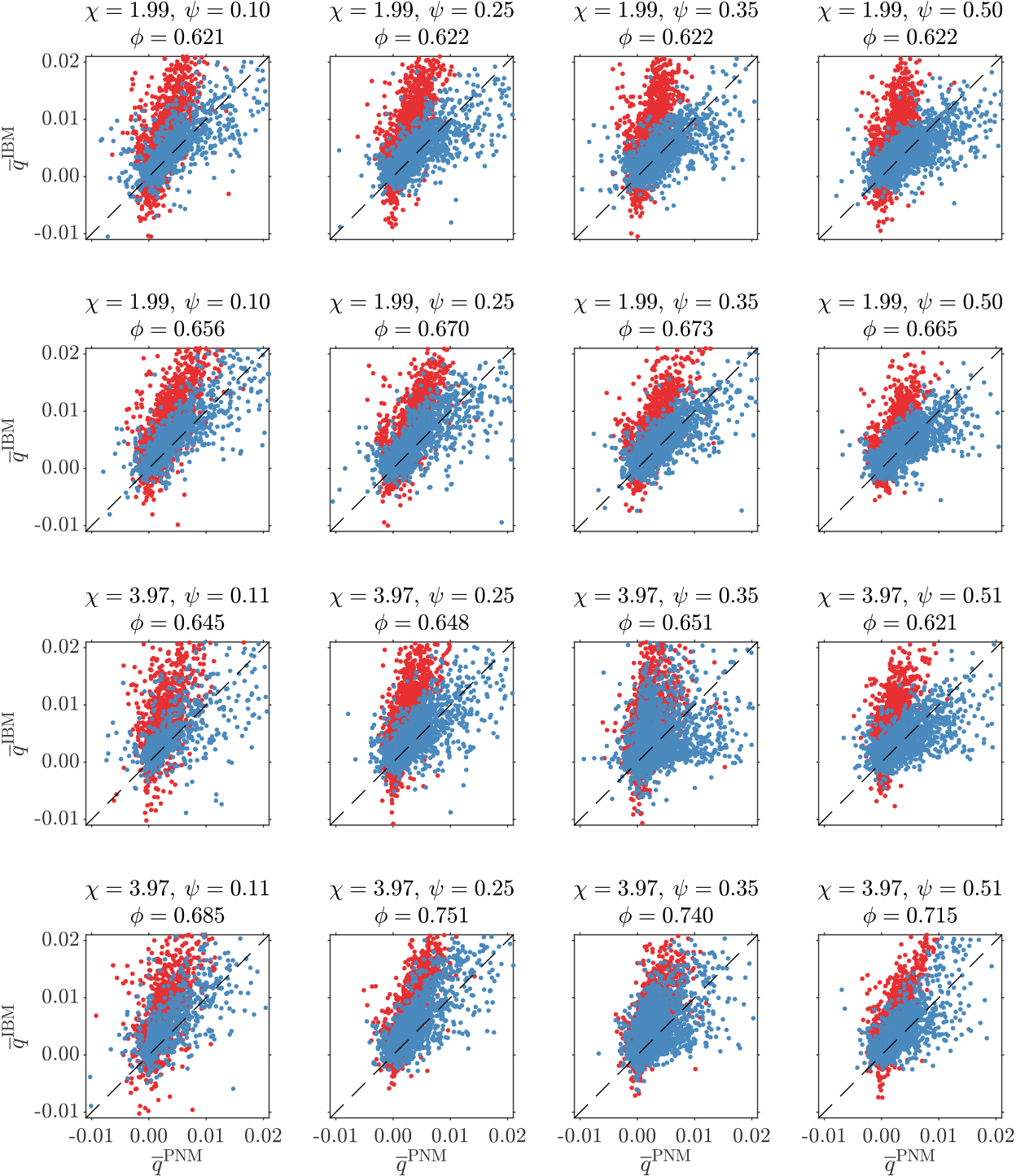}
    \caption{\label{fig:flux-corr-bimodal-ptp}Correlation in normalised local flux (flow rate) in the PNM ($\overline{q}^{\mathrm{PNM}}$) and the IBM ($\overline{q}^{\mathrm{IBM}}$) for bimodal samples using the pore-throat-pore series conductance model in Sec.~\ref{subsub:pore-throat-pore-model}. Normalisation is conducted as per Fig.~\ref{fig:flux-corr-linear-ptp}. Red markers indicate the data points for DT, while blue markers indicate data points for MDT.}
\end{figure}

\begin{figure}[H]
    \centering
    \includegraphics[scale=0.94]{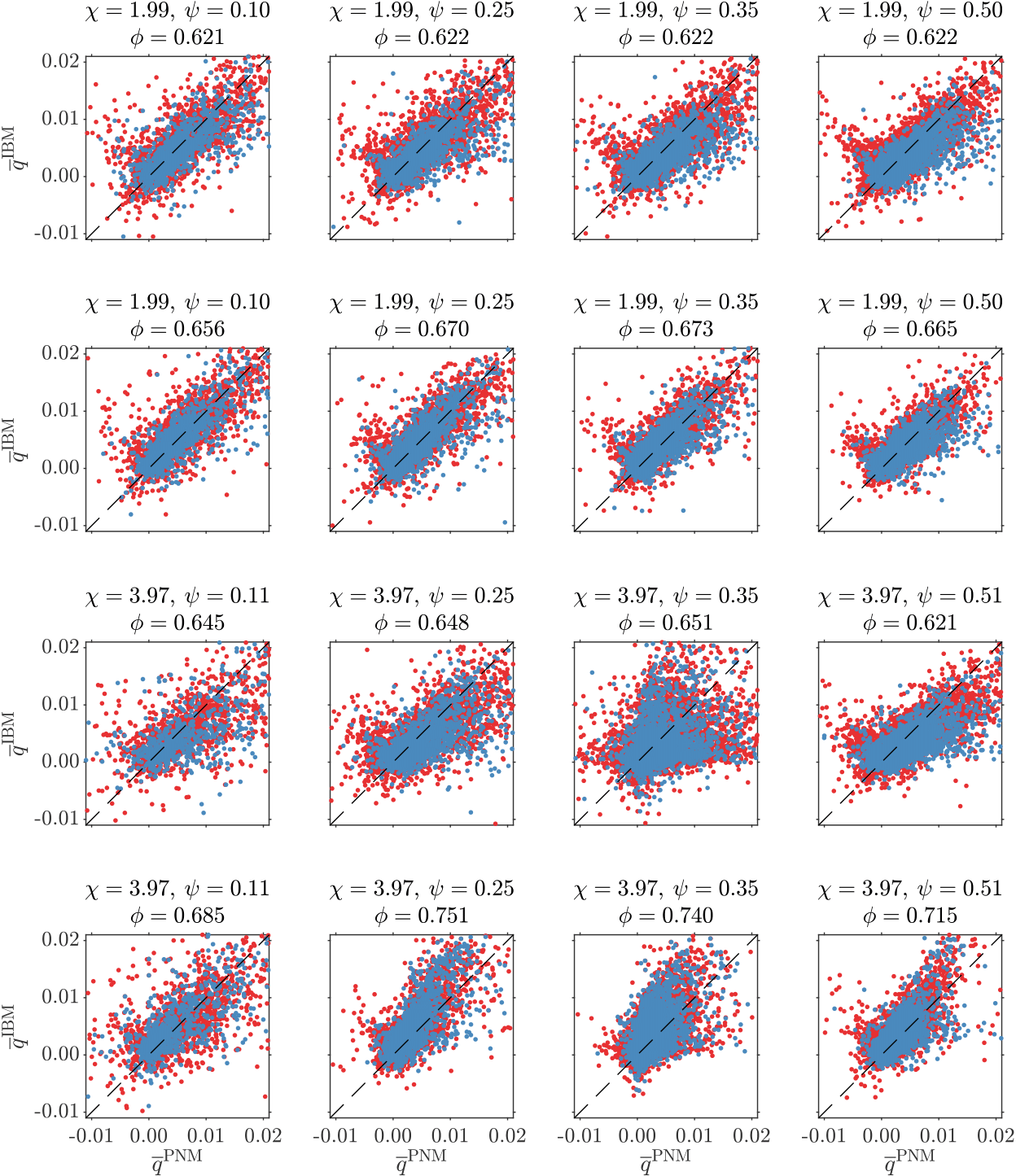}
    \caption{\label{fig:flux-corr-bimodal-hr}Correlation in normalised local flux (flow rate) in the PNM ($\overline{q}^{\mathrm{PNM}}$) and the IBM ($\overline{q}^{\mathrm{IBM}}$) for bimodal samples using the hydraulic radius conductance model in Sec.~\ref{subsub:hydraulic-radius-model}. Normalisation is conducted as per Fig.~\ref{fig:flux-corr-linear-ptp}. Red markers indicate the data points for DT, while blue markers indicate data points for MDT.}
\end{figure}

The correlation in local flux is shown in Figs.~\ref{fig:flux-corr-linear-ptp}-\ref{fig:flux-corr-linear-hr} for linear graded samples using both conductance models, and in Figs.~\ref{fig:flux-corr-bimodal-ptp}-\ref{fig:flux-corr-bimodal-hr} for the bimodal samples. Unlike the near-perfect correlation in the local pressure field, the local flux correlation shows significant scatter. 

When applying the pore-throat-pore series conductance model, Fig.~\ref{fig:flux-corr-linear-ptp} and Fig.~\ref{fig:flux-corr-bimodal-ptp} shows significant difference in the local flux correlation when the DT and MDT partitions are compared, with the MDT better capturing the local flux. In the linear graded samples, the difference between the MDT and DT predictions is most noticeable in the least dense samples ($\phi \approx 0.40$, first row in Fig.~\ref{fig:flux-corr-linear-ptp}). The difference between the DT and MDT data reduces with increasing packing fraction, as evident for the cases with $\phi \approx 0.65$ (last row in Fig.~\ref{fig:flux-corr-linear-ptp}). An increase in $C_u$ results in a wider scatter in the correlation plots for all packing fractions, which can be seen by comparing the first column in Fig.~\ref{fig:flux-corr-linear-ptp} for $C_u = 1.01$ with the last column in Fig.~\ref{fig:flux-corr-linear-ptp} for $C_u = 2.00$. 

Similar trends are observed in the bimodal samples (Fig.~\ref{fig:flux-corr-bimodal-ptp}), noting that the bimodal cases only consider stress-percolating assemblies (and have equivalent density to the last two rows in Fig.~\ref{fig:flux-corr-linear-ptp} for linear graded samples). As such, the difference in flux correlation for DT and MDT is less pronounced, but the MDT is consistently in better agreement with the IBM data. Furthermore, there is wider scatter for data points at large fluxes, particularly in the MDT. This is associated with non-physically large constrictions which have not been merged, as also recognised by \citet{shire2013} in quantifying anisotropy of constrictions.

The difference in local flux correlation due to the DT and MDT partitions is less pronounced when considering the hydraulic radius conductance model (Fig.~\ref{fig:flux-corr-linear-hr} for linear graded samples and Fig.~\ref{fig:flux-corr-bimodal-hr} for bimodal samples). The DT method exhibits slightly larger scatter, confirming that the MDT better captures local flux correlation. Moreover, the correlation appears to improve with increasing packing fraction. This can be visualised by comparing the first row (low packing fraction) with the last row (high packing fraction) in Fig.~\ref{fig:flux-corr-linear-hr} for the linear graded samples. This differed from the observations for the pore-throat-pore series model in Fig.~\ref{fig:flux-corr-linear-ptp}, which showed the opposing trend, suggesting that certain conductance models are more appropriate for certain ranges of packing fractions.  

These visual observations of moderate correlation in local flux is confirmed by the Pearson correlation coefficient in Table~\ref{tab:corr-coeff-linear} for local flux ($\rho_q^{\textrm{DT}}$ and $\rho_q^{\textrm{MDT}}$), which generally take a value between $0.6$ and $0.8$. Note that the clear visual differences in the DT and MDT for the pore-throat-pore series model (Fig.~\ref{fig:flux-corr-linear-ptp} and Fig.~\ref{fig:flux-corr-bimodal-ptp}) are not apparent when comparing the Pearson correlation coefficient, as $\rho_x$ quantifies how close the two quantities are to any linear relationship and not necessarily the linear relationship with unit gradient (as marked on Fig.~\ref{fig:flux-corr-linear-ptp}), which corresponds to a perfect match between the IBM and PNM data. The Pearson correlation coefficients for the bimodal samples are shown in  Table~\ref{tab:corr-coeff-bimodal}. In general, a wider scatter is observed with increasing particle size ratio, $\chi$, and with decreasing fines content, $\psi$, but is of a similar magnitude to that seen in the linear graded samples.

\subsection{Shortest Paths and Streamline Profiles}

\begin{figure}[t]
  \centering
  \includegraphics[width=150mm]{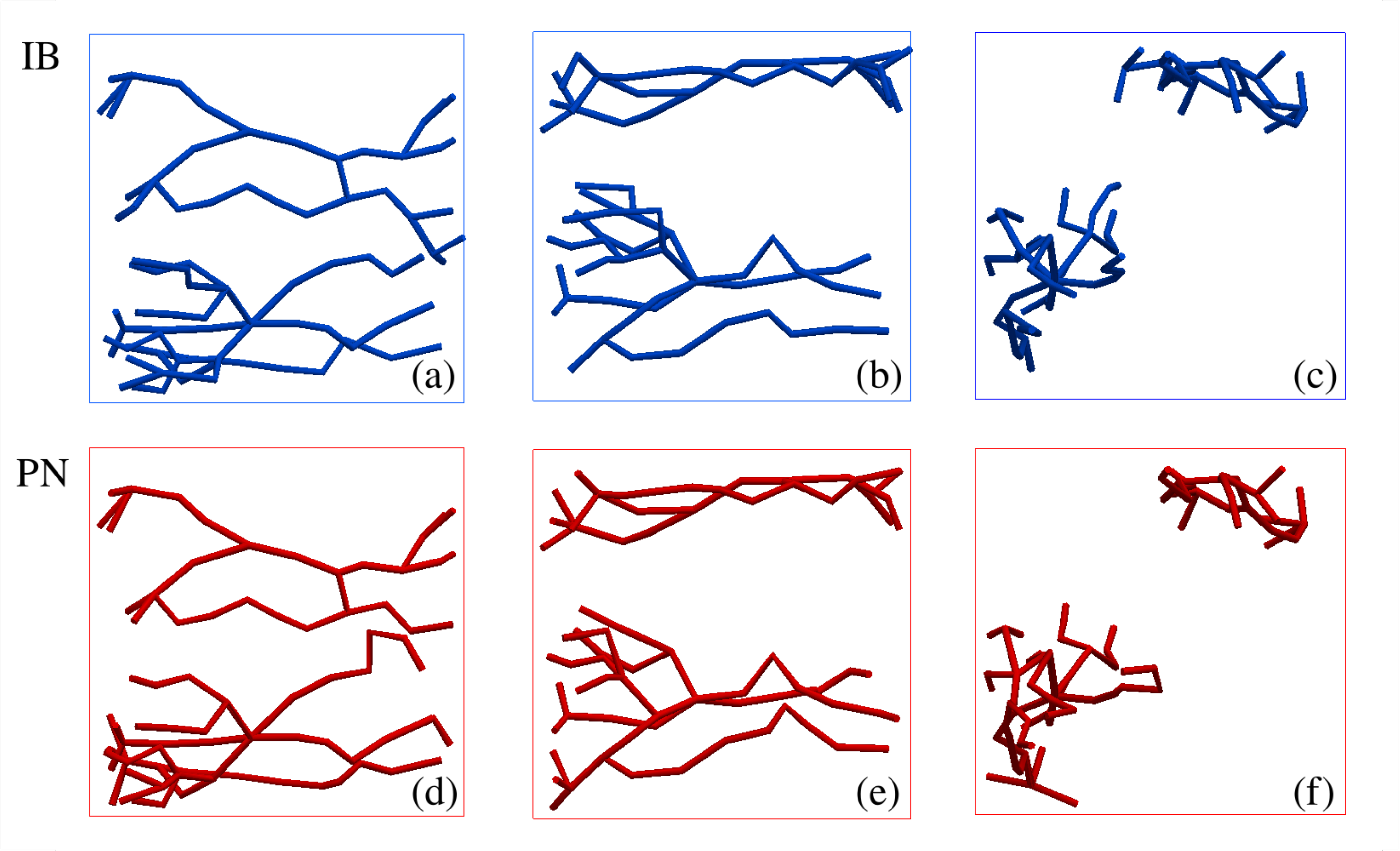}
  \caption{\label{fig:flow-paths-visual-comparison}Shortest paths (rendered with ParaView) for the $C_u = 1.01$, $\phi = 0.651$ case extracted from the IBM simulation (top row) in $x$-$y$ projection (a), $x$-$z$ projection (b), and $y$-$z$ projection, with the corresponding results from the PNM simulation (bottom row) shown in panels (d), (e) and (f).}
\end{figure}

Flow in a granular, porous medium is dominated by preferential flow channels. To evaluate the ability of PNM to capture these flow channels through the samples predicted by the IBM simulations, the shortest paths between pairs of inlet and outlet nodes were determined using Dijkstra's shortest path algorithm as implemented in the C++ software library \texttt{igraph} \citep{igraph}. A weighted, directed graph was constructed based on the PNM. Directionality was assigned by considering the fluxes, $q_{ij}$, along the edges of the graph, where edges with $q_{ij} < 0$ were reversed to ensure that all edges experienced a positive flux. A weight, $w_{ij}$ was assigned to each edge and was set to be inversely proportional to the flux, $q_{ij}$, in the corresponding constriction. The shortest path is the path which minimises the sum of the weights on the edges connecting an inlet and outlet node. This effectively captures the paths along which tracer particles would be transmitted most quickly. The shortest path from every inlet node to every outlet node was determined. Fig.~\ref{fig:flow-paths-visual-comparison} shows the ten shortest paths in the entire assembly based on the weighted, directed graph obtained from IBM and PNM fluxes. Despite the differences in local fluxes observed in Figs.~\ref{fig:flux-corr-linear-ptp}-\ref{fig:flux-corr-bimodal-ptp}, it is clear that the shortest paths predicted by the IBM and PNM are in close agreement.

The shortest paths can be described as ordered sequence of indices of the nodes traversed along the path. To quantify the similarity of the shortest paths demonstrated visually in Fig.~\ref{fig:flow-paths-visual-comparison}, a measure of the similarity, $S$, of pairs of paths between nodes indices $a$ and $b$ was defined as:
\begin{equation} \label{eqn:flow-path-similarity-measure}
  S(a, b) = \frac{|{}_{a}P^{b}_{\text{IBM}} \cap {}_{a}P^{b}_{\text{PNM}}|}{\frac{1}{2}\left(|{}_{a}P^{b}_{\text{IBM}}| + |{}_{a}P^{b}_{\text{PNM}}|\right)}
\end{equation}
An example is provided to explain the above path similarity measure. For a path starting at inlet node $0$ and ending at outlet node $5$, the set of nodal indices the path passes through in the IBM results may be ${}_{0}P^{5}_{\text{IBM}} = \{0, 1, 2, 3, 5\}$, and in the PNM results may be ${}_{0}P^{5}_{\text{PNM}} = \{0, 1, 2, 4, 5\}$. In this case, the similarity of the pair of paths from the IBM and PNM results between nodes $0$ and $5$ has a value $S(0,5) = 4/5$. This is a simplified example; flow paths in the present study typically traverse between 10 and 20 nodes.

For each inlet node, the shortest $N^{SP}$ paths are obtained from the IBM and PNM results. The $\widetilde{N^{SP}}$ pairs of paths with common outlet nodes were identified and a mean value of $S$ was calculated for each inlet node:
\begin{equation} \label{eqn:mean-flow-path-similarity-measure}
\langle S(a) \rangle =
\left\{
  \begin{array}{ll}
    \frac{1}{\widetilde{N^{SP}}}\sum^{\widetilde{N^{SP}}}_{j = 1} S(a, b_j)  & \mbox{if } \widetilde{N^{SP}} > 0 \\
    0 & \mbox{otherwise}
  \end{array}
\right.,
\end{equation}
where $b_j$ is the index of the outlet node of the $j\textsuperscript{th}$ path of the $\widetilde{N^{SP}}$ paths from the input node with index $a$ which have common output nodes in the IBM and PNM data.

\begin{figure}[t]
  \centering
  \includegraphics[width=140mm]{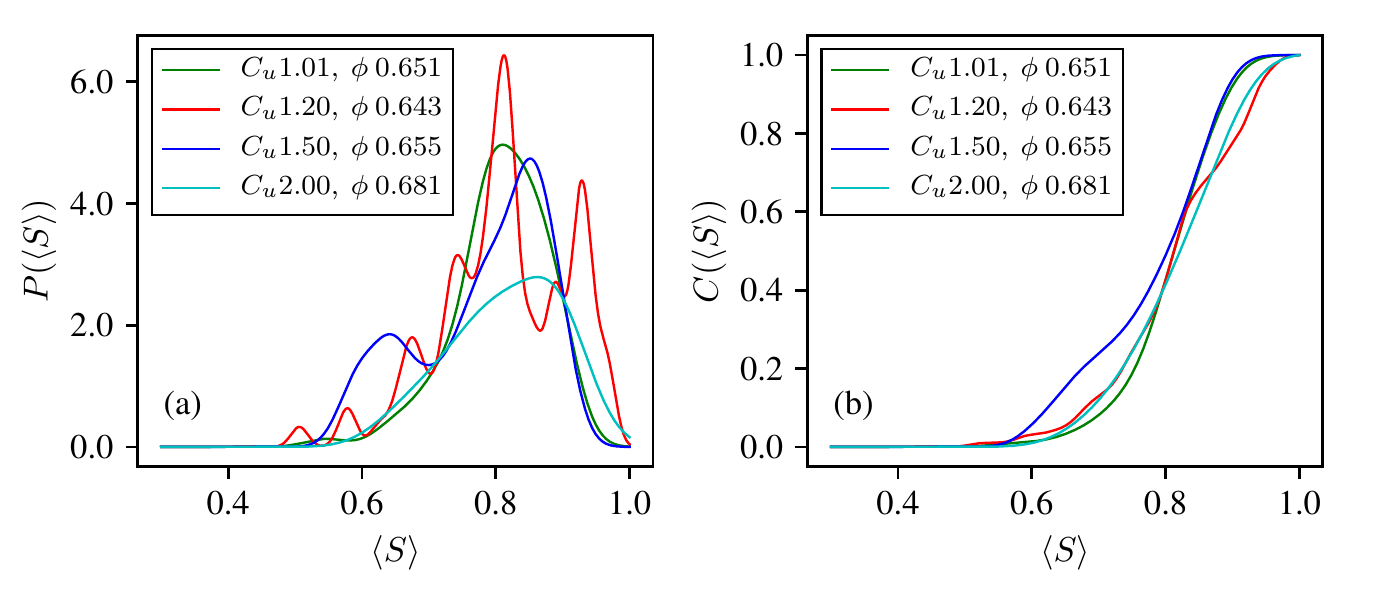}
  \caption{\label{fig:path-similarity-cdf-trim_90}(a) Probability density, $P \left( \langle S \rangle \right)$, of the path similarity measure, $\langle S \rangle$, for the densest cases considered for $C_u = 1.01$, $1.20$, $1.50$, and $2.00$ with $N^{SP} = 90$. (b) Corresponding cumulative probability distribution $C\left( \langle S \rangle \right)$.}
\end{figure}

\begin{figure}[t]
  \centering
  \includegraphics[width=140mm]{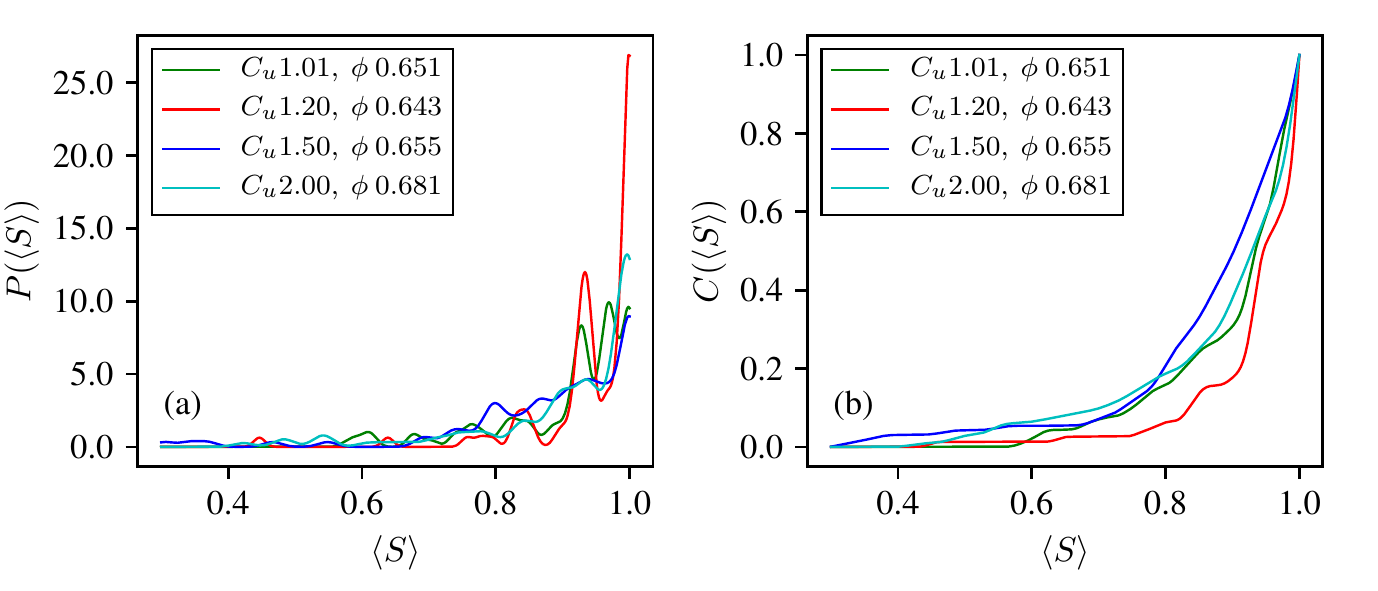}
  \caption{\label{fig:path-similarity-cdf-trim_10}(a) Probability density, $P \left( \langle S \rangle \right)$, of the path similarity measure, $\langle S \rangle$, for the densest cases considered for $C_u = 1.01$, $1.20$, $1.50$, and $2.00$ with $N^{SP} = 10$. (b) Corresponding cumulative probability distribution $C\left( \langle S \rangle \right)$.}
\end{figure}

The probability density functions and cumulative probability distributions of $\langle S \rangle$ for the densest case considered for the $C_u = 1.01$, $1.20$, $1.50$, and $2.00$ cases calculated with $N^{SP} = 90$ and $N^{SP} = 10$ are shown in Fig.~\ref{fig:path-similarity-cdf-trim_90} and Fig.~\ref{fig:path-similarity-cdf-trim_10}, respectively. The distributions of $\langle S \rangle$ shows that around half of the paths through the samples have in excess of 80\% of the nodes in common between the IBM and PNM results for all values of $C_u$ considered. This provides strong quantitative confirmation that the flow paths predicted by the PNM match very closely with those predicted by the IBM, supporting what is visually observed in Fig.~\ref{fig:flow-paths-visual-comparison}. Further, Fig.~\ref{fig:path-similarity-cdf-trim_10} shows that if only the top ten shortest paths are considered for each input node, then a better agreement is found, with the peak of the probability density function at $\langle S \rangle = 1$ for all $C_u$.

In addition to shortest paths, streamlines provide further insight into dominant flow channels. A streamline is a parametric curve which is instantaneously tangent to a vector field at all points. Streamlines tracing the paths of massless particles through the interstitial space of the samples were extracted from the highly resolved flow fields obtained from the IBM simulations. Seed points specifying the starting location of the streamlines were uniformly distributed over the inlet boundary. At steady state flow conditions a streamline is fully defined by its seed location.

\begin{figure}[t]
  \centering
  \includegraphics[width=140mm]{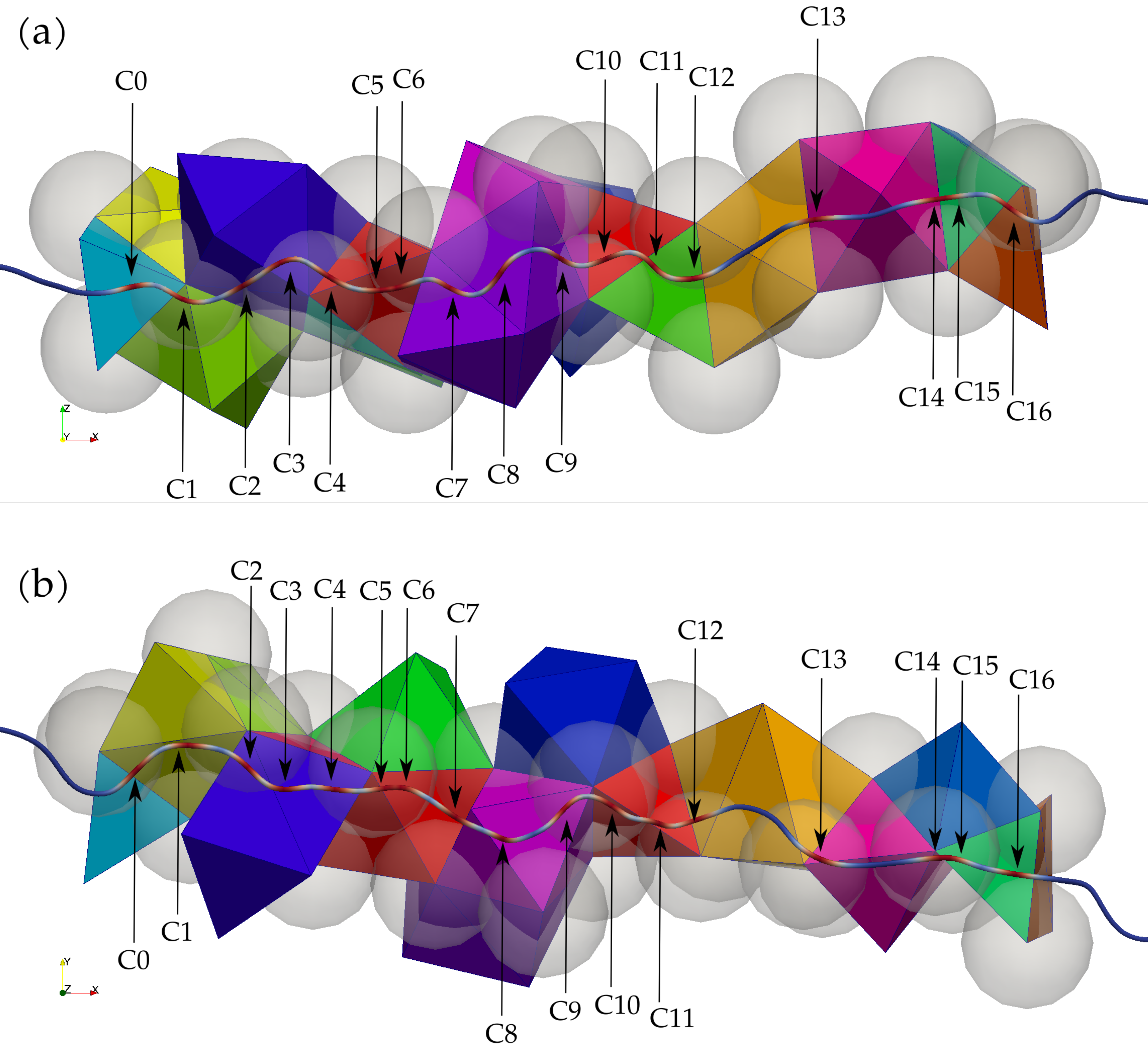}
  \caption{\label{fig:cu_1.01-conc_0.651-sl_11-constrictions-mod-cells-y-proj}A single streamline (rendered with ParaView) shown in (a) $x$-$z$ projection; and (b) $x$-$y$ projection. The corresponding cells of the MDT that the streamline passes through are shown in various colours (noting that the cells are a union of tetrahedral Delaunay cells). The particles forming the vertices of the MDT cells are shown with low opacity. The constrictions between adjacent cells are labelled and the velocity profile through the constriction is shown in Fig.~\ref{fig:constriction-velocity-profiles}. The colour of the streamline indicates the proximity to the labelled constrictions, as defined by $C_p(t)$ in Eq.~\ref{eqn:constriction-signal}.}
\end{figure}

\begin{figure}[H]
  \centering
  \includegraphics[width=140mm]{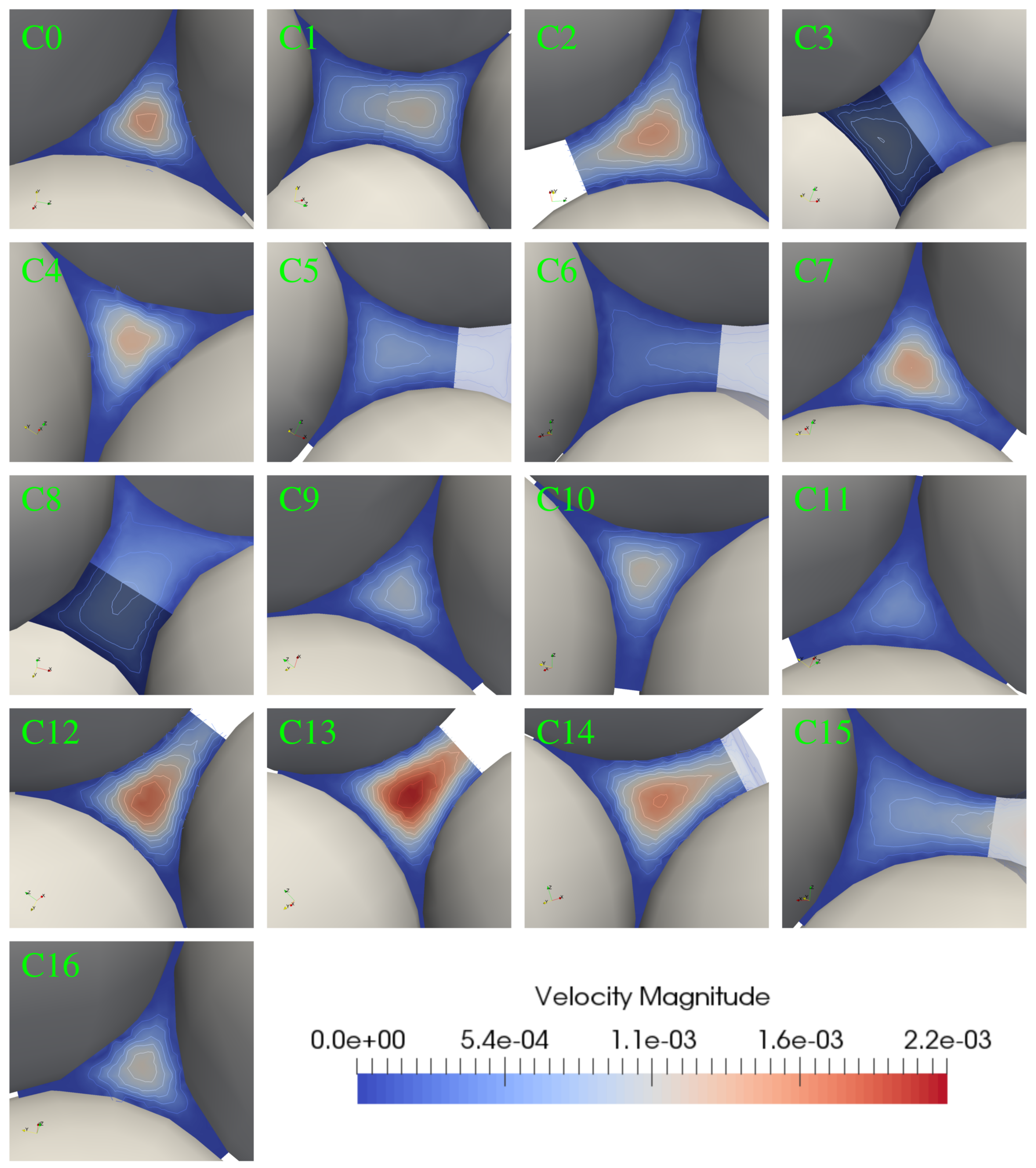}
  \caption{\label{fig:constriction-velocity-profiles}Velocity profiles for each of the constrictions consisting of one or more triangular faces (rendered with ParaView). The label at the top left of each image identifies the constriction.}
\end{figure}

The role of pore and constriction geometry in pressure loss at the micro-scale was assessed by combining the extracted streamlines with the pore geometry and topology specified by the MDT in the PNM. Streamlines are parametrised in terms of time, $t$. Starting from the seed location at time, $t=0$, as the tracer particle travels through the voids, it is enclosed within a single tetrahedral cell of the DT, and hence a single pore cell of the MDT, at all times. The distance, $C_d(t)$, along the streamline to the closest constriction plane was determined for every point comprising each streamline. A constriction proximity measure, $C_p(t)$, was defined at each point along the streamlines:
\begin{equation} \label{eqn:constriction-signal}
  C_p(t) = \exp{-\left(\dfrac{C_d(t)}{\alpha}\right)^{2}},
\end{equation}
where $\alpha$ is a characteristic length with value $\alpha = 0.1$~mm. This value was selected to obtain stiffly peaked Gaussians at the locations of the constrictions. In Fig.~\ref{fig:cu_1.01-conc_0.651-sl_11-constrictions-mod-cells-y-proj} a streamline extracted from the $C_u = 1.01$ and $\phi = 0.651$ case is shown. The traversed pore cells of the MDT are shown beneath the streamline, coloured by their cell indices to clearly delineate the boundaries between adjacent pore cells.  The streamline is coloured according to the value of $C_p$, such that points where the streamline crosses the constrictions are shown in red. Each constriction has been labelled. The velocity profile through each constriction is shown in Fig.~\ref{fig:constriction-velocity-profiles}. 

\begin{figure}[H]
  \centering
  \includegraphics[width=120mm]{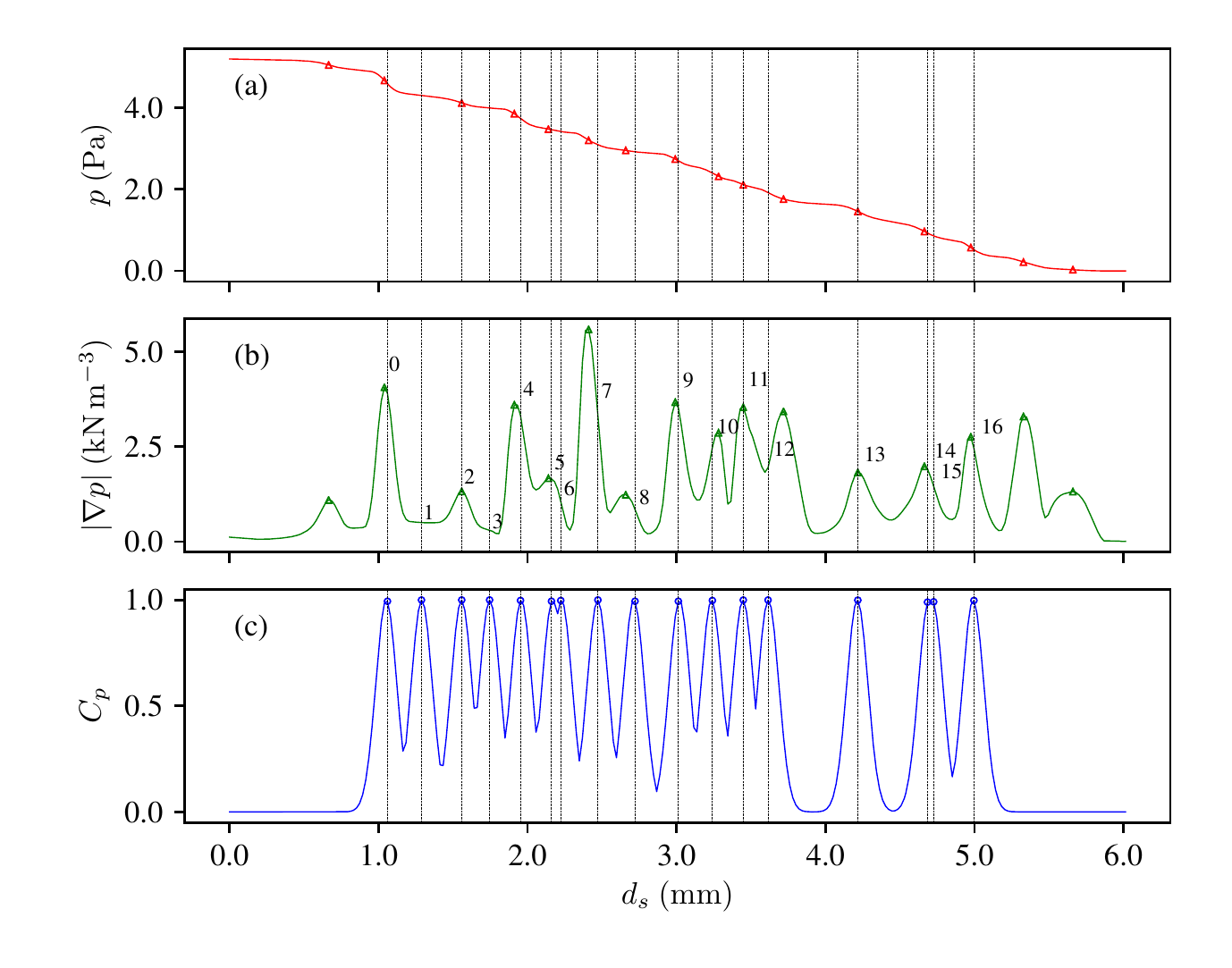}
  \caption{\label{fig:cu_1.01-conc_0.651-sl_11-press-pgrad-proxim}Pressure profile (a) and pressure gradient profile (b) against arc distance, $d_s$, along the path of a stream tracer particle for the $C_u = 1.01$, $\phi = 0.651$ case. In the lower panel (c), $C_p$, indicates the proximity to the nearest constriction.}
\end{figure}

\begin{figure}[H]
  \centering
  \includegraphics[width=120mm]{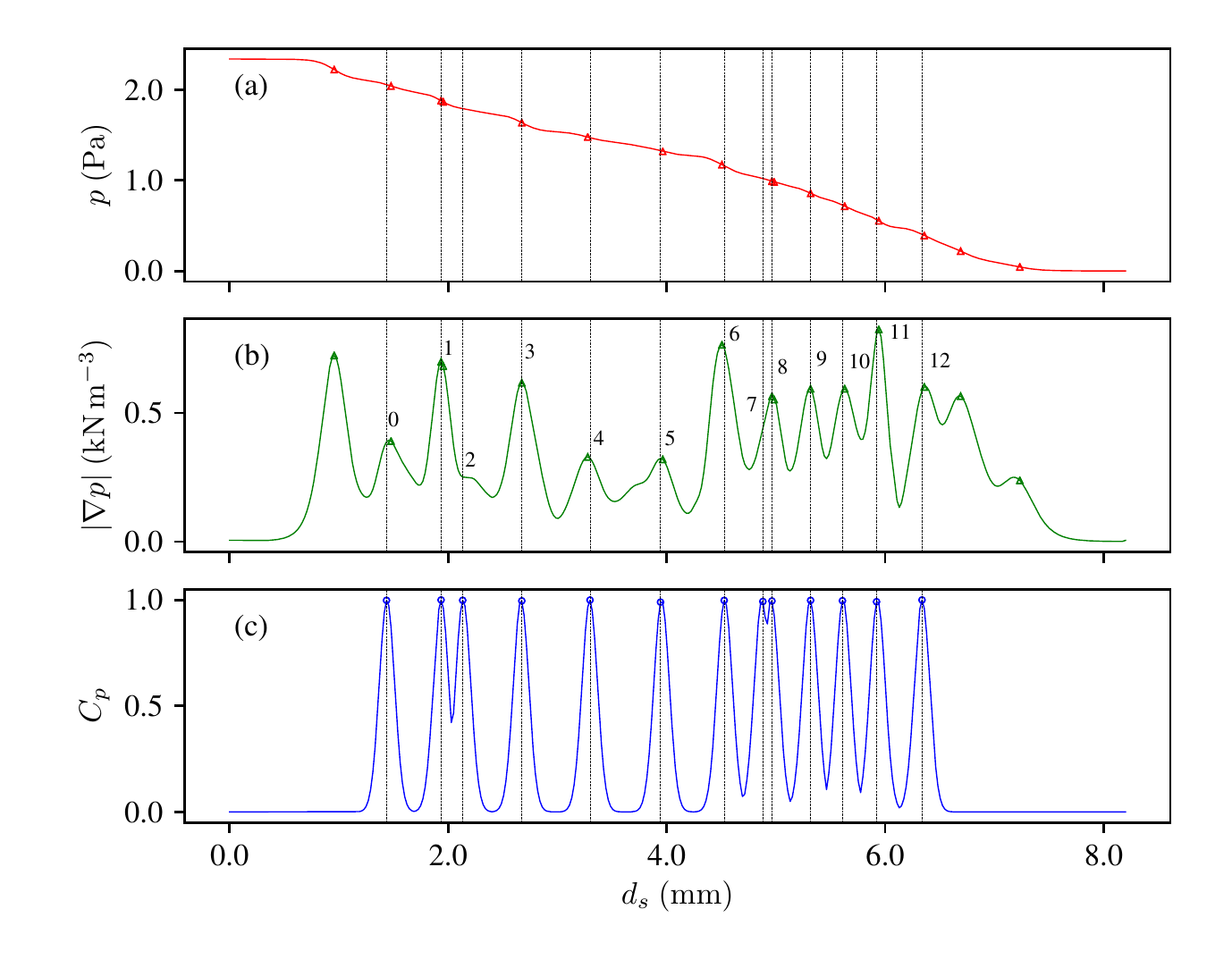}
  \caption{\label{fig:cu_1.50-conc_0.621-sl_1-press-pgrad-proxim}Pressure profile (a) and pressure gradient profile (b) against arc distance, $d_s$, along the path of a stream tracer particle for the $C_u = 1.50$, $\phi = 0.621$ case. In the lower panel (c), $C_p$, indicates the proximity to the nearest constriction.}
\end{figure}

The pressure, $p$, and magnitude of the pressure gradient, $|\nabla p|$, along with the value of $C_p$ at each point along the streamline in Fig.~\ref{fig:cu_1.01-conc_0.651-sl_11-constrictions-mod-cells-y-proj} are shown in Fig.~\ref{fig:cu_1.01-conc_0.651-sl_11-press-pgrad-proxim}. Peaks identified in the $C_p$ data are marked across the three axes with black lines. Peaks in $|\nabla p|$ are clearly observed to occur either exactly at, or very close to, a peak in $C_p$, that is, at the constriction planes. The first peak and the two last peaks in $|\nabla p|$ do not correspond to peaks in $C_p$ because these peaks occur at the boundary layers of the IBM data that are outside the domain of the MDT. In Fig.~\ref{fig:cu_1.50-conc_0.621-sl_1-press-pgrad-proxim} pressure drop data is shown for the $C_u = 1.50$, $\phi = 0.621$ case which exhibits the same behaviour seen in Fig.~\ref{fig:cu_1.01-conc_0.651-sl_11-press-pgrad-proxim}. This feature is exhibited in all other cases (not presented here). \citet{Indraratna2012} suggested that the constriction size distribution rather than the particle size distribution is the dominant factor in determining the macro-scale pressure drop. \citet{Taylor2017} conducted CFD simulations with soil samples re-created from X-ray micro-computed tomography images of real sands and found that the majority of pressure loss occurred at the constrictions. The data presented here support these earlier findings and highlights the importance of the constriction geometry in pore scale pressure loss. These results also demonstrate the ability of the Modified Delaunay Tessellation based on the inscribed sphere procedure described in \citet{sufian2015} to identify the constrictions which govern the permeability of porous media.

\subsection{Validation and Correlation for Fluid-Particle Interaction Force}

\begin{figure}[!ht]
    \centering
    \begin{subfigure}[b]{\textwidth}
        \centering
        \includegraphics[scale=0.98]{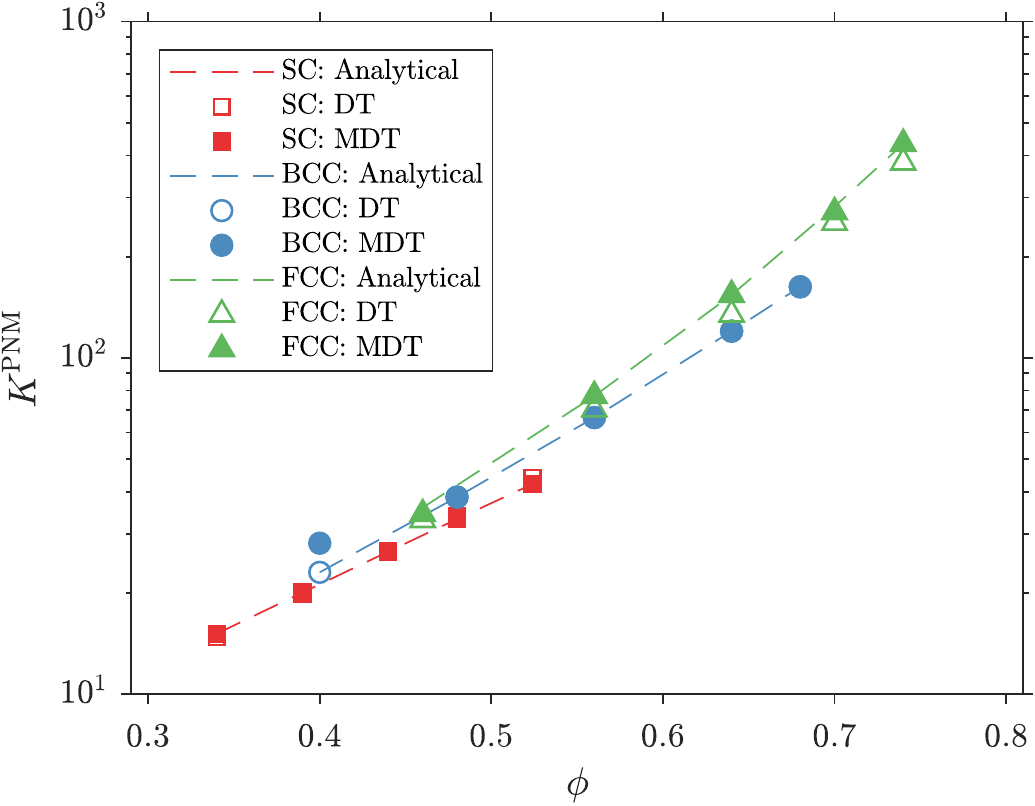}
        \caption{\label{fig:ordered-validation-ptp}Pore-throat-pore series model}
    \end{subfigure}
    \vfill
    \begin{subfigure}[b]{\textwidth}
        \centering
        \includegraphics[scale=0.98]{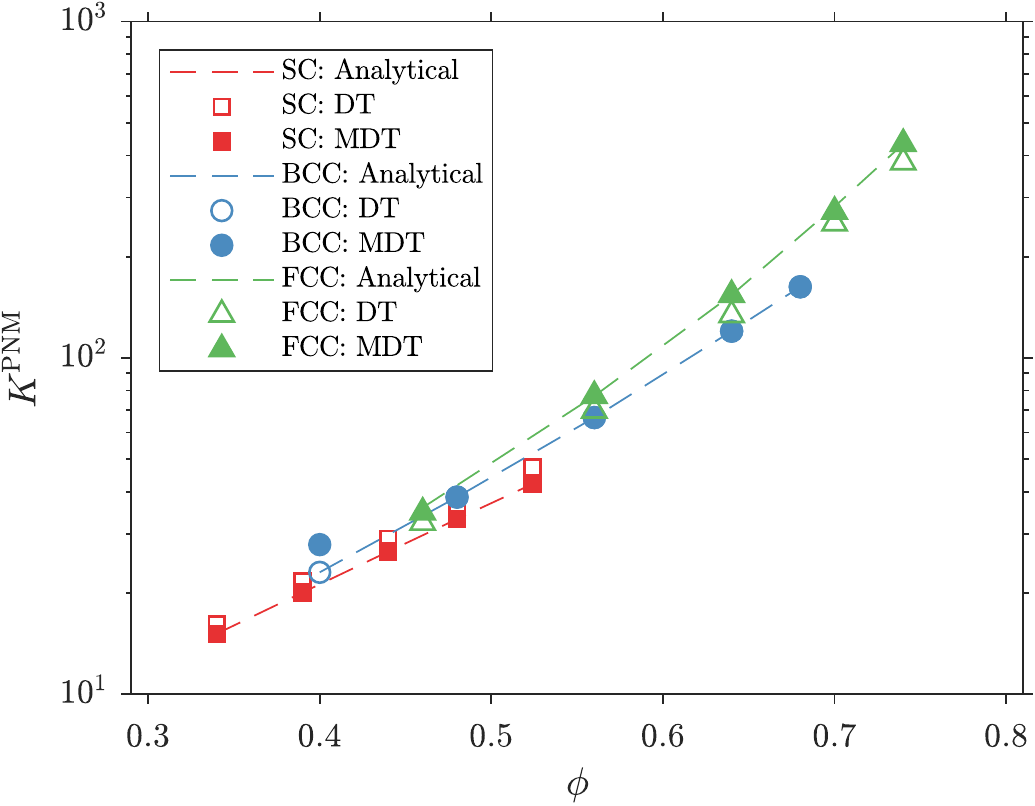}
        \caption{\label{fig:ordered-validation-hr}Hydraulic radius model}
    \end{subfigure}
    \caption{\label{fig:ordered-validation}Validation of network-based calculation of the normalised fluid-particle interaction force, for ordered assemblies, including simple cubic (SC), body centred cubic (BCC) and face centred cubic (FCC). $\Fm_{fp}$ is normalised by the Stokes drag force, which as per Eq.~\ref{eqn:zick-homsy-analytical-force}, is given by the scalar factor, $K$. Note that some of the DT markers are exactly in the same position as the MDT markers, and hence, cannot be seen in the figure.}
\end{figure}

A rigorous validation of the method proposed by \citet{chareyre2012} as outlined in Sec.~\ref{sub:Fluid-Particle-Interaction-Force} is presented here by comparing against the analytical solution in \citet{zickhomsy1982} for Stokes flow in various ordered packing configurations. The analytical solution for the (magnitude of the) fluid-particle interaction force is given by:
\begin{equation}\label{eqn:zick-homsy-analytical-force}
    \Fm^{k}_{fp} = 6\pi \mu r_p K U
\end{equation}
where $K$ represents a scalar multiple of the Stokes drag force experienced by a single particle, and $U$ is the fluid velocity, which can be calculated from the pressure gradient across the length of the sample by:
\begin{equation}
    \frac{\Delta p}{l} = -\frac{9 \mu \phi K U}{2 r_p^2}
\end{equation}
\citet{zickhomsy1982} tabulated values for the factor, $K$, for various ordered configurations and packing fractions. In this study, validation is performed for simple cubic (SC), body centred cubic (BCC) and face centred cubic (FCC) configurations, with five packing fractions considered for each configuration (Fig.~\ref{fig:ordered-validation}). In all cases, the particle radius was $r_p = 0.001$m and assemblies of different densities are formed by increasing the lattice spacing. The ordered assemblies were generated by replicating the crystal lattice for each ordered configuration $4 \times 4 \times 4$ in each direction. The number of particles ($N_p$) in each assembly can be found in Tables~\ref{tab:force-sc}-\ref{tab:force-fcc}.

\begin{table}[hb]
    \caption{\label{tab:force-sc}Comparison of fluid-particle interaction force in a simple cubic assembly from PNM simulation ($K^{\DT}$ and $K^{\MDT}$) and analytical solution ($K^{\mathrm{SC}}$) using the pore-throat-pore (PTP) and hydraulic radius (HR) conductance model}
    \centering
    \resizebox{\textwidth}{!}{
    \begin{tabular}{|c|c c c|c|c c c|c c c|c c c|c c c|}
        \hline
         & & & & & \multicolumn{6}{|c|}{PTP} &  \multicolumn{6}{|c|}{HR} \\ 
        $\phi$ & $N_p$ & $N_c^{\DT}$ & $N_c^{\MDT}$ & $K^{\mathrm{SC}}$ & 
            $K_{min}^{\DT}$ & $K_{mean}^{\DT}$ & $K_{max}^{\DT}$ & $K_{min}^{\MDT}$ & $K_{mean}^{\MDT}$ & $K_{max}^{\MDT}$ & 
            $K_{min}^{\DT}$ & $K_{mean}^{\DT}$ & $K_{max}^{\DT}$ & $K_{min}^{\MDT}$ & $K_{mean}^{\MDT}$ & $K_{max}^{\MDT}$ \\
        \hline
        \hline
        0.340 & 64 & 162 & 27 & 15.140 & 12.328 & 14.775 & 17.086 & 15.140 & 15.140 & 15.140 & 14.735 & 16.147 & 17.511 & 15.140 & 15.140 & 15.140 \\
        0.390 & 64 & 162 & 27 & 20.075 & 16.798 & 19.841 & 22.724 & 20.075 & 20.075 & 20.075 & 19.918 & 21.643 & 23.315 & 20.075 & 20.075 & 20.075 \\
        0.440 & 64 & 162 & 27 & 26.575 & 22.977 & 26.690 & 30.219 & 26.575 & 26.575 & 26.575 & 26.941 & 29.006 & 31.013 & 26.575 & 26.575 & 26.575 \\
        0.480 & 64 & 162 & 27 & 33.187 & 29.604 & 33.871 & 37.939 & 33.187 & 33.187 & 33.187 & 34.293 & 36.628 & 38.904 & 33.187 & 33.187 & 33.187 \\
        0.524 & 64 & 162 & 27 & 42.191 & 39.205 & 44.014 & 48.623 & 42.191 & 42.191 & 42.191 & 44.625 & 47.213 & 49.745 & 42.191 & 42.191 & 42.191 \\
        \hline
    \end{tabular}
    }
\end{table}

\begin{table}[hb]
    \caption{\label{tab:force-bcc}Comparison of fluid-particle interaction force in a body centred cubic assembly from PNM simulation ($K^{\DT}$ and $K^{\MDT}$) and analytical solution ($K^{\mathrm{BCC}}$) using the pore-throat-pore (PTP) and hydraulic radius (HR) conductance model}
    \centering
    \resizebox{\textwidth}{!}{
    \begin{tabular}{|c|c c c|c|c c c|c c c|c c c|c c c|}
        \hline
         & & & & & \multicolumn{6}{|c|}{PTP} &  \multicolumn{6}{|c|}{HR} \\
        $\phi$ & $N_p$ & $N_c^{\DT}$ & $N_c^{\MDT}$ & $K^{\mathrm{SC}}$ & 
            $K_{min}^{\DT}$ & $K_{mean}^{\DT}$ & $K_{max}^{\DT}$ & $K_{min}^{\MDT}$ & $K_{mean}^{\MDT}$ & $K_{max}^{\MDT}$ & 
            $K_{min}^{\DT}$ & $K_{mean}^{\DT}$ & $K_{max}^{\DT}$ & $K_{min}^{\MDT}$ & $K_{mean}^{\MDT}$ & $K_{max}^{\MDT}$ \\
        \hline
        \hline
        0.400 & 128 & 378 & 151 & 23.051 & 23.051 & 23.051 & 23.051 & 19.457 & 28.131 & 34.360 & 23.051 & 23.051 & 23.051 & 21.115 & 27.864 & 36.893 \\
        0.480 & 128 & 378 & 378 & 38.575 & 38.575 & 38.575 & 38.575 & 38.575 & 38.575 & 38.575 & 38.575 & 38.575 & 38.575 & 38.575 & 38.575 & 38.575 \\
        0.560 & 128 & 378 & 378 & 66.501 & 66.501 & 66.501 & 66.502 & 66.501 & 66.501 & 66.502 & 66.501 & 66.501 & 66.502 & 66.501 & 66.501 & 66.502 \\
        0.640 & 128 & 378 & 378 & 120.150 & 120.149 & 120.149 & 120.149 & 120.149 & 120.149 & 120.149 & 120.149 & 120.149 & 120.149 & 120.149 & 120.149 & 120.149 \\
        0.680 & 128 & 378 & 378 & 162.754 & 162.754 & 162.754 & 162.754 & 162.754 & 162.754 & 162.754 & 162.754 & 162.754 & 162.754 & 162.754 & 162.754 & 162.754 \\
        \hline
    \end{tabular}
    }
\end{table}

\begin{table}[hb]
    \caption{\label{tab:force-fcc}Comparison of fluid-particle interaction force in a face centred cubic assembly from PNM simulation ($K^{\DT}$ and $K^{\MDT}$) and analytical solution ($K^{\mathrm{FCC}}$) using the pore-throat-pore (PTP) and hydraulic radius (HR) conductance model}
    \centering
    \resizebox{\textwidth}{!}{
    \begin{tabular}{|c|c c c|c|c c c|c c c|c c c|c c c|}
        \hline
         & & & & & \multicolumn{6}{|c|}{PTP} &  \multicolumn{6}{|c|}{HR} \\
        $\phi$ & $N_p$ & $N_c^{\DT}$ & $N_c^{\MDT}$ & $K^{\mathrm{SC}}$ & 
            $K_{min}^{\DT}$ & $K_{mean}^{\DT}$ & $K_{max}^{\DT}$ & $K_{min}^{\MDT}$ & $K_{mean}^{\MDT}$ & $K_{max}^{\MDT}$ & 
            $K_{min}^{\DT}$ & $K_{mean}^{\DT}$ & $K_{max}^{\DT}$ & $K_{min}^{\MDT}$ & $K_{mean}^{\MDT}$ & $K_{max}^{\MDT}$ \\
        \hline
        \hline
        0.460 & 256 & 1006 & 580 & 35.894 & 27.039 & 32.976 & 44.678 & 31.383 & 34.362 & 38.179 & 26.505 & 32.392 & 41.706 & 31.605 & 34.750 & 38.724 \\
        0.560 & 256 & 775 & 451 & 77.064 & 59.758 & 70.273 & 92.970 & 77.065 & 77.181 & 77.299 & 59.476 & 69.772 & 95.987 & 77.065 & 77.181 & 77.299 \\
        0.640 & 256 & 775 & 451 & 153.839 & 122.997 & 134.616 & 177.084 & 153.832 & 154.063 & 154.300 & 121.388 & 134.168 & 177.705 & 153.832 & 154.063 & 154.300 \\
        0.700 & 256 & 928 & 535 & 281.829 & 170.959 & 252.439 & 324.629 & 225.921 & 271.517 & 288.436 & 170.240 & 251.381 & 326.687 & 224.840 & 272.480 & 290.999 \\
        0.740 & 256 & 775 & 451 & 432.674 & 330.260 & 380.939 & 514.843 & 432.655 & 433.305 & 433.971 & 331.249 & 381.750 & 515.121 & 432.655 & 433.305 & 433.971 \\
        \hline
    \end{tabular}
    }
\end{table}

A pore network based on the DT and MDT was generated following exactly the same procedure as the linear graded and bimodal samples using both conductance models detailed in Sec.~\ref{sub:conductance-models}. The number of unit cells in the DT ($N_c^{\DT}$) and MDT ($N_c^{\MDT}$) is listed in Tables~\ref{tab:force-sc}-\ref{tab:force-fcc}. In the SC cases, a significant amount of merging has occurred, but this remains consistent for all densities considered. In the BCC cases, no merging has taken place for all assemblies except for the least dense case, $\phi = 0.400$, which due to the larger lattice spacing would allow for inscribed spheres to intersect. Some merging is observed in the FCC cases, but unlike the SC and BCC cases, the assemblies at $\phi = 0.460$ and $\phi = 0.700$ exhibit different numbers of unit cells in the DT. This is due to boundary artefacts with the Delaunay triangulation algorithm, where additional tetrahedral cells are detected at the boundaries for the FCC configuration. In this validation study, these boundary artefacts have not been removed, to ensure that the procedure followed in the validation study is exactly the same as that undertaken for the linear graded and bimodal samples, where this issue was not observed.

Flow is simulated in these ordered networks by applying an inlet pressure of $1.00$Pa and an outlet pressure of $0.01$Pa. Fig.~\ref{fig:ordered-validation} compares the numerical results against the analytical solution for various ordered configurations and packing fractions, where the factor, $K$, is obtained from a cubic spline interpolation of the tabulated values in \citet{zickhomsy1982} and compared against the numerical value obtained from the PNM, $K^{\mathrm{PNM}}$, which is equal to the fluid-particle interaction force normalised by the equivalent Stokes force. The comparison is performed for only internal particles, as particles adjacent to the domain boundary have an incomplete force computation. The fluid-particle interaction force obtained from the MDT is more accurate compared to the DT. The only exception is the BCC case with $\phi = 0.400$, which is due to the increased degree of merging not evident in other BCC cases. Also, the increase number of unit cells in the FCC cases at $\phi = 0.460$ and $\phi = 0.700$ only resulted in a very slight discrepancy, providing further justification that the boundary artefacts had minimal influence.

While Fig.~\ref{fig:ordered-validation} shows the average fluid-particle interaction force over all particles, further details are provided in Tables~\ref{tab:force-sc}-\ref{tab:force-fcc}. In the SC cases, a range of forces is obtained for DT, as indicated by the different minimum ($K_{min}^{\DT}$) and maximum ($K_{max}^{\DT}$) values. For the pore-throat-pore conductance model, the fluid-particle interaction force was slightly underestimated ($K_{mean}^{\DT} < K^{\mathrm{SC}}$), while it was slightly overestimated for the hydraulic radius model when applying the DT. The same value is obtained for all particles when applying the MDT (as $K_{min}^{\MDT} = K_{max}^{\MDT} = K^{\mathrm{SC}}$, where $K^{\mathrm{SC}}$ is the analytical solution based on \citet{zickhomsy1982}). The same force value for all particles is also obtained in the BCC cases (with the exception of $\phi = 0.400$ for the reasons discussed above). The FCC cases exhibit similar response with a range of forces obtained for DT and similar forces in MDT (there is only a slight difference in $K_{min}^{\MDT}$ and $K_{max}^{\MDT}$ for $\phi = 0.560, \: 0.640, \: 0.740$). As per the discussion above, differences are noted for the FCC cases for $\phi = 0.460, \: 0.700$.

Therefore, the above validation provides strong support for the use of the method developed in \citet{chareyre2012} to calculate the fluid-particle interaction force. While the DT provides a reasonable estimate, the use the MDT results in a more accurate prediction of the force. This provides confidence in applying the method to linear graded and bimodal samples.

\begin{figure}[ht]
    \centering
    \begin{subfigure}[b]{0.48\textwidth}
        \includegraphics[scale=0.90]{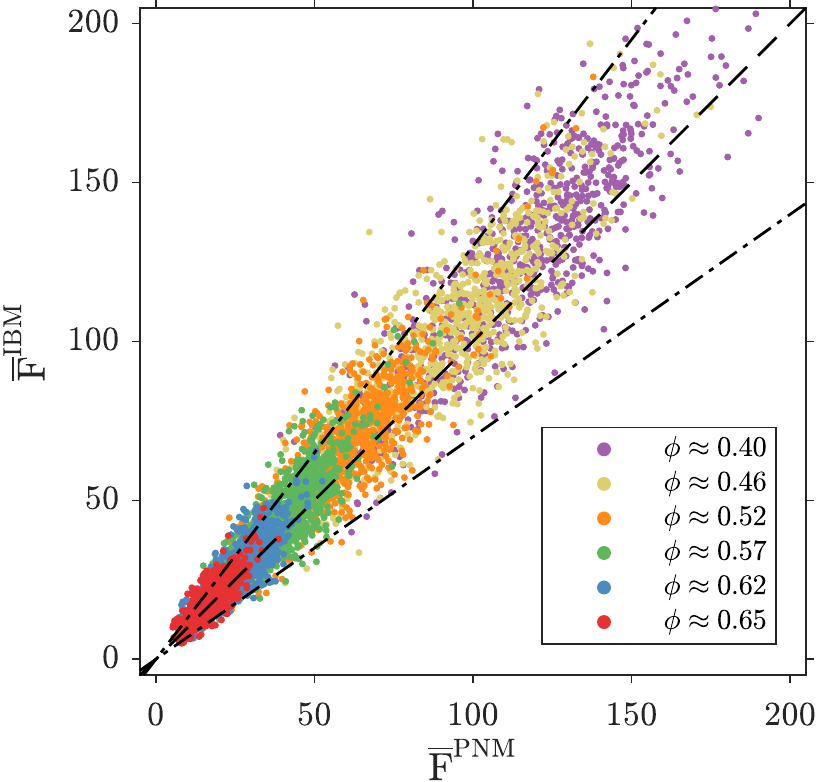}
        \caption{DT}
    \end{subfigure}
    \hfill
    \begin{subfigure}[b]{0.48\textwidth}
        \includegraphics[scale=0.90]{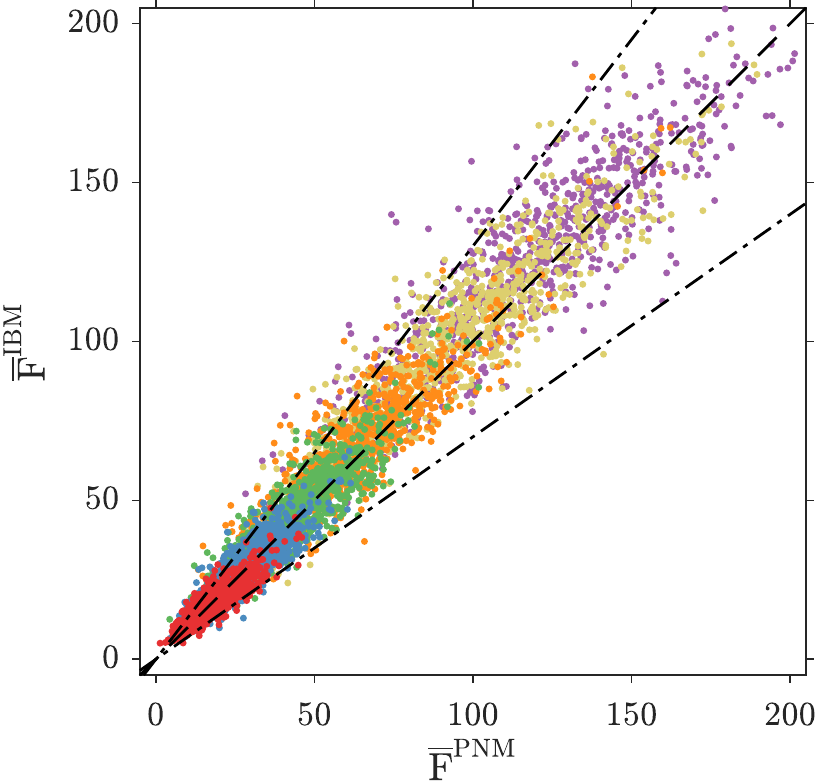}
        \caption{MDT}
    \end{subfigure}
    \caption{\label{fig:force-corr-linear-ptp}Correlation in normalised fluid-particle interaction force on individual particles in the PNM ($\overline{\mathrm{F}}^{\mathrm{PNM}}$) and the IBM ($\overline{\mathrm{F}}^{\mathrm{IBM}}$) for linear graded samples using the pore-throat-pore series conductance model in Sec.~\ref{subsub:pore-throat-pore-model}. Normalisation is conducted by $\overline{\mathrm{F}} = \frac{\mathrm{F}}{\mathrm{F}_{\mathrm{stokes}}}$, where $\mathrm{F}$ is the interaction force on a particle and $\mathrm{F}_{\mathrm{stokes}} = 3\pi \mu d U$ is the Stokes drag force on an individual particle ($\mu$ is the dynamic viscosity of the fluid, $d$ is the particle diameter and $U$ is the inlet velocity). The dashed-dotted line represents 30\% deviation from perfect correspondence between IBM and PNM.}
\end{figure}

\begin{figure}[!ht]
    \centering
    \begin{subfigure}[b]{0.48\textwidth}
        \includegraphics[scale=0.90]{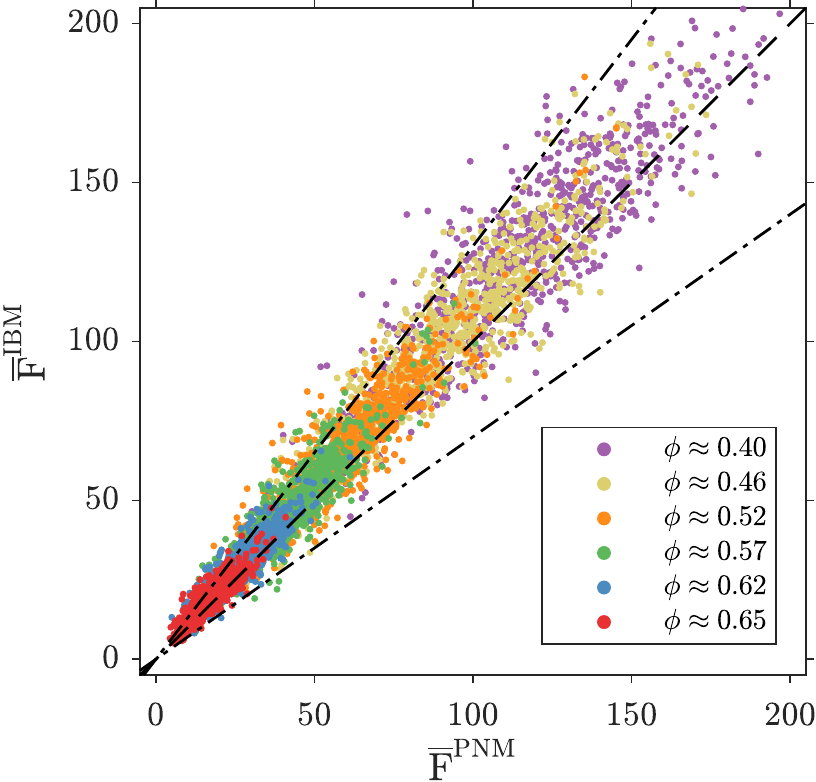}
        \caption{DT}
    \end{subfigure}
    \hfill
    \begin{subfigure}[b]{0.48\textwidth}
        \includegraphics[scale=0.90]{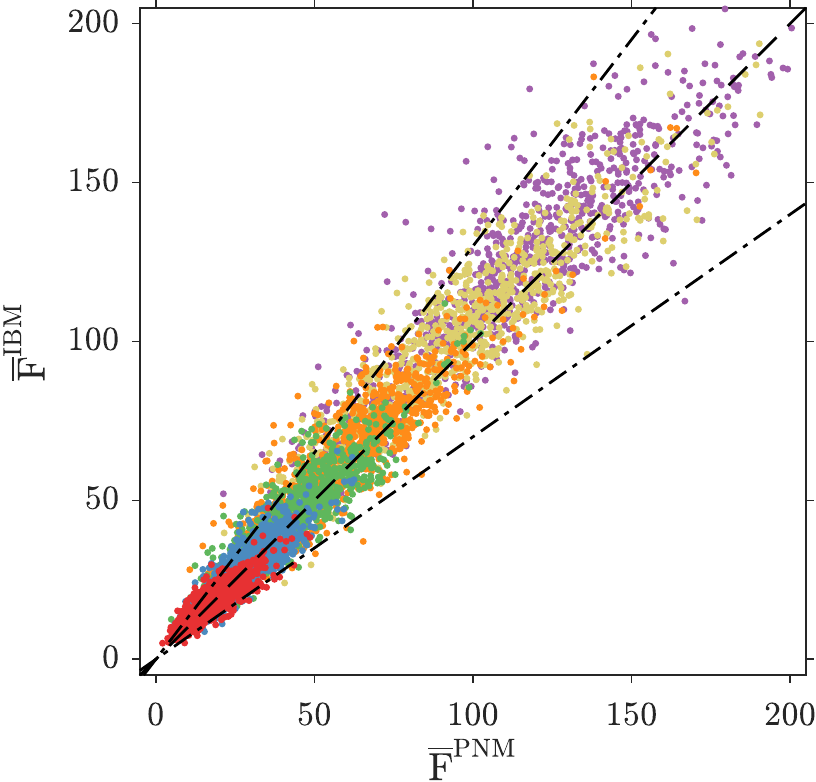}
        \caption{MDT}
    \end{subfigure}
    \caption{\label{fig:force-corr-linear-hr}Correlation in normalised fluid-particle interaction force on individual particles in the PNM ($\overline{\mathrm{F}}^{\mathrm{PNM}}$) and the IBM ($\overline{\mathrm{F}}^{\mathrm{IBM}}$) for linear graded samples using the hydraulic radius conductance model in Sec.~\ref{subsub:hydraulic-radius-model}. Normalisation is conducted as per Fig.~\ref{fig:force-corr-linear-ptp}. The dashed-dotted line represents 30\% deviation from perfect correspondence between IBM and PNM.}
\end{figure}

Comparison of normalised fluid-particle interaction force, $\overline{\Fm}^k_{fp}$, obtained from the PNM and IBM for the linear graded samples is shown in Fig.~\ref{fig:force-corr-linear-ptp}-\ref{fig:force-corr-linear-hr} for both conductance models, along with DT and MDT partitions. This demonstrates that the PNM can reasonably capture the fluid-particle interaction force with most data points within a $\pm 30\%$ error bound, where the force obtained from the PNM is generally lower than that observed in the IBM. In Figs.~\ref{fig:force-corr-linear-ptp}-\ref{fig:force-corr-linear-hr}, all cases are plotted on a single figure to show consistent trends between the samples, and Table~\ref{tab:corr-coeff-linear} provides the Pearson correlation coefficient data. For the pore-throat-pore series conductance model, the MDT provides a more accurate agreement between the IBM and PNM, as reflected by $\rho_\Fm^{\textrm{MDT}} \geq \rho_\Fm^{\textrm{DT}}$ in Tables~\ref{tab:corr-coeff-linear} with the correlation improving with increasing packing fraction. This is in line with observations for local pressure and flux correlations with this conductance model. In the hydraulic radius conductance model, both DT and MDT show similar degrees of correlation, with DT exhibiting slightly higher Pearson correlation values.

Table~\ref{tab:corr-coeff-linear} appears to suggest that the correlation improves within increasing $C_u$. However, this is an artefact of the definition of the Pearson correlation coefficient in Eq.~\ref{eqn:coeff-pearson}. At small $C_u$, there is only a small particle size ratio and hence, only a small range in $\overline{\Fm}^k_{fp}$. Consequently, small differences can have a significant affect on the Pearson correlation coefficient. This should be considered when interpreting the correlation shown in Tables~\ref{tab:corr-coeff-linear}-\ref{tab:corr-coeff-bimodal}. This could be resolved by performing simulations (PNM and IBM) at various pressure gradients (using the same particle configuration), which would in turn result in a fluid-particle interaction force over wider range. This was not possible within the computational limits of this study, but it would be anticipated that the Pearson correlation coefficient would improved substantially for lower $C_u$.

\begin{figure}[t]
    \centering
    \begin{subfigure}[b]{0.48\textwidth}
        \includegraphics[scale=0.90]{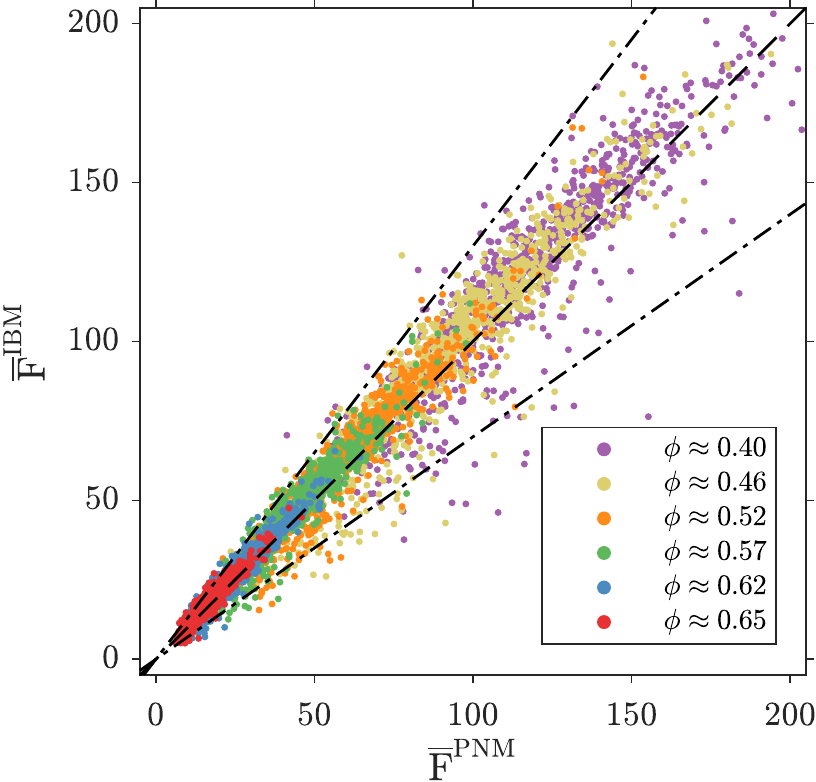}
        \caption{DT}
    \end{subfigure}
    \hfill
    \begin{subfigure}[b]{0.48\textwidth}
        \includegraphics[scale=0.90]{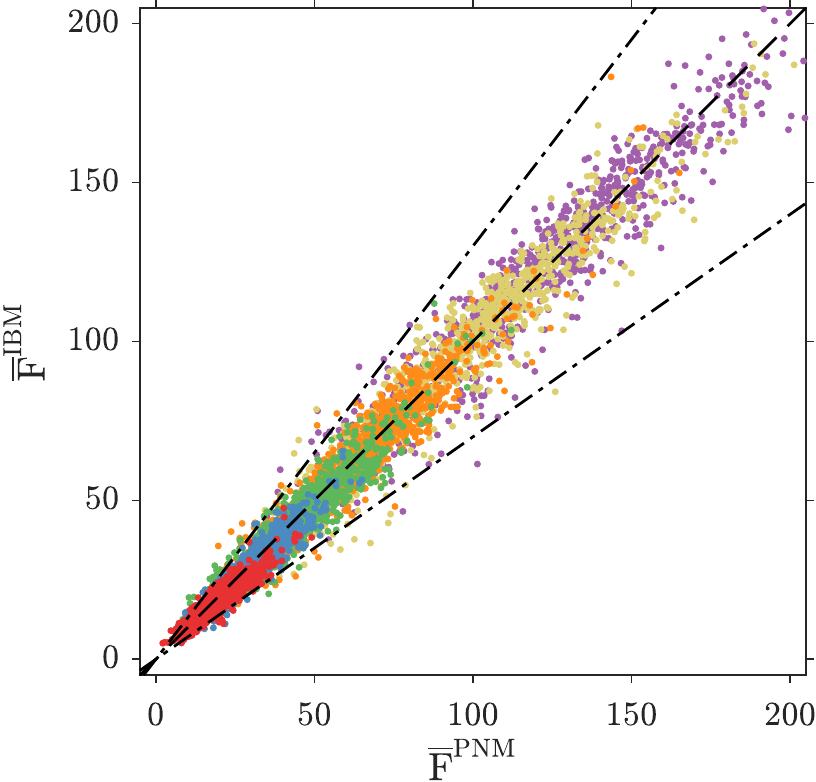}
        \caption{MDT}
    \end{subfigure}
    \caption{\label{fig:checkforce-corr-linear}Correlation in normalised fluid-particle interaction force on individual particles in the PNM ($\overline{\mathrm{F}}^{\mathrm{PNM}}$) and the IBM ($\overline{\mathrm{F}}^{\mathrm{IBM}}$) for linear graded samples using the nodal pressures obtained from the IBM simulation. Normalisation is conducted as per Fig.~\ref{fig:force-corr-linear-ptp}. The dashed-dotted line represents 30\% deviation from perfect correspondence between IBM and PNM.}
\end{figure}

The calculation of the fluid-particle interaction force using the method proposed by \citet{chareyre2012} relies on the pressure gradient across a constriction, along with the geometric properties of the constriction. In order to further test the simplifications and assumptions associated with the method, the normalised fluid-particle interaction force is recalculated by directly applying the nodal pressures obtained from the IBM simulations into the equations in Sec.~\ref{sub:Fluid-Particle-Interaction-Force} (instead of the calculated pressures obtained from the mass continuity equations in Eq.~\ref{eqn:network-continuity}). The correlation is shown in Fig.~\ref{fig:checkforce-corr-linear} and the corresponding Pearson correlations values are included in Table~\ref{tab:corr-coeff-linear} (see 'IBM' columns in the 'Force' table). Fig.~\ref{fig:checkforce-corr-linear} shows a strong correlation, particularly with the MDT partition, with data points well within the $\pm 30\%$ error bound, which is confirmed by the Pearson correlation coefficients with value generally higher than $0.85$. These observations provide support to the method developed in \citet{chareyre2012} and indicates that the underlying assumptions are appropriate for linear graded samples.

\begin{figure}[ht]
    \centering
    \begin{subfigure}[b]{0.48\textwidth}
        \includegraphics[scale=0.90]{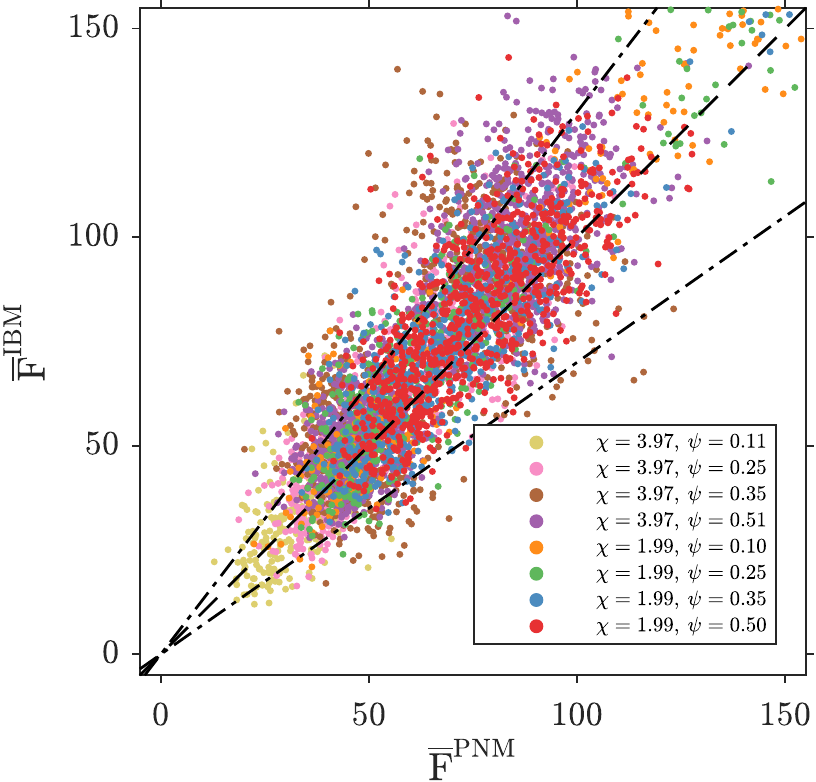}
        \caption{DT}
    \end{subfigure}
    \hfill
    \begin{subfigure}[b]{0.48\textwidth}
        \includegraphics[scale=0.90]{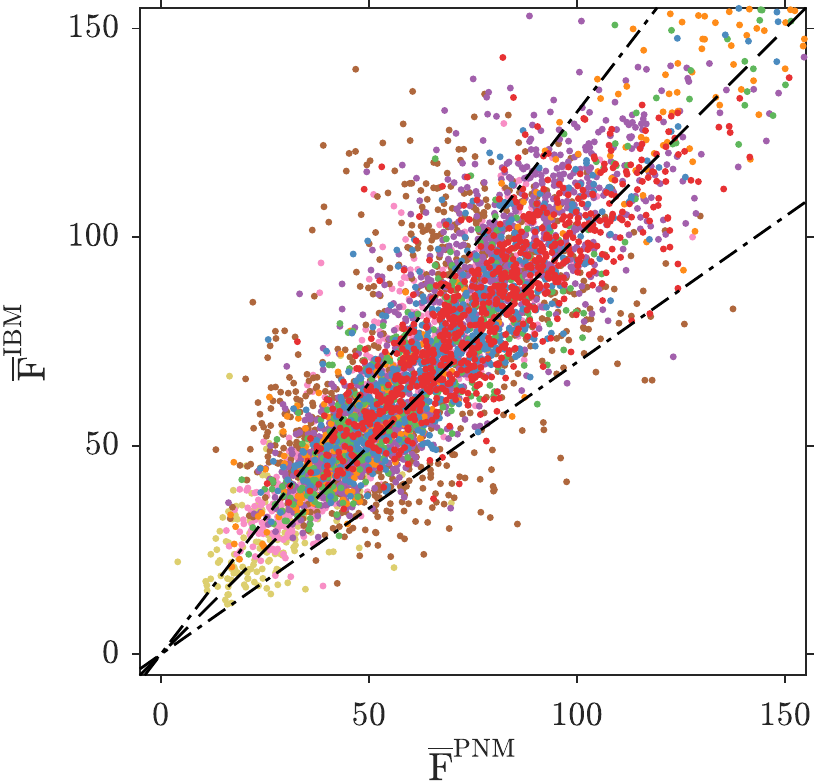}
        \caption{MDT}
    \end{subfigure}
    \caption{\label{fig:force-corr-bimodal-ptp}Correlation in normalised fluid-particle interaction force on individual particles in the PNM ($\overline{\mathrm{F}}^{\mathrm{PNM}}$) and the IBM ($\overline{\mathrm{F}}^{\mathrm{IBM}}$) for bimodal samples using the pore-throat-pore series conductance model in Sec.~\ref{subsub:pore-throat-pore-model}. Normalisation is conducted as per Fig.~\ref{fig:force-corr-linear-ptp}. The dashed-dotted line represents 30\% deviation from perfect correspondence between IBM and PNM.}
\end{figure}

\begin{figure}[!ht]
    \centering
    \begin{subfigure}[b]{0.48\textwidth}
        \includegraphics[scale=0.90]{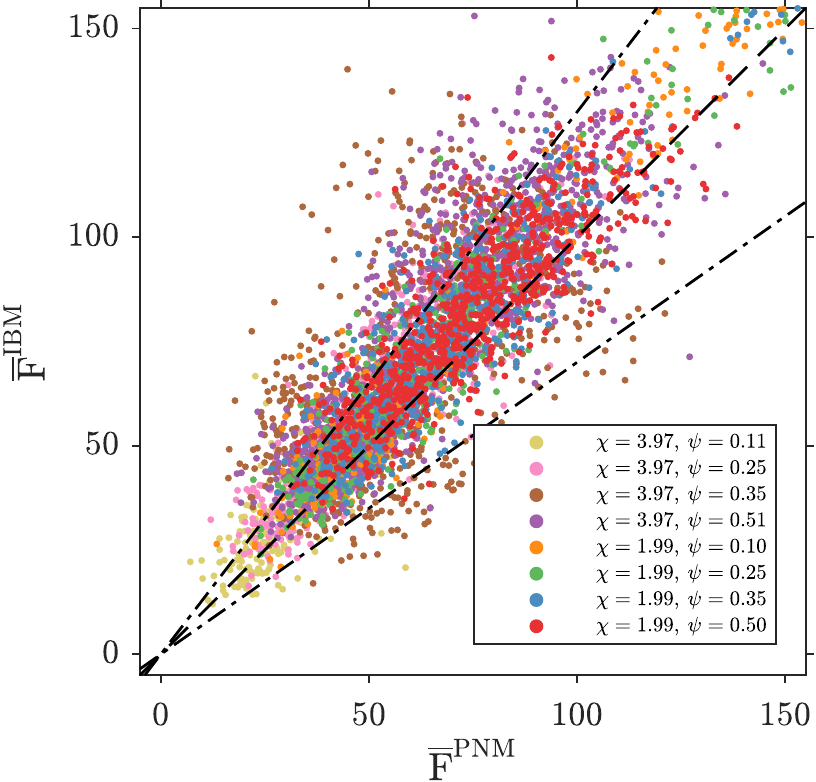}
        \caption{DT}
    \end{subfigure}
    \hfill
    \begin{subfigure}[b]{0.48\textwidth}
        \includegraphics[scale=0.90]{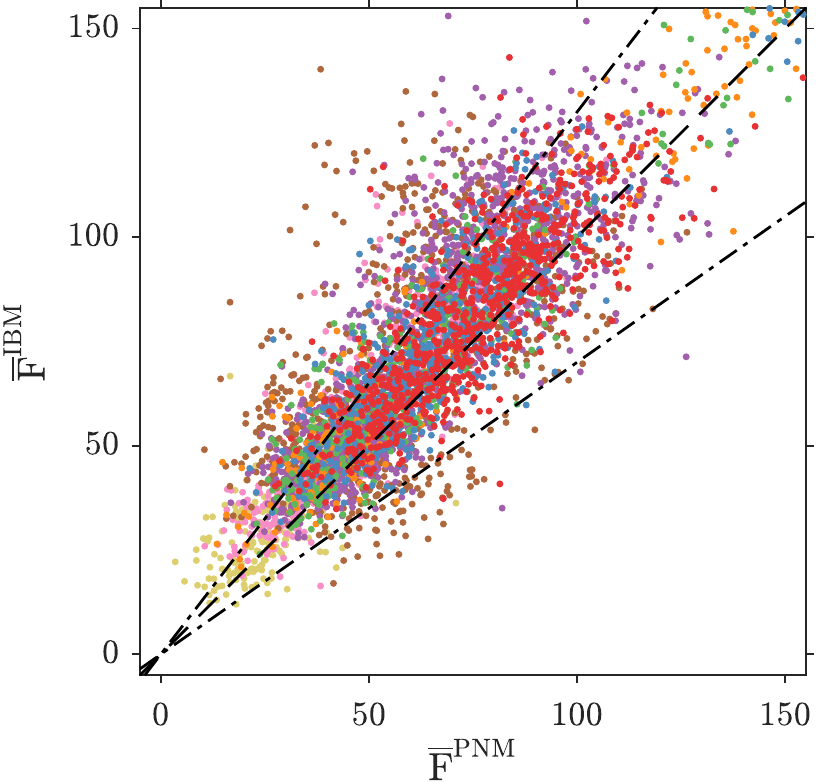}
        \caption{MDT}
    \end{subfigure}
    \caption{\label{fig:force-corr-bimodal-hr}Correlation in normalised fluid-particle interaction force on individual particles in the PNM ($\overline{\mathrm{F}}^{\mathrm{PNM}}$) and the IBM ($\overline{\mathrm{F}}^{\mathrm{IBM}}$) for bimodal samples using the hydraulic radius conductance model in Sec.~\ref{subsub:hydraulic-radius-model}. Normalisation is conducted as per Fig.~\ref{fig:force-corr-linear-ptp}. The dashed-dotted line represents 30\% deviation from perfect correspondence between IBM and PNM.}
\end{figure}

The comparison for the fluid-particle interaction force in the bimodal samples is shown in Figs.~\ref{fig:force-corr-bimodal-ptp}-\ref{fig:force-corr-bimodal-hr}. While the PNM can reasonably capture $\overline{\Fm}^k_{fp}$, there is significant scatter beyond the $\pm 30\%$ error bound shown in Figs.~\ref{fig:force-corr-bimodal-ptp}-\ref{fig:force-corr-bimodal-hr} for both tessellation methods (DT and MDT) and conductance models. The large scatter is most prevalent in the $\chi = 3.97$ samples. This is not reflected in the Pearson correlation coefficient in Table~\ref{tab:corr-coeff-bimodal}. In bimodal samples, the fine particles experience lower force, while the large particles experience higher force. This results in a very large variance in force, and consequently only very large errors will affect the correlation. Therefore, these values should be interpreted carefully. 

\begin{figure}[ht]
    \centering
    \begin{subfigure}[b]{0.48\textwidth}
        \includegraphics[scale=0.90]{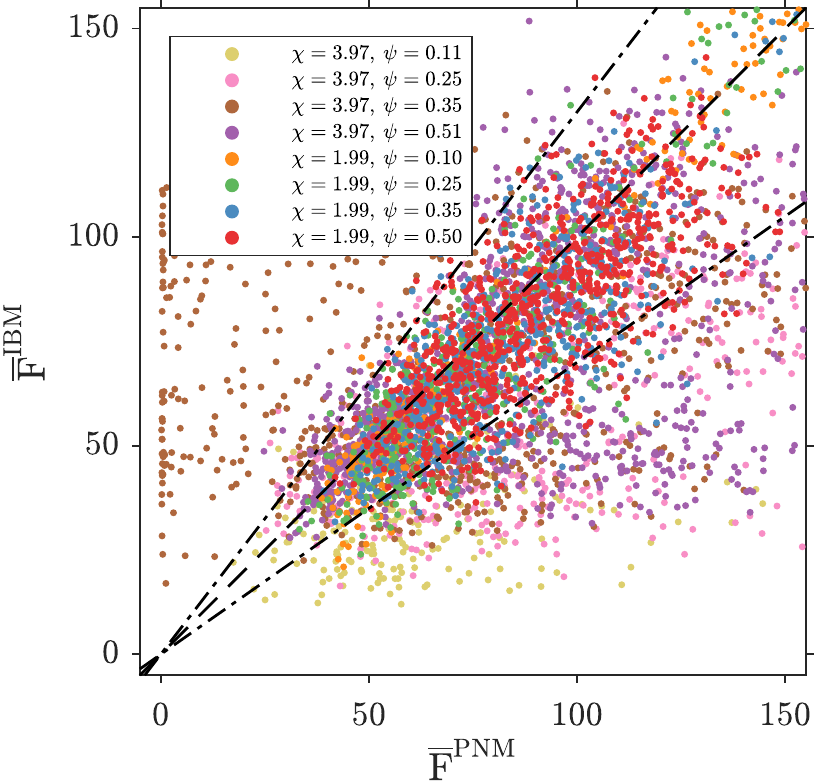}
        \caption{DT}
    \end{subfigure}
    \hfill
    \begin{subfigure}[b]{0.48\textwidth}
        \includegraphics[scale=0.90]{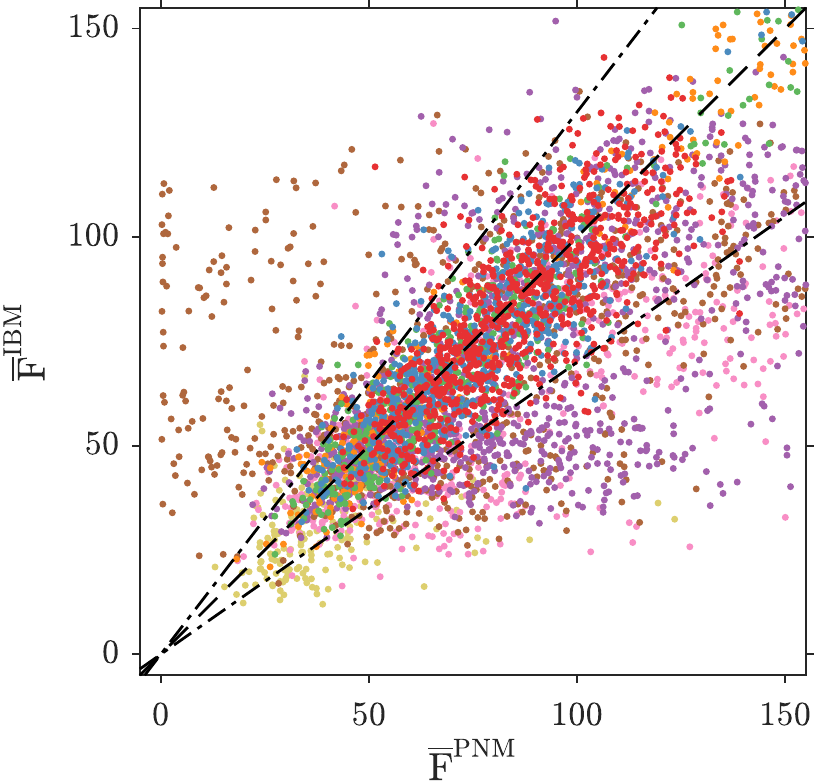}
        \caption{MDT}
    \end{subfigure}
    \caption{\label{fig:checkforce-corr-bimodal}Correlation in normalised fluid-particle interaction force on individual particles in the PNM ($\overline{\mathrm{F}}^{\mathrm{PNM}}$) and the IBM ($\overline{\mathrm{F}}^{\mathrm{IBM}}$) for bimodal samples using the nodal pressures obtained from the IBM simulation. Normalisation is conducted as per Fig.~\ref{fig:force-corr-linear-ptp}. The dashed-dotted line represents 30\% deviation from perfect correspondence between IBM and PNM.}
\end{figure}

Following Fig.~\ref{fig:checkforce-corr-linear}, the fluid-particle interaction force is recalculated using the IBM nodal pressures in the bimodal samples. The force correlation is shown in Fig.~\ref{fig:checkforce-corr-bimodal}. Quite significant fluctuations are observed for the cases with $\chi = 3.97$, which is a result of the relatively small size (particularly with the insufficient number of coarse particles) of these samples. These fluctuations are shown to alert readers of the need to have a representative element volume, which can increase significantly with increasing particle size ratio. By isolating the results for $\chi = 1.99$, it is apparent that the method developed in \citet{chareyre2012} can reasonably handle bimodal (and more generally gap-graded) samples. Nevertheless, the scatter observed in Fig.~\ref{fig:checkforce-corr-bimodal} for the bimodal cases is significantly more prominent that the minimal fluctuations observed in the linear graded samples (Fig.~\ref{fig:checkforce-corr-linear}). 

\section{Concluding Remarks}

This study addressed the ability of Pore Network Models (PNM) to capture the local pressure and flow field in granular, porous media, and the seepage induced drag (also referred to as the fluid-particle interaction force). Fully-resolved flow simulations were conducted using the Immersed Boundary Method (IBM) to obtain local pressure and flow fields at a sub-particle spatial resolution, along with the seepage induced forces acting on individual particles. The IBM simulations were conducted on linear graded samples with the coefficient of uniformity, $C_u$, ranging from $1.01$ to $2.00$, and a wide range of packing fractions were considered for each gradation. In addition, bimodal samples were considered with a particle size ratio of $1.99$ and $3.97$, and at various fines content, ranging from $0.10$ to $0.50$.

Using these linear graded and bimodal samples, PNM was generated using Delaunay-based methods. The weighted Delaunay tessellation (DT) was considered, which accounted for the different sized particles by using a weighting factor based on the particle radius. This weighted DT was essential in properly delineating the pore space in the $C_u = 2.00$ linear graded samples along with all bimodal samples, and hence, is the recommended approach for polydisperse sphere assemblies. In addition, the modified Delaunay tessellation (MDT) was considered, which merged tetrahedral cells according an inscribed sphere criterion to form polyhedral unit cells. The MDT better captures variation in pore and constriction sizes, along with removing the connectivity constraint in the DT.

Fluid flow in the network model was simulated using a Stokes flow algorithm, which ensured conservation of mass in the pore network. Two local conductance models were compared: (i) pore-throat-pore series element; and (ii) equivalent hydraulic radius approach. These conductance models considered the geometry of pores and constrictions in assigning a local conductivity to the network edge elements. The pressure at each node in the network, along with the flux (or flow rate) along each edge of the network was determined using the Stokes flow algorithm, with boundary conditions matching the fully-resolved IBM simulations.

Comparison of the sample permeability showed that the PNM based on the hydraulic radius conductance model was able to closely match the permeability obtained from the IBM for both DT and MDT partitions. In the pore-throat-pore series conductance model, the MDT accurately captured the sample permeability, while the DT showed significant differences. The local pressure and flow field were also compared. Across the wide range of packing fractions considered in both linear-graded and bimodal samples, the local pressure field could be accurately captured by the PNM using either conductance models and for both the DT or MDT. However, the local flux showed more sensitivity to the choice of the conductance model and the method of partitioning the pore space (i.e. DT or MDT). In the pore-throat-pore series conductance model, local flux was reasonably captured using the MDT, while the DT showed significant differences with the IBM. In the hydraulic radius conductance model, DT and MDT methods exhibited similar behaviour (with the MDT being slightly more accurate). 

Preferential flow channels were compared in the IBM and PNM by obtaining the shortest paths from the inlet to outlet nodes. A high degree of correspondence was observed between the IBM and PNM shortest paths, despite the differences in local fluxes. As flow characteristics of a granular, porous medium are dominated by the shortest paths, as they carry the majority of the flux of fluid through the sample, these results provide strong evidence supporting the use of PNM. Furthermore, streamline profiles demonstrated that pressure drops in the granular, porous medium coincided with the pore constrictions identified in the MDT, providing further evidence that the MDT provides a better representation of the pore space and should be considered when generating pore networks.

In addition, the fluid-particle interaction force was obtained from the PNM using the methodology proposed in \citet{chareyre2012}. This method was rigorously validated by comparing against the analytical solutions in \citet{zickhomsy1982} for simple cubic, body centred cubic and face centred cubic array of spheres, which showed that the MDT was essential in accurately capturing the fluid-particle interaction force, although the DT provided a reasonable estimate. Following this validation, the force obtained from the PNM and IBM were compared for the linear graded and bimodal samples. The PNM provided a reasonable estimate of the interaction force for linear graded samples with most particles having an error bound of $\pm 30\%$. However, the comparison was less promising for the bimodal samples, which exhibited a wider scatter. Interestingly, calculation of the fluid-particle interaction force was relatively insensitive to the network generation method with the MDT being slightly more accurate than the DT.

The differences in the local flux field could be attributed to the conductance model implemented, which may be adjusted to provide a better correlation. For example, different cross-sectional shapes could be considered in the hydraulic radius model, including a popular approach known as the circle-triangle-square (CTS) method \citep{patzek2001}, along with the consideration of adjustable parameters in the pore-throat-pore series conductance model as suggested by \citet{sholokhova2009}. Nevertheless, the choice of this conductance model is quite arbitrary, and this remains an unresolved feature of modelling pore networks. The findings in this study suggests that conductance models calibrated to match the overall permeability will reasonably capture the local pressure and flow field, as is the case for the PNM based on the MDT, but further research is required to support this statement. An alternate approach is to use machine learning techniques to obtain a conductance model without shape simplifications, as suggested by \citet{miao2017}, and this is a promising direction to resolve the issues surrounding the choice of the conductance model.

Differences in the calculated fluid-particle interaction force with the IBM results (particularly in the bimodal samples) may be attributed to the pore space discretisation, choice of the conductance model, or the method proposed in \citet{chareyre2012}. The force computation was relatively insensitive to the choice of the partitioning method (i.e. DT or MDT) or the choice of the conductance models considered in this study. The methodology in \citet{chareyre2012} was further tested by directly applying the IBM nodal pressures in the force calculation, which exhibited excellent correlation in the linear graded samples. However, the bimodal samples exhibited some fluctuations despite the fact that the local pressure field was well captured. Therefore, this study provides strong support for the method developed in \citet{chareyre2012} which is capable of capturing the fluid-particle interaction force, but further research is required to improve the accuracy of the method (particularly in bimodal or gap-graded particle assemblies) by further investigating discretisation methods, local conductance models and the underlying assumptions and simplifications in the force computation. 

\section*{Acknowledgements}
AS was funded by EPSRC grant EP/P010393/1. CK was funded through a studentship in the Centre for Doctoral Training on Theory and Simulation of Materials at Imperial College London funded by the EPSRC (EP/G036888/1). DD acknowledges support via the EPSRC Established Career Fellowship (EP/N025954/1). This work used the ARCHER UK National Supercomputing Service (\url{http://www.archer.ac.uk}) through grant e549.

\bibliographystyle{ms}
\bibliography{ms}

\end{document}